# Slice-Less Optical Arbitrary Waveform Measurement (OAWM) in a Bandwidth of More than 600 GHz Using Soliton Microcombs

Daniel Drayss[1,2,+], Dengyang Fang[1], Christoph Füllner[1], Grigory Lihachev[3],
Thomas Henauer[4], Yung Chen[1], Huanfa Peng[1], Pablo Marin-Palomo[1], Thomas Zwick[4],
Wolfgang Freude[2], Tobias J. Kippenberg[3], Sebastian Randel[1] and Christian Koos[1,2,++]

[1]*Institute of Photonics and Quantum Electronics (IPQ), Karlsruhe Institute of Technology (KIT),*
*76131 Karlsruhe, Germany*
[2]*Institute of Microstructure Technology (IMT), Karlsruhe Institute of Technology (KIT),*
*76344 Eggenstein-Leopoldshafen, Germany*
[3]*Institute of Physics, Swiss Federal Institute of Technology Lausanne (EPFL),*
*CH-1015 Lausanne, Switzerland*
[4]*Institute of Radio Frequency Engineering and Electronics (IHE), Karlsruhe Institute of Technology (KIT),*
*76131 Karlsruhe, Germany*
[+]*daniel.drayss@kit.edu* [++]*christian.koos@kit.edu*

**Abstract:** We propose and demonstrate a novel scheme for optical arbitrary waveform measurement (OAWM) that exploits chip-scale Kerr soliton combs as highly scalable multi-wavelength local oscillators (LO) for ultra-broadband full-field waveform acquisition. In contrast to earlier concepts, our approach does not require any optical slicing filters and thus lends itself to efficient implementation on state-of-the-art high-index-contrast integration platforms such as silicon photonics. The scheme allows to measure truly arbitrary waveforms with high accuracy, based on a dedicated system model which is calibrated by means of a femtosecond laser with known pulse shape. We demonstrated the viability of the approach in a proof-of-concept experiment by capturing an optical waveform that contains multiple 16 QAM and 64 QAM wavelength-division multiplexed (WDM) data signals with symbol rates of up to 80 GBd, reaching overall line rates of up to 1.92 Tbit/s within an optical acquisition bandwidth of 610 GHz. To the best of our knowledge, this is the highest bandwidth that has so far been demonstrated in an OAWM experiment.

## 1. Introduction

Optical arbitrary waveform measurement (OAWM) based on frequency combs [1-7] has the potential to unlock a wide variety of applications, ranging from reception of high-speed data signals [3-11] and elastic optical networking [6] to investigation of ultra-short events and photonic-electronic analog-to-digital conversion [12-15]. Previous demonstrations of OAWM relied on spectrally sliced coherent detection [1-7], where optical filters are used to decompose a broadband input signal into several spectral slices. These slices are then individually detected by an array of in-phase and quadrature (IQ) receivers using local-oscillator (LO) tones derived from a phase-locked frequency comb as a common reference, and the optical waveform is reconstructed by stitching of the received spectral slices through digital signal processing (DSP). Based on this concept, OAWM of a 228 GHz-wide signal was demonstrated using discrete components [2], and further work demonstrated a 320 GHz photonic-electronic analog-to-digital converter that combines spectrally sliced OAWM with high-speed electro-optic modulators [14]. However, all these schemes crucially rely on high-quality optical filters for spectral slicing of the optical signal and for separating the comb tones. While the IQ receivers can be efficiently integrated using readily available platforms such as silicon photonics (SiP) [16,17] or indium phosphide (InP) [18], high-quality slicing filters are much more challenging to implement on high index-contrast photonic integration platforms. Specifically, previous demonstrations of integrated

OAWM receivers either relied on InP-based arrayed waveguide gratings (AWG), that required individual phase correction in the various arms [6], or on SiP coupled-resonator optical waveguide (CROW) structures [7,19], that need sophisticated control schemes for thermal tuning. In addition, most of the previously demonstrated OAWM schemes [1-7] were based on frequency combs generated by RF modulation of a narrowband laser tone. This approach requires complex sequences of broadband electro-optic modulators, each driven with a dedicated radio-frequency signal, and limits the number of achievable comb tones and thus the acquisition bandwidth of the OAWM scheme.

In this paper, we propose and experimentally demonstrate an OAWM scheme that does not require optical slicing filters – neither for the signal nor for the LO [10,11] – and that exploits chip-scale Kerr soliton combs as highly scalable multi-wavelength LO. For detection, our scheme relies on an array of IQ receivers, which are fed by the full optical waveform as well as by time-delayed copies of the full LO comb. The electrical output signals of the IQ receivers then contain superimposed mixing products of the various LO tones with the respective adjacent portions of the signal spectrum and allow to reconstruct the full-field information of the incoming waveform using advanced DSP. In a proof-of-concept experiment, we implement the scheme using a chip-scale dissipative Kerr soliton comb as multi-wavelength LO. Our experiment relies on a precise frequency-domain model of the OAWM system that accounts for the complex-valued transfer functions of the various detection paths and that is instrumental for high-fidelity reconstruction of truly arbitrary waveforms. We calibrate our model by measuring these transfer functions using a femtosecond laser with a known pulse shape. The viability of the scheme is demonstrated by simultaneously acquiring multiple 16 QAM and 64 QAM wavelength-division multiplexed (WDM) data signals with symbols rates of up to 80 GBd, reaching overall line rates of up to 1.92 Tbit/s within an optical acquisition bandwidth of 610 GHz [10]. To the best of our knowledge, this is the highest bandwidth that has so far been demonstrated in an OAWM experiment.

## 2. Concept of slice-less optical arbitrary waveform measurement (OAWM)

*Hardware and signal acquisition*

The fundamental concept of a slice-less OAWM system is illustrated in Fig. 1. The scheme combines a chip-scale frequency comb generator such as a Kerr soliton comb [20-27], see Inset 2, with a receiver system, that does not contain any slicing filters and that is thus amenable to implementation on state-of-the-art high-index-contrast integration platforms such as silicon photonics [11]. The incoming optical arbitrary waveform $\underline{a}_S(t)$ with Fourier spectrum $\underline{\tilde{a}}_S(f)$ is split into $N$ copies, which are routed to an array of IQ receivers, labeled by subscripts $\nu = 1,\ldots,N$. The IQ receivers are also fed with time-delayed copies (delays $\tau_\nu$) of the LO comb comprising $M$ tones with complex amplitudes $\underline{A}_{\mathrm{LO},\mu}$ at optical frequencies $f_\mu, \mu = 1,\ldots,M$, separated by the free spectral range (FSR) of the comb,

$$\underline{a}_{\mathrm{LO}}(t) = \sum_{\mu=1}^{M} \underline{A}_{\mathrm{LO},\mu} \exp(j2\pi f_\mu t). \tag{1}$$

The $2N$ output signals of all IQ receivers are digitized with an array of analog-to-digital converters (ADC), and an estimate of the envelope of the arbitrary optical waveform $\underline{a}_S(t)$ is reconstructed via DSP. Each IQ receiver (IQR) comprises a 90° optical hybrid (OH) and two pairs of balanced photodetectors (BPD) and is read out by two corresponding ADC, to obtain the in-phase component $I_\nu(t)$ and the quadrature component $Q_\nu(t)$ of the respective baseband signal, see Inset 1 in Fig. 1. All IQ receivers and ADC have a bandwidth of $B$, within which mixing products of the various LO tones with the respective adjacent portions of the signal spectrum can be detected. For simplicity, we assume for now that the overall transfer function for the in-phase $I_\nu(t)$ and quadrature $Q_\nu(t)$ components are identical for each IQR $\nu$, and that they have an ideal 90° phase relationship. In real systems, this is not necessarily the case, such that a more advanced system model is needed, see Section 1.1 of Supplement 1 for a more detailed description of a general system model.

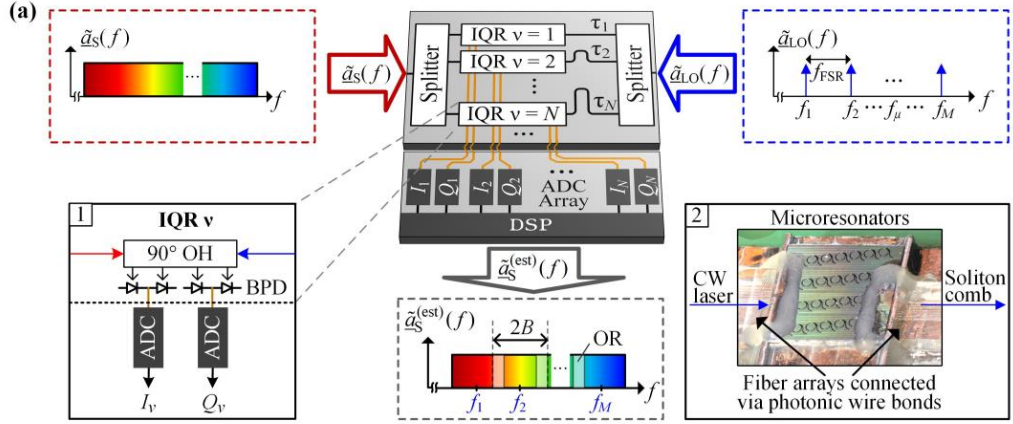

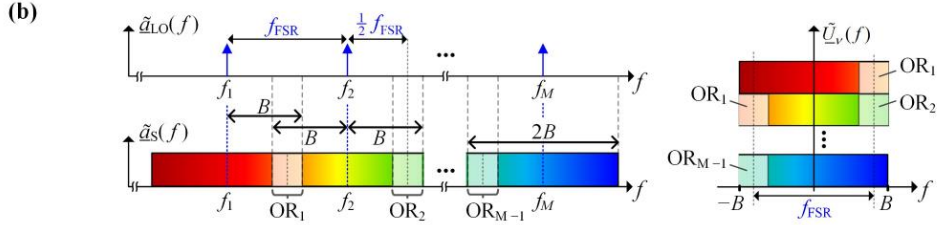

**Fig. 1**. Concept of slice-less optical arbitrary waveform measurement (OAWM) using a chip-scale frequency comb generator such as a Kerr soliton microcomb as a multi-wavelength LO. The scheme does not require any slicing filters and thus lends itself to integration on high-index-contrast photonic platforms such as silicon photonics. **(a)** The optical input signal with spectrum $\tilde{a}_S(f)$ and the LO comb with spectrum $\tilde{a}_{LO}(f)$ are split into $N$ copies. The $N$ copies of the LO comb, each comprising $M$ narrowband tones spaced by a free spectral range $f_{FSR}$, are then delayed by distinct delays $\tau_\nu$ before being fed to $N$ in-phase and quadrature receivers (IQR), where they act as LO for coherent detection of the respective signal portion. The in-phase ($I_\nu$) and quadrature ($Q_\nu$) outputs of the IQ receivers are read out by $2N$ corresponding analog-to-digital converters (ADC). Digital signal processing (DSP) is applied to recover an estimate $\tilde{a}_S^{(est)}(f)$ of the optical input spectrum. **Inset 1:** Schematic of IQR $\nu$ with a 90° optical hybrid (OH), a pair of balanced photodetectors (BPD) and corresponding ADC (bandwidth $B$), providing the in-phase and quadrature components $I_\nu(t)$ and $Q_\nu(t)$ of the respective baseband signals. **Inset 2:** Illustration of a chip-scale Kerr soliton comb source, comprising a continuous-wave pump laser and a high-Q Kerr-nonlinear microresonator. Kerr soliton microcombs provide a multitude of optical tones with narrow linewidths that are typically spaced by for free-spectral range $f_{FSR}$ of tens of GHz, thus offering a highly compact attractive option for scaling the bandwidth of the scheme. **(b)** The composite baseband signals $\tilde{U}_\nu(f) = \tilde{I}(f) + j\tilde{Q}(f)$ contain a superposition of the individual mixing products of each LO tone with the spectrally adjacent portion of the signal within a bandwidth $B$ to either side of the respective LO tone $f_1, f_2, \ldots, f_M$. Exploiting the fact that the various portions of the LO are subject to different time delays $\tau_\nu$ and that individual LO tones thus feature distinct phase differences, we can separate the superimposed mixing products by DSP. Note that the receiver bandwidth $B$ is chosen slightly larger than half the FSR, $B > f_{FSR}/2$, leading to $M-1$ overlap regions $OR_1, OR_2, \ldots, OR_{M-1}$ (shaded stripes). Within these overlap regions, the same spectral portion of the signal is transferred to two distinct portions at the edges of the baseband signal spectrum close to $f = \pm\frac{1}{2}f_{FSR}$. This leads to redundant information that can be used to estimate random phase drifts along the various detection paths.

*Waveform reconstruction*

Based on the acquired in-phase $I_\nu(t)$ and quadrature $Q_\nu(t)$ components obtained from the various IQ receivers, we construct the complex-valued composite baseband signals $\underline{U}_\nu(t)$,

$$\underline{U}_\nu(t) = I_\nu(t) + \mathrm{j}Q_\nu(t). \tag{2}$$

Importantly, as the optical arbitrary waveform $a_\mathrm{S}(t)$ is simultaneously down-converted with all LO tones, each complex-valued composite baseband signal $\underline{U}_\nu(t)$ contains a superposition of independent components, generated by mixing each LO tone with the spectrally adjacent portion of the signal, see Fig. 1 (b) for a graphical illustration. The baseband spectra $\underline{\tilde{U}}_\nu(f)$ can hence be expressed by a superposition of various contributions that are directly related to frequency-shifted copies $\underline{\tilde{a}}_\mathrm{S}(f+f_\mu)$ of the signal spectrum $\underline{\tilde{a}}_\mathrm{S}(f)$, where $f_\mu$ denotes the LO tone frequency used to down-convert the respective signal portion. The finite bandwidth $B$ of the corresponding IQ receivers leads to a low-pass filtering of these signal portions. By merging the low-pass filters and the optical transfer characteristics of the various detection paths with index $\nu=1,\ldots,N$ as well as the complex-valued amplitudes $\underline{A}_{\mathrm{LO},\mu}$ of the individual LO tones with index $\mu=1,\ldots,M$ into equivalent transfer functions $\underline{\tilde{H}}_{\nu\mu}(f)$, the relationship between the frequency-shifted portions of the optical input signal and the $N$ recorded electrical baseband spectra $\underline{\tilde{U}}_\nu(f)$ can be written in matrix form. Accounting additionally for electrical noise $\tilde{G}_\nu(f)$ of the various receivers $\nu=1,\ldots,N$, this leads to a relation of the form

$$\begin{pmatrix}\underline{\tilde{U}}_1(f)\\ \vdots \\ \underline{\tilde{U}}_N(f)\end{pmatrix} = \begin{pmatrix}\underline{\tilde{H}}_{1,1}(f) & \cdots & \underline{\tilde{H}}_{1,M}(f)\\ \vdots & \ddots & \vdots \\ \underline{\tilde{H}}_{N,1}(f) & \cdots & \underline{\tilde{H}}_{N,M}(f)\end{pmatrix} \begin{pmatrix}\underline{\tilde{a}}_\mathrm{S}(f+f_1)\\ \vdots \\ \underline{\tilde{a}}_\mathrm{S}(f+f_M)\end{pmatrix} + \begin{pmatrix}\tilde{G}_1(f)\\ \vdots \\ \tilde{G}_N(f)\end{pmatrix}. \tag{3}$$

$$\underline{\tilde{\mathbf{U}}}(f) = \qquad\qquad \underline{\tilde{\mathbf{H}}}(f)\underline{\tilde{\mathbf{A}}}_\mathrm{S}(f) \qquad + \quad \tilde{\mathbf{G}}(f)$$

Note that the expressions $\tilde{G}_\nu(f)$ only account for electrical noise at the respective receiver, whereas the optical noise entering the receiver is regarded as part of the input signal $\underline{\tilde{a}}_\mathrm{S}(f)$. The electrical receiver noise spectra $\tilde{G}_\nu(f)$ of the various receivers comprise thermal noise, shot noise, and quantization noise, and are assumed to have identical power spectral densities and to be statistically independent. These assumptions allow to use a simple least-square estimator as a special case of a minimum-variance unbiased estimator for reconstructing the incoming signal, see Section 1.2 of Supplement 1 for a more detailed explanation. In essence, our signal reconstruction technique relies on calculating a least-squares estimate $\underline{\tilde{\mathbf{A}}}_\mathrm{S}^{(\mathrm{est})}(f)$ of the various frequency-shifted copies $\underline{\tilde{a}}_\mathrm{S}(f+f_\mu)$ in a first step, which are then merged into the overall reconstructed waveform $\underline{\tilde{a}}_\mathrm{S}^{(\mathrm{est})}(f)$ in a second step. To calculate the least-squares estimate $\underline{\tilde{\mathbf{A}}}_\mathrm{S}^{(\mathrm{est})}(f)$, we multiply the vector $\underline{\tilde{\mathbf{U}}}(f)$ of the received baseband signals with the pseudo-inverse $\underline{\tilde{\mathbf{H}}}^+ = (\underline{\tilde{\mathbf{H}}}^\dagger\underline{\tilde{\mathbf{H}}})^{-1}\underline{\tilde{\mathbf{H}}}^\dagger$ of the transfer matrix $\underline{\tilde{\mathbf{H}}}(f)$ [28];

$$\underline{\tilde{\mathbf{A}}}_\mathrm{S}^{(\mathrm{est})}(f) = \underline{\tilde{\mathbf{H}}}^+(f)\underline{\tilde{\mathbf{U}}}(f) = \underline{\tilde{\mathbf{A}}}_\mathrm{S}(f) + \underline{\tilde{\mathbf{A}}}_\mathrm{G}(f) \quad \text{for } |f|<B, \tag{4}$$

where

$$\underline{\tilde{\mathbf{A}}}_\mathrm{G}(f) = \underline{\tilde{\mathbf{H}}}^+(f)\mathbf{G}(f). \tag{5}$$

The reconstructed signal vector $\underline{\tilde{\mathbf{A}}}_\mathrm{S}^{(\mathrm{est})}(f)$ is thus impaired by a noise term $\underline{\tilde{\mathbf{A}}}_\mathrm{G}(f)$ representing the receiver noise $\tilde{\mathbf{G}}(f)$, modified by the pseudo inverse $\underline{\tilde{\mathbf{H}}}^+(f)$. In these relations, we disregard noise impairments of the incoming optical signal – they are considered part of the signal – and we assume perfectly balanced photodiodes as well as an LO with a sufficiently large optical carrier-to-noise ratio (OCNR) such that optical noise contributions from the LO can be neglected. If this condition is not fulfilled then the reconstructed signal will additionally be impaired by multiplicative noise from the LO, see Section 7 of Supplement 1 for a detailed noise and distortion characterization. Note that the pseudo-inverse of the $(M, N)$-transfer matrix $\underline{\tilde{\mathbf{H}}}(f)$ can be calculated if the number $N$ of IQ receivers corresponds at least to the number $M$ of comb tones, $N \geq M$, which leads to an overall optical acquisition bandwidth of $B_{\text{opt}} = (M-1)f_{\text{FSR}} + 2B$.
Once the estimated signal vector $\underline{\tilde{\mathbf{A}}}_S^{(\text{est})}(f) = \left[\underline{\tilde{a}}_{S,1}^{(\text{est})}(f+f_1),\ldots,\underline{\tilde{a}}_{S,M}^{(\text{est})}(f+f_M)\right]^T$ is digitally reconstructed, its components $\underline{\tilde{a}}_{S,1}^{(\text{est})}(f+f_1),\ldots,\underline{\tilde{a}}_{S,M}^{(\text{est})}(f+f_M)$ are frequency-shifted back to their relative original position according to the respective LO frequencies $f_1,\ldots,f_M$ and are spectrally stitched by DSP, resulting in an estimated signal spectrum, $\underline{\tilde{a}}_S^{(\text{est})}(f) = \underline{\tilde{a}}_S(f) + \underline{\tilde{a}}_G(f).$, where $\underline{\tilde{a}}_G(f)$ represents again the reconstructed receiver noise in analogy to Eq. (5) above. To keep the sampling rate in the DSP algorithms manageable, we avoid working at optical carrier frequencies and introduce a global frequency shift of the entire signal spectrum by a reference frequency $f_{\text{ref}}$, effectively reconstructing the frequency-shifted spectrum $\underline{\tilde{a}}_S^{(\text{est})}(f+f_{\text{ref}})$ of the envelope of the arbitrary input waveform $\underline{a}_S(t)$.

Importantly, the bandwidth $B$ of the IQ receivers slightly exceeds half of the comb FSR $f_{\text{FSR}}$ such that after frequency-shifting according to the respective LO frequencies neighboring spectral slices $a_{S,\mu}^{(\text{est})}(f)$ and $a_{S,\mu+1}^{(\text{est})}(f)$, $\mu = 0,\ldots,M-1$ exhibit an overlap region (OR) $f_\mu^{(\text{OR})} \in [f_{\mu+1}-B, f_\mu+B]$, see shaded stripes in Fig. 1 (a) and (b). The redundant signal information in these overlap regions can be exploited to digitally compensate for phase drifts within the optical system and for phase, amplitude, and FSR fluctuations of the LO comb, see Section 1.3 and 1.4 of Supplement 1 as well as Section "System calibration" below. Once these impairments are compensated, the redundant signal portions can be merged by calculating a weighted average of the spectral components $a_{S,\mu}^{(\text{est})}(f)$ and $a_{S,\mu+1}^{(\text{est})}(f)$ within the overlap regions, where the weights are chosen according to the noise background of the respective spectral component as to maximize the SNR of the resulting superposition, see Section 1.2 in Supplement 1 for details.

*System calibration and compensation of phase drifts*

For a practical implementation of the slice-less OAWM system, the frequency-dependent transfer matrix $\underline{\tilde{\mathbf{H}}}(f)$ must be determined in a separate calibration step. Importantly, the transfer functions $\underline{\tilde{H}}_{\nu\mu}(f)$ are impacted by amplitude and phase drifts of the comb lines and by random phase drifts in the setup. For Kerr combs pumped with highly stable lasers, the optical linewidth typically amounts to a few kHz, and mechanical vibrations in our setup are also limited to a similar frequency range [29]. The associated amplitude and phase changes hence occur on a time scale of hundreds of microseconds, and the transfer function $\underline{\tilde{\mathbf{H}}}(f)$ may be considered constant during one recording with a typical length of a few microseconds. Therefore, we split each transfer-matrix element $\underline{\tilde{H}}_{\nu\mu}(f)$ into a time-invariant, but frequency-dependent part $\underline{\tilde{H}}_{\nu\mu}^{(f)}(f)$, and into a time-variant, but frequency-independent part $\underline{H}_{\nu\mu}^{(t)}$,

$$\underline{\tilde{H}}_{\nu\mu}(f) = \underline{\tilde{H}}_{\nu\mu}^{(f)}(f) \times \underline{H}_{\nu\mu}^{(t)}, \qquad \underline{H}_{\nu\mu}^{(t)} = \underline{H}_{F,\nu}^{(t)} \times \underline{H}_{LO,\mu}^{(t)}. \qquad (6)$$

The time-variant, but frequency-independent matrix elements $\underline{H}_{\nu\mu}^{(t)}$ can be expressed by a product of a first $\nu$-dependent complex-valued parameter $\underline{H}_{F,\nu}^{(t)}$, which accounts for the time-dependent phase drift in the fiber leading to IQR $\nu$, and of a second $\mu$-dependent complex-valued parameter $\underline{H}_{LO,\mu}^{(t)}$, which represents the amplitude and phase fluctuation of the $\mu$-th comb tone. These parameters are estimated by using the redundant information contained in the overlap regions of the various slices, see Section 1.4 of Supplement 1 for details. For measuring the time-invariant frequency-dependent parts $\underline{\tilde{H}}_{\nu\mu}^{(f)}(f)$ of the overall transfer functions, we perform a one-time calibration by feeding the system with a known optical reference waveform (ORW). The ORW is derived from an ultra-stable femtosecond laser (MENHIR-1550) with well-defined pulse shape and an FSR $f_{\text{ORW}} \ll f_{\text{FSR}}$, which is not an integer fraction of $f_{\text{FSR}}$. A step-by-step description of the calibration procedure can be found in Section 4 of Supplement 1.

*System conditioning and relation to time-domain optical sampling*

To allow for a high-quality signal reconstruction according to Eq. (4), the transfer matrix $\underline{\tilde{\mathbf{H}}}(f)$ must be well-conditioned, which can be achieved by choosing approximately equidistant time delays, see Section 1.5 of Supplement 1 for a more detailed discussion,

$$\tau_\nu \approx \frac{(\nu-1)T_{\text{LO}}}{N}, \qquad \nu = 1, 2, \ldots N, \quad T_{\text{LO}} = \frac{1}{f_{\text{FSR}}}. \tag{7}$$

This relation can be intuitively understood when considering the special case of strictly equidistant time delays, $\tau_\nu = (\nu-1)T_{\text{LO}}/N$, in combination with chirp-free LO signals, for which all comb lines have the same initial phases, $\arg(\underline{A}_{\text{LO},\mu}) = \text{const.} \ \forall \mu$. In this case, the LO signals arriving at the various IQ receivers can be interpreted as short transform-limited pulses, and the mixing process with the signal can be understood as a time-interleaved linear optical sampling process [8,9,30], see Section 2 of Supplement 1 for details. In contrast to those techniques, however, the frequency-domain method presented in this paper inherently compensates for imperfectly chosen time delays $\tau_\nu$, for chirped LO pulses, for relative drifts of the optical phase between IQ receivers or LO comb lines, and for the exact frequency-dependent transfer characteristics of the various detection channels. This allows for measuring truly arbitrary waveforms without restriction to data signals with known structure, as was the case for previously demonstrated time-interleaved linear optical sampling techniques [8,9].

**3. Experimental demonstration**

To demonstrate the viability of the proposed OAWM scheme, we perform a proof-of-concept experiment using a setup based on discrete fiber-optic components, see Fig. 2 (a). The LO comb source (blue box) comprises a Kerr soliton comb with a pump laser L8, an erbium-doped fiber amplifier (EDFA) to boost the comb power, and a wavelength-selective switch (WSS) to select four adjacent comb lines. The free spectral range (FSR) of the comb is 110 GHz, see Fig. 2 (b) for the associated spectrum taken at point Ⓑ. The optical carrier-to-noise ratio (OCNR) of the individual LO comb lines was measured with respect to a reference bandwidth of 12.5 GHz, corresponding to a wavelength interval of 0.1 nm at a center wavelength of 1550 nm, and ranges from 23 dB to 24 dB. Variable optical delay lines are used to adjust the time delays $\tau_\nu$ to the IQ receiver array. For the first IQ receiver (IQR 1) we use balanced photodiodes (BPD) with a nominal 3 dB bandwidth of 43 GHz and an actual 12 dB bandwidth of 80 GHz, while the nominal BPD bandwidth amounts to 100 GHz for the remaining IQ receivers (IQR 2 … IQR 4). The outputs of IQR 1 and IQR 2 are digitized with a 100 GHz oscilloscope (Keysight UXR1004A), while an 80 GHz oscilloscope (Keysight UXR0804A) was used for IQR 3 and IQR 4. Both oscilloscopes are synchronized. For the signal reconstruction, we digitally limit the RF frequency range to $B = 80$ GHz for all eight channels. The input signal is either given by a known optical reference waveform (ORW) Ⓒ, used for calibrating the system, or by a broadband test signal Ⓐ for demonstrating the viability of the scheme. The ORW is derived from a highly stable mode-locked laser (Menhir 1550, Menhir Photonics AG) and features a smooth spectral amplitude and phase profile, see Fig. 2 (b) Ⓒ. The broadband test signal comprises seven wavelength-division multiplexing (WDM) data channels and is configured such that the full optical acquisition bandwidth $B_{\text{opt}}$ is occupied. The test signal is generated by modulating four optical carriers (L1 to L4) with a first and three optical carriers (L5 to L7) with a second IQ modulator. The driving signals are generated by a 120 GSa/s arbitrary-waveform generator (AWG, Keysight M8194A) and amplified by electrical amplifiers.

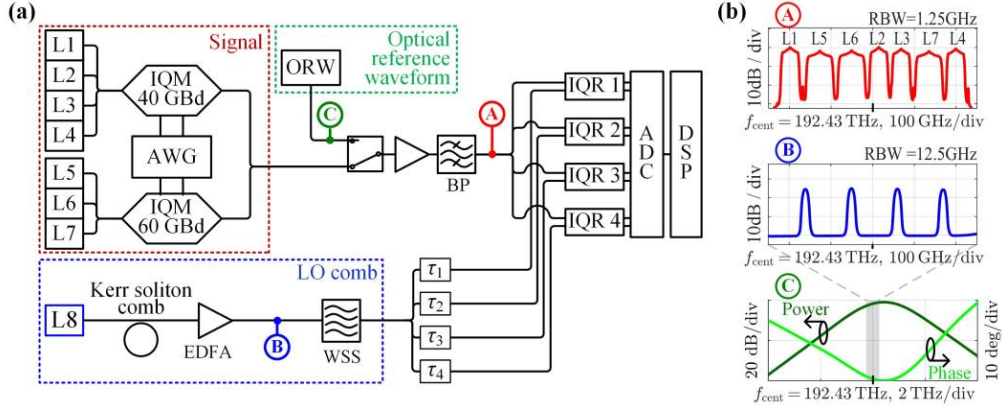

**Fig. 2**. Proof-of-concept experiment demonstrating slice-less optical arbitrary waveform measurement (OAWM) using Kerr soliton combs as multi-wavelength local oscillator (LO). **(a)** Experimental setup based on discrete fiber-optic components: A broadband optical signal is generated by modulating seven spectrally interleaved lasers (L1 to L7) with two different IQ modulators (IQM) that are driven with different signals, derived from a common arbitrary-waveform generator (AWG, Keysight M8194A). The output signals of the IQM are combined, and the resulting broadband optical test signal is then amplified and routed through a band-pass filter (BP) to the IQ receiver (IQR) array of the OAWM system. At the LO side, four coherent tones are isolated from a Kerr frequency comb (LO comb), split into four paths, delayed by distinct time intervals $\tau_1,\ldots\tau_4$, and routed to the IQ receiver array. For calibrating the OAWM system, a known optical reference waveform (ORW) derived from a highly stable mode-locked laser is fed to the input of the OAWM system. The ORW features a smooth spectral amplitude and phase, see Subfigure (b) **(b)** Exemplary optical spectra of the broadband test signal at point Ⓐ, of the LO comb with FSR $f_{FSR} = 110\,\text{GHz}$ at point Ⓑ, and of the ORW at point Ⓒ. The ORW was generated by a solid-state mode-locked femtosecond laser (Menhir 1550), and the spectral amplitude and phase were measured by Menhir Photonics using frequency-resolved optical gating (FROG). All plots are shown with respect to the same center frequency $f_{cent} = 192.43\,\text{THz}$.

*System calibration*

For calibration our OAWM system, we independently measure the frequency-dependent part $\underline{\tilde{\mathbf{H}}}^{(f)}(f)$ of the overall transfer matrix $\underline{\tilde{\mathbf{H}}}(f)$, see Eq. (6), using a known reference waveform. Note that the implementation of the OAWM scheme in our experiment deviates slightly from the simplified notation introduced in Sect. 2 above, in which the in-phase and quadrature components of the each IQ receiver were merged into a common complex-valued baseband signal, see Eq. (2). Specifically, this merge requires identical transfer functions and an ideal 90° phase relationship of the in-phase and the quadrature detection path, which, in practice, is usually not fulfilled since the underlying balanced detectors typically have slightly different characteristics. We therefore reformulate Eq. (4) in terms of the signals $I_\nu(t)$ obtained from the in-phase detector and the signals $Q_\nu(t)$ obtained from the quadrature detector, see Section 1.1 of Supplement 1 for details. This reformulation leaves the overall concept unchanged but allows for compensating slight differences of the in-phase and quadrature detection paths during system calibration. In the following, the complex-valued frequency-dependent and time-invariant transfer functions of the in-phase and quadrature components are referred to as $\tilde{\underline{H}}_{\nu\mu}^{(I)}(f)$ and $\tilde{\underline{H}}_{\nu\mu}^{(Q)}(f)$, respectively. These transfer functions are measured by feeding an ORW with a comb spectrum of known amplitude and phase to the OAWM system and by extracting the amplitudes and phases of the beat notes with the LO comb at the in-phase and the quadrature detectors of the various IQ receivers, see

Section 4.3 of Supplement 1. Figure 3 (a) shows a measured example of the magnitude ($\left|\tilde{\underline{H}}_{1,2}^{(\mathrm{I})}(f)\right|$, red) and the phase ($\varphi_{1,2}^{(\mathrm{I})}$, blue) of the transfer function $\tilde{\underline{H}}_{1,2}^{(\mathrm{I})}(f) = \left|\tilde{\underline{H}}_{1,2}^{(\mathrm{I})}(f)\right|\exp\left(\mathrm{j}\varphi_{1,2}^{(\mathrm{I})}\right)$, that is associated with the in-phase detection path of the first IQ receiver (IQR 1) and the second LO comb line at frequency $f_2$. To validate our calibration technique for the magnitude, we replace the ORW in Fig. 2 (a) by a tunable external cavity lacer (ECL), ECL1, and we substitute the LO comb by another fixed-frequency ECL2 emitting at frequency $f_2$, see Section 4.1 of Supplement 1. We sweep ECL1 in discrete frequency steps over the full detection bandwidth of the IQ receivers and record the amplitudes of the generated sinusoidal signals. This allows to extract the magnitude of the frequency response, indicated by a green trace in Fig. 3 (a), which is only partially visible since it coincides very well with the results of the ORW calibration. Note that the quality the calibration measurements crucially relies on the linearity of the underlying photodetectors and that detector saturation should hence be avoided, see Sections 4.3 and 4.4 of Supplement 1 for a more detailed discussion. In Fig. 3 (b), we show zoomed-in sections of the amplitude and phase transfer functions between 70 GHz and 78 GHz, indicated by a red and blue box in Fig. 3 (a), and visualize the results from 58 individual calibration recordings that were taken over a period of 2.5 h. We observe stable frequency-dependent transfer characteristics, where the frequency-dependent fluctuations are indeed a time-invariant property of the OAWM detection system. This validates our assumptions associated with Eq. (6).

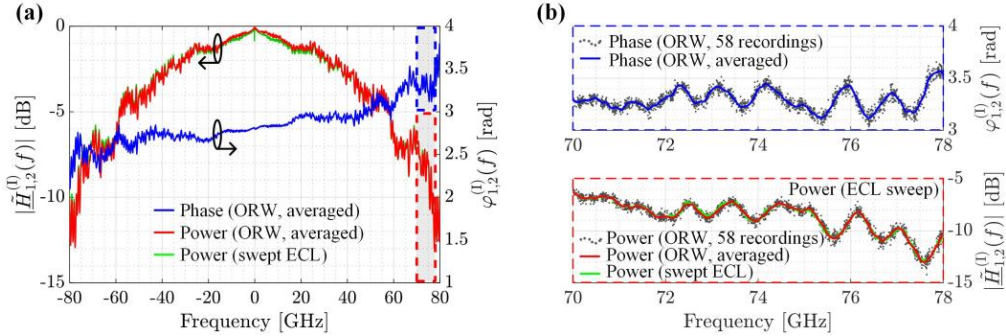

**Fig. 3.** System-calibration measurements. **(a)** Example for the measured magnitude $\left|\tilde{\underline{H}}_{1,2}^{(\mathrm{I})}(f)\right|$ (red) and the corresponding phase $\varphi_{1,2}^{(\mathrm{I})}(f)$ (blue) of the complex-valued transfer function $\tilde{\underline{H}}_{1,2}^{(\mathrm{I})}(f)$ as obtained from a calibration measurement with a known ORW. The transfer function $\tilde{\underline{H}}_{1,2}^{(\mathrm{I})}(f)$ is associated with the in-phase detection path of IQR 1 and the down-conversion via second LO comb line at frequency $f_2$. For verification, we re-measure the power transfer function by sweeping an external-cavity laser across a single-tone LO at frequency $f_2$ - the results are indicated by a green trace and reproduce the ORW-based measurement very well. **(b)** Zoom-in into the phase (blue) and magnitude (red, green) of the transfer function $\tilde{\underline{H}}_{1,2}^{(\mathrm{I})}(f)$ for the frequency region from 70 GHz to 78 GHz, as indicated by a frame in Subfigure (a). The results from 58 individual calibration recordings taken over a period of 2.5 hours are plotted as dark dots. We observe stable frequency dependent transfer characteristics, where the frequency-dependent fluctuations are indeed a time-invariant property of the OAWM detection system. This validates our assumptions associated with Eq. (6). Note that the transfer function $\tilde{\underline{H}}_{1,2}^{(\mathrm{I})}(f)$ shown here is associated with IQR 1 and has a rather low nominal bandwidth of 43 GHz. The IQ receivers IQR 2, IQR 3, and IQR 4 have nominal bandwidths of 100 GHz, leading to a much smaller frequency-dependent roll-off, see Fig. S13 in Section 4.3 of Supplement 1 for the power transfer functions associated with IQR 2, IQR 3, and IQR 4.

*Demonstration of broadband waveform measurements*

We finally demonstrate the capability of the slice-less OAWM system by measuring broadband test signals that comprise seven wavelength-division multiplexed (WDM) data channels. In a first experiment, we use an LO comb with a free spectral range of $f_{\mathrm{FSR}} = 110$ GHz and a test signal that consists of four 40 GBd 64 QAM signals modulated on carriers L1 to L4 and three 60 GBd 16 QAM signals modulated on the carriers L5 to L7, see Ⓐ in Fig. 2 (b). We successfully separate the superimposed mixing products associated with the various LO tones and reconstruct the optical spectrum $\tilde{\underline{a}}_S^{(\mathrm{est})}(f)$, the normalized magnitude of which is plotted in Fig. 4 (a) with a resolution bandwidth of 100 MHz. The overlapping spectral slices associated with the $\mu$-th LO tone $f_\mu$ are annotated with braces labeled $\mu = 1,\ldots,4$, and the data channels are labeled according to the respective optical carrier L1 to L7. For reference, we separately record the acquisition noise $\tilde{\underline{G}}_{\mathrm{acq}}(f)$ comprising the thermal noise and the quantization noise contributed by the various ADC by disconnecting all optical receiver inputs. The recordings of the acquisition noise are processed in the same way as the signal recordings. Note that this processing introduces a frequency dependence of the reconstructed noise power spectral density $\left|\tilde{\underline{a}}_G(f)\right|^2$, which is caused by two effects: First, the photodetector responses are equalized by multiplying the spectrally white acquisition noise $\tilde{\underline{G}}_{\mathrm{acq}}(f)$ with the frequency-dependent pseudo inverse $\tilde{\underline{H}}^+(f)$, see Eq. (4), which increases the noise for high frequencies, i.e., further away from the center frequency of each slice. Second, the complex-valued spectral portions in the overlap regions are weighted and averaged. Because the noise contributions of adjacent reconstructed slices $\tilde{\underline{a}}_{S,\mu}^{(\mathrm{est})}(f)$ and $\tilde{\underline{a}}_{S,\mu+1}^{(\mathrm{est})}(f)$ are uncorrelated, the resulting noise power within the overlap regions is reduced by up to 3 dB [31]. This effect does not occur at the lower and upper edge of the overall spectral acquisition range, because no overlap with an adjacent slice exists, thus leading to a slightly increased noise floor, Fig. 4(a). Note that the shot noise, which is unavoidably present in the signal recordings, is significantly smaller than measured acquisition noise, such that we can assume with good approximation that $\tilde{\underline{G}}(f) \approx \tilde{\underline{G}}_{\mathrm{acq}}(f)$ in Eq. (3).

The constellation diagrams as well as the estimated constellation SNR (CSNR) of all seven WDM data channels are shown in the bottom row of Fig. 4 (a). The CNSR corresponds to the square of the reciprocal error-vector magnitude (EVM) normalized to the average signal power $\mathrm{EVM_m}$ [32,33], $\mathrm{CSNR_{dB}} = 10 \times \log_{10}\left(1/\mathrm{EVM_m}^2\right)$. We find that the reconstructed data signals are slightly impaired by amplified spontaneous emission (ASE) noise in the LO signal, which leads to a pronounced blurring of the outer constellation points since the LO noise affects the signal in multiplicative form [34]. This ASE noise originates from the EDFA that is used to boost the power of the soliton comb, see Fig. 2 (a), which originally has a power of only -31 dBm per comb line, see Section 5 of Supplement 1 for an in-depth analysis of the LO comb. Because we use different filter bandwidths to suppress this ASE noise in between neighboring comb lines, the noise is more pronounced for the LO tones $f_1$ and $f_4$ and the corresponding data channels L1 and L4. This problem can be overcome by using dark or bright Kerr soliton combs with higher conversion efficiency and correspondingly higher per-line power, as demonstrated in recent experiments [35,36].

In a second experiment, we increase the optical acquisition bandwidth to $B_{\mathrm{opt}} = 610$ GHz by using an LO comb with an FSR of 150 GHz and again $M = 4$ comb tones for coherent detection. We also increase the bandwidth of the individual data channels such that the broadband test signal fills the full optical acquisition bandwidth. All data channels can again be recovered with good SNR, Fig. 4(b). Note that the periodic increase of the reconstructed and stitched acquisition noise $\tilde{\underline{a}}_G(f)$ (gray) in the reconstructed spectrum is now more pronounced due to a stronger roll-off of approximately 10 dB of the bandwidth-limited IQR 1 (43 GHz photodetectors) at half the FSR, $f_{FSR}/2 = 75$ GHz, see Fig. 3 for the corresponding transfer function. Still, these results correspond to the highest bandwidth that has so far been demonstrated in an OAWM experiment [1-7] while offering a greatly simplified scheme that does not require any slicing filters [10].

In a last step, we perform an in-depth analysis of various impairments that are relevant for the proposed OAWM scheme. In this analysis, we benchmark the proposed OAWM system against channel-by-channel reception of WDM signals with a series of independent IQ receivers, investigate the impact of the peak-to-average power ratio (PAPR) of different signals, estimate the crosstalk associated with the digital separation of the superimposed mixing products in the detected composite baseband spectra $\tilde{U}_\nu(f)$, and finally quantify the impact of multiplicative noise originating from the ASE noise in the LO spectrum. Further details and results of this study can be found in Sections 6-9 of Supplement 1.

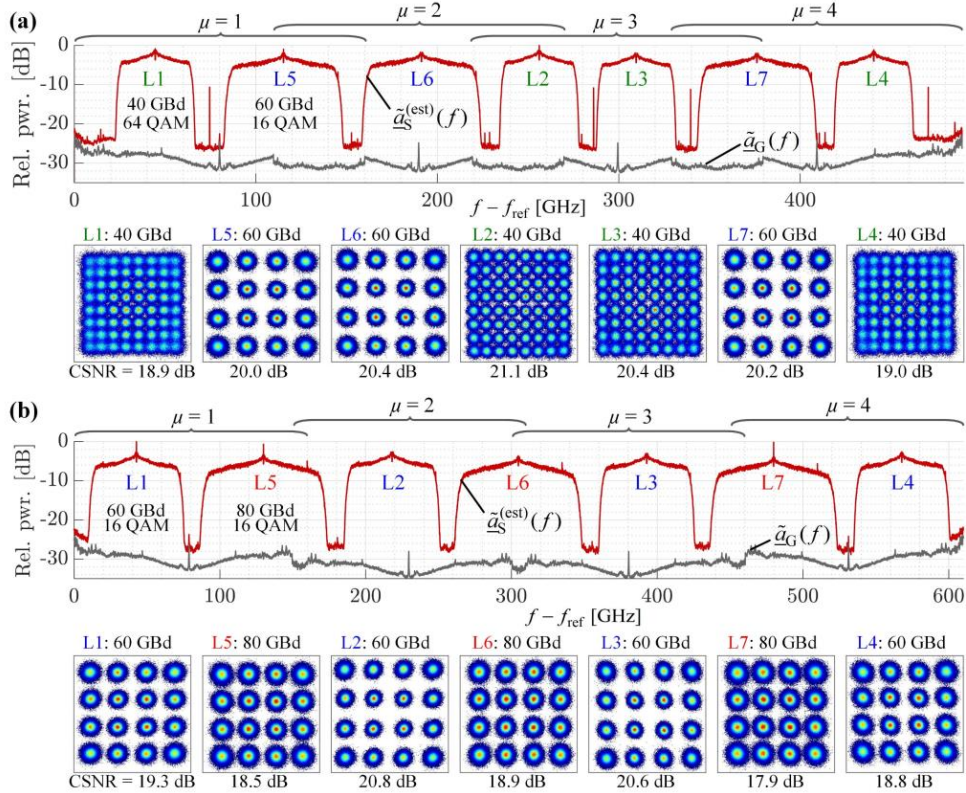

**Fig. 4**. Experimental results for demonstration experiments with optical acquisition bandwidths of 490 GHz and 610 GHz. **(a)** Normalized 490 GHz-wide power spectrum of a reconstructed waveform $|\tilde{a}_S^{(est)}(f)|^2$ (top, red trace) along with the constellation diagrams of the individual WDM channels (bottom) and the constellation signal-to-noise ratios (CSNR). The horizonal axis indicates the offset of the optical frequency $f$ from a reference frequency $f_{ref}$, which, without loss of generality, was chosen to correspond to the lower frequency edge of the first spectral slice, $f_{ref} = f_1 - B \approx 192.52$ THz. The reconstructed optical waveform contains four 40 GBd 64 QAM signals modulated on carriers L1 to L4 and three 60 GBd 16 QAM signals modulated on carriers L5 to L7, see Inset Ⓐ of Fig. 2. The power spectrum $|\tilde{a}_G(f)|^2$ of the stitched acquisition noise is indicated in gray. **(b)** Normalized power spectrum of a reconstructed 610 GHz-wide waveform $|\tilde{a}_S^{(est)}(f)|^2$ (red) along with the corresponding constellation diagrams (bottom row). The reconstructed optical waveform contains four 60 GBd 16 QAM signals (L1 to L4) and three 80 GBd 16 QAM signals (L5 to L7). The power spectrum $|\tilde{a}_G(f)|^2$ of the stitched acquisition noise is again indicated in gray. The horizontal axis indicates the offset from the lower-frequency edge of the first spectral slice, $f_{ref} \approx 192.41$ THz. All shown power spectra are smoothed with a 100 MHz wide moving average filter.

## 4. Summary

We demonstrated a technique for optical arbitrary waveform measurement (OAWM) that exploits chip-scale Kerr soliton combs as highly scalable multi-wavelength local oscillator (LO) and that does not require any optical slicing filters – neither for the signal nor for the LO. This greatly simplifies the implementation and the operation of the scheme and paves a path towards efficient integration on readily available platforms such as silicon photonics (SiP). Our scheme allows for precise reconstruction of truly arbitrary waveforms and relies on an accurate frequency-domain model of the detection system that is measured by means of a highly stable femtosecond laser with known pulse shape. We demonstrated the viability of the approach in a proof-of-concept experiment by capturing an optical waveform that contains multiple 16 QAM and 64 QAM wavelength-division multiplexed (WDM) data signals with symbol rates of up to 80 GBd, reaching overall line rates of up to 1.92 Tbit/s within a detection bandwidth of 610 GHz. To the best of our knowledge, this is the highest bandwidth so far been demonstrated in an OAWM experiment.

**Funding.** This work was supported by the ERC Consolidator Grant TeraSHAPE (# 773248), by the EU H2020 project TeraSlice (# 863322), by the DFG projects PACE (# 403188360) and GOSPEL (# 403187440) within the Priority Programme SPP 2111, by the joint DFG-ANR project HybridCombs (# 491234846), by the DFG Collaborative Research Center (CRC) WavePhenomena (SFB 1173, Project-ID 258734477), by the BMBF project Open6GHub (# 16KISK010), by the EU H2020 Marie Skłodowska-Curie Innovative Training Network MICROCOMB (# 812818), by the Alfried Krupp von Bohlen und Halbach Foundation, by the MaxPlanck School of Photonics (MPSP), and by the Karlsruhe School of Optics & Photonics (KSOP).

# Supplementary Document

This document provides supplementary information to "Slice-Less Optical Arbitrary Waveform Measurement (OAWM) in a Bandwidth of More than 600 GHz Using Soliton Microcombs". In Section 1, we derive the detailed frequency-domain system model that forms the base for digital signal reconstruction and discuss the compensation of optical phase drifts at the receiver. Sections 2 and 3 compare the slices-less concept to previous demonstrations of time-interleaved optical sampling and spectrally sliced OAWM, respectively. The system calibration techniques are explained in Section 4, which also details the measured transfer functions of all IQ receiver channels used in the experiments. Details of our setup for generation of dissipative Kerr soliton combs and associated optical carrier-to-noise ratios (OCNR) are given in Section 5. In Section 6, we use our OAWM system to simultaneously acquire a multitude of wavelength-division multiplexed (WDM) data signals of different optical signal-to-noise ratios (OSNR), and we benchmark the resulting signal quality against channel-by-channel reception of individual WDM signals. Section 7 presents single-tone measurements along with an analysis of the associated distortions and noise in the setup, and Section 8 gives more details on the noise of the oscilloscopes used to digitize all waveforms. Section 9 describes how the simultaneous down-conversion of a broadband signal with various LO tones affects the peak-to-average power ratio and how we can use amplitude clipping to maximize the overall constellation signal-to-noise ratio (CSNR) of acquired data signals.

## 1. System model and signal reconstruction

### 1.1. System model

In this section we derive a mathematical description for the OAWM system illustrated in Fig.1 of the main manuscript, which is the basis for the signal reconstruction. Throughout our documents, we use lowercase letters $\psi(t)$ for signals oscillating at optical carrier frequencies and uppercase letters $\Psi(t)$ for the associated complex-valued envelopes (baseband signals), which in many cases correspond to electrical signals obtained from in-phase/quadrature (IQ) receivers or fed to IQ modulators. Fourier transforms $\tilde{\psi}(f)$ are indicated with a tilde, and complex-valued time-domain signals $\underline{\psi}(t)$ and their Fourier transforms $\underline{\tilde{\psi}}(f)$ have an underscore. Vectors and matrices are written using bold symbols $\boldsymbol{\Psi}$ and a superscript "T" denotes the transpose $\boldsymbol{\Psi}^{\mathrm{T}}$ of the respective vector or matrix. The optical input signal $\underline{a}_{\mathrm{S}}(t)$ with envelope $\underline{A}_{\mathrm{S}}(t)$, its Fourier transform $\underline{\tilde{a}}_{\mathrm{S}}(f)$, and the local oscillator (LO) $\underline{a}_{\mathrm{LO}}(t)$, its Fourier transform $\underline{\tilde{a}}_{\mathrm{LO}}(f)$, and the associated complex-valued amplitudes $\underline{A}_{\mathrm{LO},\mu}$ of the various comb tones are given by,

$$\underline{a}_{\mathrm{S}}(t) = \underline{A}_{\mathrm{S}}(t)\exp(\mathrm{j}2\pi f_c t) \quad \circ\!\!-\!\!\bullet \quad \underline{\tilde{a}}_{\mathrm{S}}(f) = \underline{\tilde{A}}_{\mathrm{S}}(f-f_c), \qquad (\mathrm{S1})$$

$$\underline{a}_{\mathrm{LO}}(t) = \sum_{\mu=1}^{M}\underline{A}_{\mathrm{LO},\mu}\exp(\mathrm{j}2\pi f_\mu t) \quad \circ\!\!-\!\!\bullet \quad \underline{\tilde{a}}_{\mathrm{LO}}(f) = \sum_{\mu=1}^{M}\underline{A}_{\mathrm{LO},\mu}\delta(f-f_\mu). \qquad (\mathrm{S2})$$

We assume that the recording length of our system is much shorter than the coherence time of the LO so that phase and amplitude noise of the LO can be neglected, i.e., $\underline{A}_{\mathrm{LO},\mu}$ is assumed constant within a given recording.

The slice-less optical arbitrary waveform measurement (OAWM) scheme described in the main manuscript relies on simultaneous down-conversion of different spectral portions of the broadband signal using an LO comb with distinct tones at frequencies $f_\mu$, $\mu = 1,...,M$, in combination with a multitude of optical receiver channels, characterized by their respective delay $\tau_\nu$, $\nu = 1,...,N$ see Fig. 1 of the main manuscript. The underlying system concept is illustrated in the block diagram depicted in Fig. S1, describing the reception of the broadband signal $\underline{a}_S(t)$ by IQ receiver $\nu$ (IQR $\nu$). The power splitter and the characteristic propagation delays $\tau_\nu$ accumulated by the signal and LO on their ways to the 90° optical hybrid are modeled by the optical impulse responses $h_{S,\nu}(t)/\sqrt{N}$ and $h_{LO,\nu}(t)/\sqrt{N}$, respectively, where the factor $1/\sqrt{N}$ accounts for the splitting of the signal and LO comb power into $N$ receiver paths. The signals at the input of the $\nu$-th optical hybrid are given by

$$\underline{a}_{S,\nu}(t) = \frac{1}{\sqrt{N}} h_{S,\nu}(t) * \underline{a}_S(t) \qquad \circ\!\!-\!\!\bullet \qquad \underline{\tilde{a}}_{S,\nu}(f) = \frac{1}{\sqrt{N}} \tilde{h}_{S,\nu}(f) \underline{\tilde{a}}_S(f),$$

$$\underline{a}_{LO,\nu}(t) = \frac{1}{\sqrt{N}} h_{LO,\nu}(t) * \underline{a}_{LO}(t) \qquad \circ\!\!-\!\!\bullet \qquad \underline{\tilde{a}}_{LO,\nu}(f) = \frac{1}{\sqrt{N}} \tilde{h}_{LO,\nu}(f) \underline{\tilde{a}}_{LO}(f). \qquad (S3)$$

The optical hybrid splits the incoming signals $\underline{a}_{S,\nu}(t)$ and $\underline{a}_{LO,\nu}(t)$ into four identical copies $\underline{a}_{S,\nu}(t)/\sqrt{4}$ and $\underline{a}_{LO,\nu}(t)/\sqrt{4}$, which are then superimposed with four distinct phases. For simplicity, all these phases are assigned to the LO paths and described by the complex-valued coefficients $\underline{C}_\nu^{(I+)}$, $\underline{C}_\nu^{(I-)}$, $\underline{C}_\nu^{(Q+)}$, and $\underline{C}_\nu^{(Q-)}$, see Fig. S1. For an ideal optical hybrid, these four coefficients are given by

$$\underline{C}_\nu^{(I+)} = 1, \ \underline{C}_\nu^{(I-)} = -1, \ \underline{C}_\nu^{(Q+)} = j, \ \underline{C}_\nu^{(Q-)} = -j. \qquad (S4)$$

The wavelength-dependent transfer functions of the optical hybrids and of the optical fibers to the subsequent in-phase (I) and quadrature (Q) balanced photodetectors (BPD) are modeled by the optical impulse responses $h_\nu^{(I)}(t)$ and $h_\nu^{(Q)}(t)$, respectively. The model for the two BPD in each IQ receiver includes a pair of optical impulse responses $h_{BPD,\nu}^{(I)}(t)$ and $h_{BPD,\nu}^{(Q)}(t)$ that describe the optical properties such as wavelength-dependent responsivities, as well as a pair of electrical impulse responses $H_{BPD,\nu}^{(I)}(t)$ and $H_{BPD,\nu}^{(Q)}(t)$ that model the electrical characteristics such as the RF transfer functions. As we only measure the difference current at the output of the BPD, we can neither digitally compensate for the imbalance between the paired outputs of the optical hybrid nor for the imbalance of the BPD. Therefore, we assume a common impulse response $h_\nu^{(I)}(t)$ ($h_\nu^{(Q)}(t)$) for the paired balanced outputs of the optical hybrid, as well as common optical and electrical impulse responses, $h_{BPD,\nu}^{(I)}(t)$ ($h_{BPD,\nu}^{(Q)}(t)$) and $H_{BPD,\nu}^{(I)}(t)$ ($H_{BPD,\nu}^{(Q)}(t)$), respectively, for the two photodetectors inside each BPD. This simplification is well justified because the optical fibers connecting the optical hybrid to the subsequent BPD in our setup are well length matched and because the BPD are well balanced. According to Fig. S1 and the above simplifications, we calculate the signals $a_\nu^{(I+)}(t)$, $a_\nu^{(I-)}(t)$, $a_\nu^{(Q+)}(t)$ and $a_\nu^{(Q-)}(t)$ at the various inputs of the BPD. To simplify the subsequent derivation, these signals are already convolved with the optical impulse responses $h_{BPD,\nu}^{(I)}(t)$ ($h_{BPD,\nu}^{(Q)}(t)$) that model the wavelength dependence of the BPD. Thus, $a_\nu^{(I+)}(t)$, $a_\nu^{(I-)}(t)$, $a_\nu^{(Q+)}(t)$ and $a_\nu^{(Q-)}(t)$ do not exist as physically accessible optical signals in the real system.

$$\underline{a}_{\nu}^{(I+)}(t) = \frac{1}{\sqrt{4}} h_{BPD,\nu}^{(I)}(t) * h_{\nu}^{(I)}(t) * \left[\underline{a}_{S,\nu}(t) + \underline{C}_{\nu}^{(I+)} \underline{a}_{LO,\nu}(t)\right],$$

$$\underline{a}_{\nu}^{(I-)}(t) = \frac{1}{\sqrt{4}} h_{BPD,\nu}^{(I)}(t) * h_{\nu}^{(I)}(t) * \left[\underline{a}_{S,\nu}(t) + \underline{C}_{\nu}^{(I-)} \underline{a}_{LO,\nu}(t)\right],$$

$$\underline{a}_{\nu}^{(Q+)}(t) = \frac{1}{\sqrt{4}} h_{BPD,\nu}^{(Q)}(t) * h_{\nu}^{(Q)}(t) * \left[\underline{a}_{S,\nu}(t) + \underline{C}_{\nu}^{(Q+)} \underline{a}_{LO,\nu}(t)\right],$$

$$\underline{a}_{\nu}^{(Q-)}(t) = \frac{1}{\sqrt{4}} h_{BPD,\nu}^{(Q)}(t) * h_{\nu}^{(Q)}(t) * \left[\underline{a}_{S,\nu}(t) + \underline{C}_{\nu}^{(Q-)} \underline{a}_{LO,\nu}(t)\right].$$

(S5)

After balanced detection and analog-to-digital conversion, we obtain the digital in-phase signal $I_{\nu}(t)$ and the digital quadrature signal $Q_{\nu}(t)$, which contain additional noise contributions such as shot noise, thermal noise, or quantization noise. We summarize all such noise sources and model them by additive voltage noises $G_{\nu}^{(I)}(t)$ and $G_{\nu}^{(Q)}(t)$, respectively.

$$I_{\nu}(t) = H_{ADC,\nu}^{(I)}(t) * H_{BPD,\nu}^{(I)}(t) * \left(\left|a_{\nu}^{(I+)}(t)\right|^2 - \left|a_{\nu}^{(I-)}(t)\right|^2\right) + G_{\nu}^{(I)}(t),$$

$$Q_{\nu}(t) = H_{ADC,\nu}^{(Q)}(t) * H_{BPD,\nu}^{(Q)}(t) * \left(\left|a_{\nu}^{(Q+)}(t)\right|^2 - \left|a_{\nu}^{(Q-)}(t)\right|^2\right) + G_{\nu}^{(Q)}(t).$$

(S6)

In the next steps, we make use of Eqs. (S5), (S3), (S2) and of the Fourier transformation to relate the electrical signals $I_{\nu}(t)$ and $Q_{\nu}(t)$ in Eq. (S6) to the optical signal $\underline{a}_S(t)$ at the system input. As we already assumed perfectly balanced photodetectors with infinite common-mode rejection, we also neglect any self-beating terms of signal and LO. By substituting Eq. (S5) in Eq. (S6) and expanding the expression, we obtain for the in-phase component,

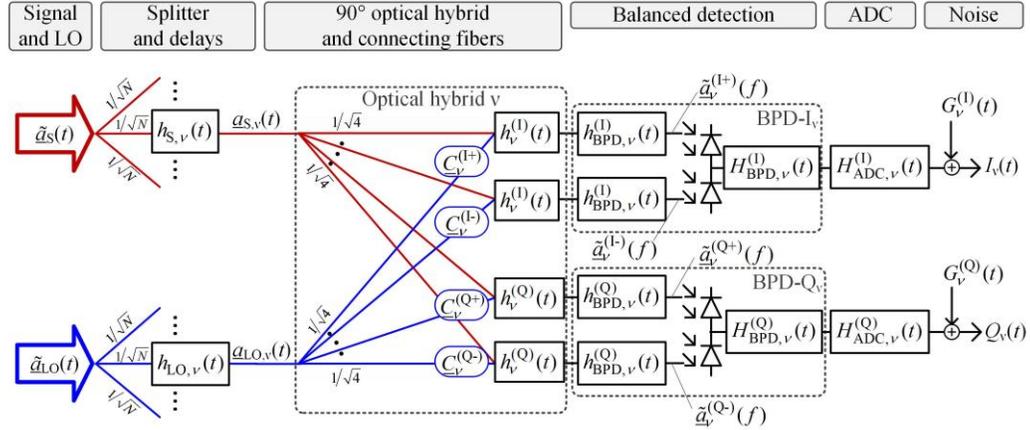

**Fig. S1**. Block diagram of a specific detection channel (index $\nu$) of the slice-less optical arbitrary waveform measurement (OAWM) system. The optical input signal $\underline{a}_S(t)$ and the LO comb $\underline{a}_{LO}(t)$ are split into $N$ copies, which are subject to different delays $\tau_{\nu}$, modeled by $h_{S,\nu}(t)$ and $h_{LO,\nu}(t)$ respectively, see also Fig. 1 of the main manuscript. Optical signals and corresponding impulse responses are denoted with lowercase letters, whereas the associated baseband signals and the corresponding impulse responses are represented by uppercase symbols.

$$I_\nu(t) = H^{(I)}_{ADC,\nu}(t) * H^{(I)}_{BPD,\nu}(t)$$

$$* \frac{1}{4}\left\{\begin{array}{l}+\left[h^{(I)}_{BPD,\nu}(t) * h^{(I)}_\nu(t) * \underline{a}_{S,\nu}(t)\right]\left[h^{(I)*}_{BPD,\nu}(t) * h^{(I)*}_\nu(t) * \underline{C}^{(I+)*}_\nu \underline{a}^*_{LO,\nu}(t)\right] \\ +\left[h^{(I)*}_{BPD,\nu}(t) * h^{(I)*}_\nu(t) * \underline{a}^*_{S,\nu}(t)\right]\left[h^{(I)}_{BPD,\nu}(t) * h^{(I)}_\nu(t)) * \underline{C}^{(I+)}_\nu \underline{a}_{LO,\nu}(t)\right] \\ -\left[h^{(I)}_{BPD,\nu}(t) * h^{(I)}_\nu(t) * \underline{a}_{S,\nu}(t)\right]\left[h^{(I)*}_{BPD,\nu}(t) * h^{(I)*}_\nu(t) * \underline{C}^{(I-)*}_\nu \underline{a}^*_{LO,\nu}(t)\right] \\ -\left[h^{(I)*}_{BPD,\nu}(t) * h^{(I)*}_\nu(t) * \underline{a}^*_{S,\nu}(t)\right]\left[h^{(I)}_{BPD,\nu}(t) * h^{(I)}_\nu(t) * \underline{C}^{(I-)}_\nu \underline{a}_{LO,\nu}(t)\right]\end{array}\right\} + G^{(I)}_\nu(t). \quad (S7)$$

By Fourier-transforming (S7), we eliminate most convolutions

$$\tilde{I}_\nu(f) = \tilde{H}^{(I)}_{ADC,\nu}(f)\tilde{H}^{(I)}_{BPD,\nu}(f)$$

$$\times \frac{1}{4}\left\{\begin{array}{l}\left[\tilde{h}^{(I)}_{BPD,\nu}(f)\tilde{h}^{(I)}_\nu(f)\underline{\tilde{a}}_{S,\nu}(f)\right] * \left[\tilde{h}^{(I)*}_{BPD,\nu}(-f)\tilde{h}^{(I)*}_\nu(-f)\underline{\tilde{a}}^*_{LO,\nu}(-f)\right]\left[\underline{C}^{(I+)*}_\nu - \underline{C}^{(I-)*}_\nu\right] \\ +\left[\tilde{h}^{(I)*}_{BPD,\nu}(-f)\tilde{h}^{(I)*}_\nu(-f)\underline{\tilde{a}}^*_{S,\nu}(-f)\right] * \left[\tilde{h}^{(I)}_{BPD,\nu}(f)\tilde{h}^{(I)}_\nu(f)\underline{\tilde{a}}_{LO,\nu}(f)\right]\left[\underline{C}^{(I+)}_\nu - \underline{C}^{(I-)}_\nu\right]\end{array}\right\} + \tilde{G}^{(I)}_\nu(f). \quad (S8)$$

We then insert the Fourier transforms of the signal and the LO according to Eq. (S2) as well as the splitting and propagation of the optical signals according to Eq. (S3) into Eq. (S8) and simplify the remaining convolutions by making use of the property $y(x) * \delta(x-x_0) = y(x-x_0)$ of the Dirac delta distribution,

$$\tilde{I}_\nu(f) = \frac{1}{4N}\tilde{H}^{(I)}_{ADC,\nu}(f)\tilde{H}^{(I)}_{BPD,\nu}(f)$$

$$\times \sum_{\mu=1}^{M}\left\{\begin{array}{l}\tilde{h}^{(I)}_{BPD,\nu}(f+f_\mu)\tilde{h}^{(I)*}_{BPD,\nu}(f_\mu)\tilde{h}^{(I)}_\nu(f+f_\mu)\tilde{h}^{(I)*}_\nu(f_\mu)\left[\underline{C}^{(I+)*}_\nu - \underline{C}^{(I-)*}_\nu\right] \\ \times \tilde{h}_{S,\nu}(f+f_\mu)\tilde{h}^*_{LO,\nu}(f_\mu)\underline{A}^*_{LO,\mu}\underline{\tilde{a}}_S(f+f_\mu) \\ + \tilde{h}^{(I)*}_{BPD,\nu}(-f+f_\mu)\tilde{h}^{(I)}_{BPD,\nu}(f_\mu)\tilde{h}^{(I)*}_\nu(-f+f_\mu)\tilde{h}^{(I)}_\nu(f_\mu)\left[\underline{C}^{(I+)}_\nu - \underline{C}^{(I-)}_\nu\right] \\ \times \tilde{h}^*_{S,\nu}(-f+f_\mu)\tilde{h}_{LO,\nu}(f_\mu)\underline{A}_{LO,\mu}\underline{\tilde{a}}^*_S(-f+f_\mu)\end{array}\right\} \quad (S9)$$

$$+ \tilde{G}^{(I)}_\nu(f).$$

As a next step, we simplify Eq. (S9). To this end, we define a baseband transfer function $\underline{\tilde{H}}^{(I,t)}_{\nu\mu}(f)$ that includes all the individual transfer functions and complex-valued factors in Eq. (S9) – the associated expressions are marked in blue. This leads to

$$\underline{\tilde{H}}^{(I,t)}_{\nu\mu}(f) = \frac{1}{4N}\tilde{H}^{(I)}_{ADC,\nu}(f)\tilde{H}^{(I)}_{BPD,\nu}(f)\tilde{h}^{(I)}_{BPD,\nu}(f+f_\mu)\tilde{h}^{(I)*}_{BPD,\nu}(f_\mu)\tilde{h}^{(I)}_\nu(f+f_\mu)\tilde{h}^{(I)*}_\nu(f_\mu)$$

$$\times \left[\underline{C}^{(I+)*}_\nu - \underline{C}^{(I-)*}_\nu\right]\tilde{h}_{S,\nu}(f+f_\mu)\tilde{h}^*_{LO,\nu}(f_\mu)\underline{A}^*_{LO,\mu}. \quad (S10)$$

The same derivation is used for the quadrature component, which leads to the baseband transfer function $\underline{\tilde{H}}^{(Q,t)}_{\nu\mu}(f)$,

$$\underline{\tilde{H}}^{(Q,t)}_{\nu\mu}(f) = \frac{1}{4N}\tilde{H}^{(Q)}_{ADC,\nu}(f)\tilde{H}^{(Q)}_{BPD,\nu}(f)\tilde{h}^{(Q)}_{BPD,\nu}(f+f_\mu)\tilde{h}^{(Q)*}_{BPD,\nu}(f_\mu)\tilde{h}^{(Q)}_\nu(f+f_\mu)\tilde{h}^{(Q)*}_\nu(f_\mu)$$

$$\times \left[\underline{C}^{(Q+)*}_\nu - \underline{C}^{(Q-)*}_\nu\right]\tilde{h}_{S,\nu}(f+f_\mu)\tilde{h}^*_{LO,\nu}(f_\mu)\underline{A}^*_{LO,\mu}. \quad (S11)$$

Note that $\underline{\tilde{H}}^{(I,t)}_{\nu\mu}(f)$ and $\underline{\tilde{H}}^{(Q,t)}_{\nu\mu}(f)$ are written with an underbar to indicate that the inverse Fourier transforms are complex-valued. This is a consequence of the fact that $\underline{\tilde{H}}^{(I,t)}_{\nu\mu}(f)$ and $\underline{\tilde{H}}^{(Q,t)}_{\nu\mu}(f)$ comprise frequency-shifted versions of optical transfer functions such as $\tilde{h}^{(I)}_\nu(f+f_\mu)$, which do not feature Hermitian symmetry such that $\underline{\tilde{H}}^{(I,t)}_{\nu\mu}(f) \neq \underline{\tilde{H}}^{(I,t)*}_{\nu\mu}(-f)$ and $\underline{\tilde{H}}^{(Q,t)}_{\nu\mu}(f) \neq \underline{\tilde{H}}^{(Q,t)*}_{\nu\mu}(-f)$. We insert $\underline{\tilde{H}}^{(I,t)}_{\nu\mu}(f)$ from Eq. (S11) in Eq. (S9) and find a simplified expression for the in-phase components $\tilde{I}_\nu(f)$ by additionally making use of the Hermitian symmetry of the Fourier transforms $\tilde{H}_{\text{BPD},\nu}^{(\text{I})}(f) = \tilde{H}_{\text{BPD},\nu}^{(\text{I})*}(-f)$ and $\tilde{H}_{\text{ADC},\nu}^{(\text{I})}(f) = \tilde{H}_{\text{ADC},\nu}^{(\text{I})*}(-f)$ of the real-valued electrical impulse responses for the BPD and ADC,

$$\tilde{I}_\nu(f) = \sum_{\mu=1}^{M}\left[\underline{\tilde{H}}_{\nu\mu}^{(\text{I,t})}(f)\underline{\tilde{a}}_{\text{S}}(f+f_\mu) + \underline{\tilde{H}}_{\nu\mu}^{(\text{I,t})*}(-f)\underline{\tilde{a}}_{\text{S}}^*(-f+f_\mu)\right] + \tilde{G}_\nu^{(\text{I})}(f). \tag{S12}$$

Similarly, the quadrature components $\tilde{Q}_\nu(f)$ are found by using the corresponding transfer functions $\underline{\tilde{H}}_{\nu\mu}^{(\text{Q,t})}(f)$ in Eq. (S11),

$$\tilde{Q}_\nu(f) = \sum_{\mu=1}^{M}\left[\underline{\tilde{H}}_{\nu\mu}^{(\text{Q,t})}(f)\underline{\tilde{a}}_{\text{S}}(f+f_\mu) + \underline{\tilde{H}}_{\nu\mu}^{(\text{Q,t})*}(-f)\underline{\tilde{a}}_{\text{S}}^*(-f+f_\mu)\right] + \tilde{G}_\nu^{(\text{Q})}(f). \tag{S13}$$

Equations (S12) and (S13) can be reformulated in matrix-vector form. To this end, we define the in-phase vector $\tilde{\mathbf{I}}(f)$ comprising all measured in-phase components $\tilde{I}_\nu(f)$, the vector $\tilde{\mathbf{Q}}(f)$ of the measured quadrature-phase components $\tilde{Q}_\nu(f)$, as well as the respective noise vectors $\tilde{\mathbf{G}}^{(\text{I})}(f)$ and $\tilde{\mathbf{G}}^{(\text{Q})}(f)$. We further define a signal vector $\underline{\tilde{\mathbf{A}}}_{\text{S}}(f)$ that contains the frequency shifted signal spectra $\underline{\tilde{a}}_{\text{S}}(f+f_\mu)$. Additionally, we construct the in-phase transfer matrix a $\underline{\tilde{\mathbf{H}}}^{(\text{I,t})}(f)$ and the quadrature transfer matrix $\underline{\tilde{\mathbf{H}}}^{(\text{Q,t})}(f)$ from the respective transfer functions $\underline{\tilde{H}}_{\nu\mu}^{(\text{I,t})}(f)$ and $\underline{\tilde{H}}_{\nu\mu}^{(\text{Q,t})}(f)$, see Eq. (S11). This leads to

$$\begin{bmatrix}\tilde{\mathbf{I}}(f)\\ \tilde{\mathbf{Q}}(f)\end{bmatrix} = \underbrace{\begin{bmatrix}\underline{\tilde{\mathbf{H}}}^{(\text{I,t})}(f) & \underline{\tilde{\mathbf{H}}}^{(\text{I,t})*}(-f)\\ \underline{\tilde{\mathbf{H}}}^{(\text{Q,t})}(f) & \underline{\tilde{\mathbf{H}}}^{(\text{Q,t})*}(-f)\end{bmatrix}}_{\underline{\tilde{\mathbf{H}}}_{\text{IQ}}^{(\text{t})}(f)}\begin{bmatrix}\underline{\tilde{\mathbf{A}}}_{\text{S}}(f)\\ \underline{\tilde{\mathbf{A}}}_{\text{S}}^*(-f)\end{bmatrix} + \begin{bmatrix}\tilde{\mathbf{G}}^{(\text{I})}(f)\\ \tilde{\mathbf{G}}^{(\text{Q})}(f)\end{bmatrix}, \tag{S14}$$

where

$$\tilde{\mathbf{I}}(f) = \begin{bmatrix}\tilde{I}_1(f)\\ \vdots\\ \tilde{I}_N(f)\end{bmatrix}, \quad \tilde{\mathbf{Q}}(f) = \begin{bmatrix}\tilde{Q}_1(f)\\ \vdots\\ \tilde{Q}_N(f)\end{bmatrix}, \quad \tilde{\mathbf{G}}^{(\text{I})}(f) = \begin{bmatrix}\tilde{G}_1^{(\text{I})}(f)\\ \vdots\\ \tilde{G}_N^{(\text{I})}(f)\end{bmatrix}, \quad \tilde{\mathbf{G}}^{(\text{Q})}(f) = \begin{bmatrix}\tilde{G}_1^{(\text{Q})}(f)\\ \vdots\\ \tilde{G}_N^{(\text{Q})}(f)\end{bmatrix},$$

$$\underline{\tilde{\mathbf{A}}}_{\text{S}}(f) = \begin{bmatrix}\underline{\tilde{a}}_{\text{S}}(f+f_1)\\ \vdots\\ \underline{\tilde{a}}_{\text{S}}(f+f_M)\end{bmatrix}, \tag{S15}$$

$$\underline{\tilde{\mathbf{H}}}^{(\text{I,t})}(f) = \begin{bmatrix}\underline{\tilde{H}}_{11}^{(\text{I,t})}(f) & \cdots & \underline{\tilde{H}}_{1M}^{(\text{I,t})}(f)\\ \vdots & \ddots & \vdots\\ \underline{\tilde{H}}_{N1}^{(\text{I,t})}(f) & \cdots & \underline{\tilde{H}}_{NM}^{(\text{I,t})}(f)\end{bmatrix}, \quad \underline{\tilde{\mathbf{H}}}^{(\text{Q,t})}(f) = \begin{bmatrix}\underline{\tilde{H}}_{11}^{(\text{I,t})}(f) & \cdots & \underline{\tilde{H}}_{1M}^{(\text{I,t})}(f)\\ \vdots & \ddots & \vdots\\ \underline{\tilde{H}}_{N1}^{(\text{I,t})}(f) & \cdots & \underline{\tilde{H}}_{NM}^{(\text{I,t})}(f)\end{bmatrix}.$$

Note that for the simplified explanation used in Eq. (3) of the main manuscript, we additional assumed that the in-phase and quadrature transfer functions $\underline{\tilde{H}}_{\nu\mu}^{(\text{I,t})}(f)$ and $\underline{\tilde{H}}_{\nu\mu}^{(\text{Q,t})}(f)$ are identical and have an ideal 90° phase relationship, $\underline{\tilde{\mathbf{H}}}^{(\text{Q,t})}(f) = -\text{j}\underline{\tilde{\mathbf{H}}}^{(\text{I,t})}(f)$. This allows to simplify Eq. (S14) by constructing the composite baseband spectra $\underline{\tilde{U}}_\nu(f) = \tilde{I}_\nu(f) + \text{j}\tilde{Q}_\nu(f)$, $\nu = 1,...,N$, that are the Fourier transforms of the complex-valued baseband signals $\underline{U}_\nu(t) = I_\nu(t) + \text{j}Q_\nu(t)$, see Eq. (2) in the main manuscript. For the composite baseband spectra $\underline{\tilde{\mathbf{U}}}(f)$ we can reformulate Eq. (S14) to obtain a relation that does not contain the frequency-inverted complex-conjugate counterpart $\underline{\tilde{\mathbf{A}}}_{\text{S}}^*(-f)$,

$$\underbrace{\tilde{\mathbf{I}}(f) + \text{j}\tilde{\mathbf{Q}}(f)}_{\underline{\tilde{\mathbf{U}}}(f)} = \underbrace{2\underline{\tilde{\mathbf{H}}}^{(\text{I,t})}(f)}_{\underline{\tilde{\mathbf{H}}}(f)}\underline{\tilde{\mathbf{A}}}_{\text{S}}(f) + \underbrace{\mathbf{G}^{(\text{I})}(f) + \text{j}\mathbf{G}^{(\text{Q})}(f)}_{\underline{\tilde{\mathbf{G}}}(f)}. \tag{S16}$$

*1.2. Signal reconstruction*

We use the system described in the previous section to calculate an estimate $\tilde{\underline{a}}_S^{(est)}(f)$ of the spectrum $\tilde{\underline{a}}_S(f)$ of the original complex-valued input signal $\underline{a}_S(t)$. To this end, we first find the best estimate $\tilde{\underline{\mathbf{A}}}_S^{(est)}(f) = \left[ a_{S,1}^{(est)}(f+f_1) \cdots a_{S,M}^{(est)}(f+f_M) \right]^T$ of the true signal vector $\tilde{\underline{\mathbf{A}}}_S(f)$ within the receiver bandwidth $B$ by minimizing the quadratic norm $\left\| \tilde{\underline{\mathbf{A}}}_S(f) - \tilde{\underline{\mathbf{A}}}_S^{(est)}(f) \right\|^2$. As the signal $\tilde{\underline{\mathbf{A}}}_S(f)$ and the noise $\tilde{\mathbf{G}}^{(I)}(f)$ and $\tilde{\mathbf{G}}^{(Q)}(f)$ are statistically independent, this can be achieved by using the minimum-variance unbiased estimator [1],

$$\begin{bmatrix} \tilde{\underline{\mathbf{A}}}_S^{(est)}(f) \\ \tilde{\underline{\mathbf{A}}}_S^{(est)*}(-f) \end{bmatrix} = \left[ \tilde{\mathbf{H}}_{IQ}^{\dagger}(f) \mathbf{C}_{GG}^{-1}(f) \tilde{\mathbf{H}}_{IQ}(f) \right]^{-1} \tilde{\mathbf{H}}_{IQ}^{\dagger}(f) \mathbf{C}_{GG}^{-1}(f) \begin{bmatrix} \tilde{\mathbf{I}}(f) \\ \tilde{\mathbf{Q}}(f) \end{bmatrix}, \quad \text{for } |f| < B. \quad (S17)$$

where $\mathbf{C}_{GG}(f)$ is the autocovariance matrix of the noise vector $\left[ \left[ \tilde{\mathbf{G}}^{(I)}(f) \right]^T, \left[ \tilde{\mathbf{G}}^{(Q)}(f) \right]^T \right]^T$, and where superscript "T" denotes the transpose of the respective vector. Note that Eq. (S17) is only valid for frequencies $f$ within the receiver bandwidth $B$, $f \in [-B, B]$, because the in-phase $\tilde{\mathbf{I}}(f)$ and quadrature spectra $\tilde{\mathbf{Q}}(f)$ are only measured within this frequency range. The components $\tilde{\underline{a}}_{S,1}^{(est)}(f+f_1), \ldots, \tilde{\underline{a}}_{S,M}^{(est)}(f+f_M)$ of the estimated signal vector $\tilde{\underline{\mathbf{A}}}_S^{(est)}(f)$ represent frequency-shifted spectral portions of the signal $\tilde{\underline{a}}_S(f)$ of interest and are referred to as spectral slices. Because the electrical noise contributions $\tilde{G}_1^{(I)}(f)$, $\tilde{G}_1^{(Q)}(f)$, ..., $\tilde{G}_N^{(I)}(f)$, $\tilde{G}_N^{(Q)}(f)$ in all receiver channels are uncorrelated and because all receiver channels add approximately the same average noise power $\sigma_G^2(f)$, we may approximate $\mathbf{C}_{GG}(f) \approx \sigma_G^2(f) \mathbf{D}$, where $\mathbf{D}$ is the identity matrix. Equation (S17) then turns into the relation for the least-square estimator $\left[ \left[ \tilde{\underline{\mathbf{A}}}_S^{(est)}(f) \right]^T, \left[ \tilde{\underline{\mathbf{A}}}_S^{(est)*}(-f) \right]^T \right]^T$ [1],

$$\begin{bmatrix} \tilde{\underline{\mathbf{A}}}_S^{(est)}(f) \\ \tilde{\underline{\mathbf{A}}}_S^{(est)*}(-f) \end{bmatrix} = \underbrace{\left[ \tilde{\underline{\mathbf{H}}}_{IQ}^{(t)\dagger}(f) \tilde{\underline{\mathbf{H}}}_{IQ}^{(t)}(f) \right]^{-1} \tilde{\underline{\mathbf{H}}}_{IQ}^{(t)\dagger}(f)}_{\text{pseudo inverse of } \tilde{\underline{\mathbf{H}}}_{IQ}^{(t)}(f)} \begin{bmatrix} \tilde{\mathbf{I}}(f) \\ \tilde{\mathbf{Q}}(f) \end{bmatrix}, \quad \text{for } |f| < B. \quad (S18)$$

The reconstructed vector comprises the signal vector $\tilde{\underline{\mathbf{A}}}_S^{(est)}(f)$ can hence either be determined by evaluating the upper half of Eq. (S18) for all frequencies $f \in [-B, B]$ or by evaluating the full equation for non-negative frequencies $f \in [0, B]$, i.e., the relation is overdetermined when evaluated in the full range of positive and negative frequencies, $f \in [-B, B]$. However, because of the special structure of the pseudo-inverse of $\tilde{\underline{\mathbf{H}}}_{IQ}^{(t)}(f)$, and because $\tilde{\mathbf{I}}(f)$ and $\tilde{\mathbf{Q}}(f)$ are spectra of real-valued time domain signals featuring Hermitian symmetry, $\tilde{\mathbf{I}}(f) = \tilde{\mathbf{I}}^*(-f)$, $\tilde{\mathbf{Q}}(f) = \tilde{\mathbf{Q}}^*(-f)$, both results are consistent. This can be better understood from the corresponding forward relation (S14), where the internal structure of the transfer matrix $\tilde{\underline{\mathbf{H}}}_{IQ}^{(t)}(f)$ is visible. To calculate $\tilde{\mathbf{I}}(f)$ and $\tilde{\mathbf{Q}}(f)$ from $\tilde{\underline{\mathbf{A}}}_S(f)$ via Eq. (S14), we can either evaluate the relation in the full range $f \in [-B, B]$ or we can evaluate Eq. (S14) only for positive frequencies $f \in [0, B]$ and infer the negative-frequency components of $\tilde{\mathbf{I}}(f)$ and $\tilde{\mathbf{Q}}(f)$ from the symmetry relations for real-valued time-domain signals, $\tilde{\mathbf{I}}(-f) = \tilde{\mathbf{I}}^*(f)$, $\tilde{\mathbf{Q}}(-f) = \tilde{\mathbf{Q}}^*(f)$. Because of the structure of $\tilde{\underline{\mathbf{H}}}_{IQ}^{(t)}(f)$, both results are consistent, which becomes apparent if we replace $f$ by $-f$ in Eq. (S14) and take the complex conjugate. These operations leave the right-hand side of Eq. (S14) unchanged and thereby reproduce the symmetry relations, $\tilde{\mathbf{I}}^*(-f) = \tilde{\mathbf{I}}(f)$, and $\tilde{\mathbf{Q}}^*(-f) = \tilde{\mathbf{Q}}(f)$. Evaluating Eq. (S14) for all frequencies is hence equivalent to assuming real-valued time-domain signals for $I_\nu(t)$ and $Q_\nu(t)$ and exploiting the symmetry relations for the associated spectra. As a consequence, it is also sufficient to either consider Eq. (S18) only for non-negative baseband frequencies $f \in [0, B]$, which yields both $\tilde{\underline{\mathbf{A}}}_S^{(est)}(f)$ and $\tilde{\underline{\mathbf{A}}}_S^{(est)*}(-f)$, or, alternatively, to restrict the evaluation of Eq. (S18) to the upper half, i.e., the calculation of $\tilde{\underline{\mathbf{A}}}_S^{(est)}(f)$, in the full range $f \in [-B, B]$. In the following, we pursue the second approach.

To simplify the subsequent notation, we define a reconstruction matrix $\tilde{\underline{\mathbf{H}}}_{\text{rec}}^{(t)}(f)$ that consists of the upper half of the pseudo inverse $\left[\tilde{\underline{\mathbf{H}}}_{\text{IQ}}^{(t)\dagger}(f)\tilde{\underline{\mathbf{H}}}_{\text{IQ}}^{(t)}(f)\right]^{-1}\tilde{\underline{\mathbf{H}}}_{\text{IQ}}^{(t)\dagger}(f)$ in Eq. (S18). The reconstruction rule (S18) then simplifies to

$$\tilde{\underline{\mathbf{A}}}_{\text{S}}^{(\text{est})}(f) = \tilde{\underline{\mathbf{H}}}_{\text{rec}}^{(t)}(f)\begin{bmatrix}\tilde{\mathbf{I}}(f)\\ \tilde{\mathbf{Q}}(f)\end{bmatrix}, \quad \text{for } |f| < B, \tag{S19}$$

where $\tilde{\underline{\mathbf{A}}}_{\text{S}}^{(\text{est})}(f) = \left[\tilde{\underline{a}}_1^{(\text{est})}(f+f_1) \;\cdots\; \tilde{\underline{a}}_M^{(\text{est})}(f+f_M)\right]^{\text{T}}$. As the transfer matrix $\tilde{\underline{\mathbf{H}}}_{\text{IQ}}^{(t)}(f)$ will eventually be determined from a calibration measurement that is influenced by noise, the pseudo-inverse $\left[\tilde{\underline{\mathbf{H}}}_{\text{IQ}}^{(t)\dagger}(f)\tilde{\underline{\mathbf{H}}}_{\text{IQ}}^{(t)}(f)\right]^{-1}\tilde{\underline{\mathbf{H}}}_{\text{IQ}}^{(t)\dagger}(f)$ and thus the reconstruction matrix $\tilde{\underline{\mathbf{H}}}_{\text{rec}}^{(t)}(f)$ is only known with limited accuracy. We therefore consider an additional error term $\Delta\tilde{\underline{\mathbf{H}}}_{\text{rec}}^{(t)}(f)$ for the reconstruction matrix, which describes the deviation of the measured matrix from its ideal counterpart $\tilde{\underline{\mathbf{H}}}_{\text{rec}}^{(t)}(f)$ and which leads to some remaining cross-talk $\tilde{\mathbf{A}}_{\text{X}}(f)$ between the various spectral slices during signal reconstruction,

$$\begin{aligned}\tilde{\underline{\mathbf{A}}}_{\text{S}}^{(\text{est})}(f) &= \left[\tilde{\underline{\mathbf{H}}}_{\text{rec}}^{(t)}(f) + \Delta\tilde{\underline{\mathbf{H}}}_{\text{rec}}^{(t)}(f)\right]\begin{bmatrix}\tilde{\mathbf{I}}(f)\\ \tilde{\mathbf{Q}}(f)\end{bmatrix}, \quad \text{for } |f| < B.\\ &= \left[\tilde{\underline{\mathbf{H}}}_{\text{rec}}^{(t)}(f) + \Delta\tilde{\underline{\mathbf{H}}}_{\text{rec}}^{(t)}(f)\right]\left\{\tilde{\underline{\mathbf{H}}}_{\text{IQ}}^{(t)}(f)\begin{bmatrix}\tilde{\underline{\mathbf{A}}}_{\text{S}}(f)\\ \tilde{\underline{\mathbf{A}}}_{\text{S}}^*(-f)\end{bmatrix} + \begin{bmatrix}\tilde{\mathbf{G}}^{(\text{I})}(f)\\ \tilde{\mathbf{G}}^{(\text{Q})}(f)\end{bmatrix}\right\} \\ &= \underbrace{\tilde{\underline{\mathbf{H}}}_{\text{rec}}^{(t)}(f)\tilde{\underline{\mathbf{H}}}_{\text{IQ}}^{(t)}(f)\begin{bmatrix}\tilde{\underline{\mathbf{A}}}_{\text{S}}(f)\\ \tilde{\underline{\mathbf{A}}}_{\text{S}}^*(-f)\end{bmatrix}}_{\tilde{\underline{\mathbf{A}}}_{\text{S}}(f)} + \underbrace{\Delta\tilde{\underline{\mathbf{H}}}_{\text{rec}}^{(t)}(f)\tilde{\underline{\mathbf{H}}}_{\text{IQ}}^{(t)}(f)\begin{bmatrix}\tilde{\underline{\mathbf{A}}}_{\text{S}}(f)\\ \tilde{\underline{\mathbf{A}}}_{\text{S}}^*(-f)\end{bmatrix}}_{\tilde{\underline{\mathbf{A}}}_{\text{X}}(f):\text{ Crosstalk.}} + \underbrace{\left[\tilde{\underline{\mathbf{H}}}_{\text{rec}}^{(t)}(f) + \Delta\tilde{\underline{\mathbf{H}}}_{\text{rec}}^{(t)}(f)\right]\begin{bmatrix}\tilde{\mathbf{G}}^{(\text{I})}(f)\\ \tilde{\mathbf{G}}^{(\text{Q})}(f)\end{bmatrix}}_{\tilde{\underline{\mathbf{A}}}_{\text{G}}(f):\text{ Receiver noise modified by reconstuction matrix.}}.\end{aligned} \tag{S20}$$

The estimated signal vector $\tilde{\underline{\mathbf{A}}}_{\text{S}}^{(\text{est})}(f)$ is thus given by the true signal vector $\tilde{\underline{\mathbf{A}}}_{\text{S}}(f)$, impaired by an additive noise vector $\tilde{\mathbf{A}}_{\text{G}}(f)$ with components $\tilde{\underline{a}}_{\text{G},1}(f),\ldots\tilde{\underline{a}}_{\text{G},M}(f)$ and by a crosstalk vector $\tilde{\mathbf{A}}_{\text{X}}(f)$, whose components are a superposition of all spectral slices $\tilde{\underline{a}}_{\text{S}}(f+f_1), \ldots, \tilde{\underline{a}}_{\text{S}}(f+f_M)$ and their conjugate counterparts $\tilde{\underline{a}}_{\text{S}}^*(-f+f_1), \ldots, \tilde{\underline{a}}_{\text{S}}^*(-f+f_M)$ as dictated by the error $\Delta\tilde{\underline{\mathbf{H}}}_{\text{rec}}^{(t)}(f)$ of the reconstruction matrix. Note that the crosstalk among slices vanishes completely for an ideal calibration, $\Delta\tilde{\underline{\mathbf{H}}}_{\text{rec}}^{(t)}(f) = 0 \Rightarrow \tilde{\mathbf{A}}_{\text{X}}(f) = 0$. Note also that Eq. (S20) assumes a perfect local oscillator according to Eq. (S2). If this condition is not fulfilled, the reconstructed signal vector $\tilde{\underline{\mathbf{A}}}_{\text{S}}^{(\text{est})}(f)$ will be additionally impaired by multiplicative noise origination from mixing of LO noise with the signal, see the blue noise spectrum in Fig. S22(a),(b) in Section 7 below. Further note that Eq. (4) in the main manuscript represents a simplified version of Eq. (S20) that does neither consider calibration errors nor the resulting crosstalk term $\tilde{\underline{\mathbf{A}}}_{\text{X}}(f)$.

In a final step, the full signal spectrum $\tilde{\underline{a}}_{\text{S}}^{(\text{est})}(f)$ needs to be reconstructed from the frequency-shifted spectral slices $\tilde{\underline{a}}_{\text{S},1}^{(\text{est})}(f+f_1), \ldots, \tilde{\underline{a}}_{\text{S},M}^{(\text{est})}(f+f_M)$ that are obtained from the components of the reconstructed signal vector $\tilde{\underline{\mathbf{A}}}_{\text{S}}^{(\text{est})}(f)$. To this end, the slices are shifted back to their relative original position according to the respective LO frequency $f_\mu$. To keep the sampling rate in the DSP algorithms manageable, we avoid working at optical carrier frequencies and introduce a global frequency shift of the entire signal spectrum by a reference frequency $f_{\text{ref}}$. This reference frequency may, e.g., be chosen at the center of the optical signal spectrum or at the lower-frequency edge of the first spectral slice, as done in Fig. S2.

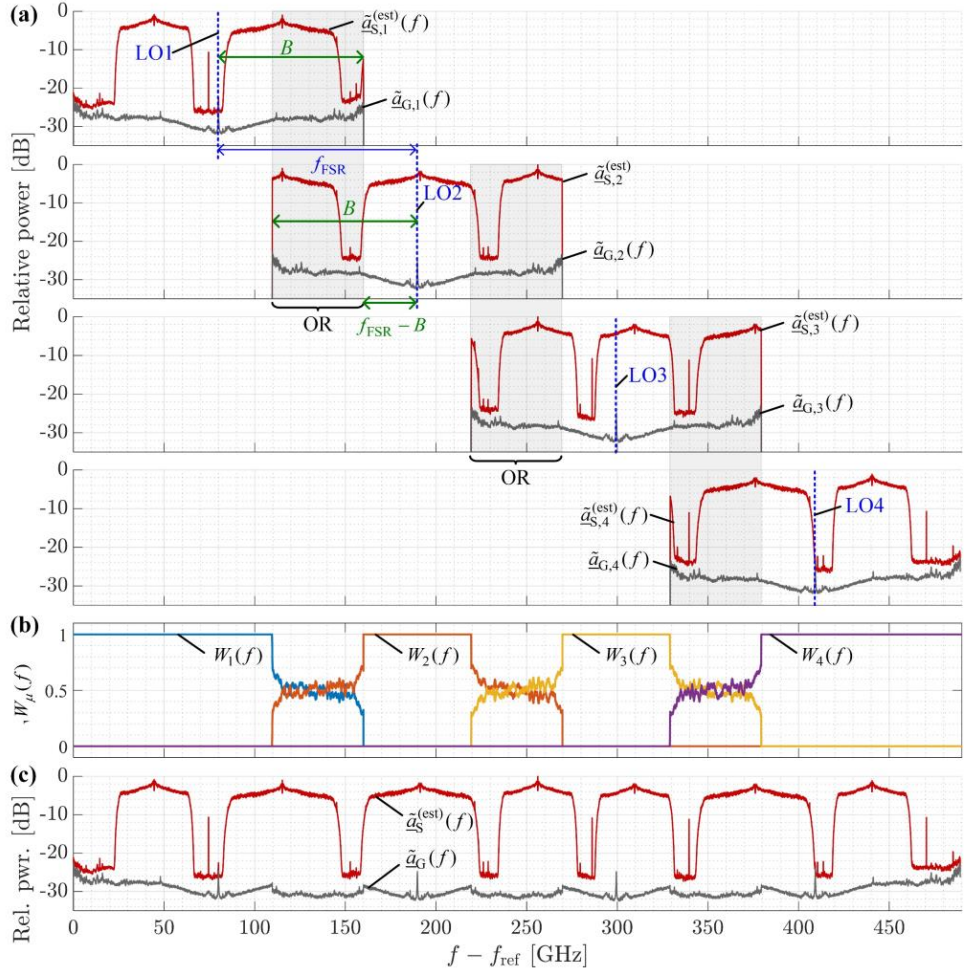

**Fig. S2.** Visualization of the stitching process for a measurement using an LO comb with $f_{\text{FSR}} = 110\,\text{GHz}$ and a receiver with bandwidth $B = 80\,\text{GHz}$. The horizonal axis indicates the offset of the optical frequency $f$ from a reference frequency $f_{\text{ref}}$, which, without loss of generality, was chosen to correspond to the lower-frequency edge of the first spectral slice, $f_{\text{ref}} = f_1 - B \approx 192.52\,\text{THz}$. **(a)** Power spectra of the reconstructed and frequency-shifted signal slices $\tilde{\underline{a}}_{S,\mu}^{(\text{est})}(f) = \tilde{\underline{A}}_{S,\mu}^{(\text{est})}(f - f_\mu)$ (red) and the corresponding separately recorded and fully processed frequency-shifted acquisition-noise contributions $\tilde{\underline{a}}_{G,\mu}(f)$ (gray) see Sect. 8 for details. Blue dotted lines show the spectral position of the LO comb lines LO1,..., LO4 with frequencies $f_1$,..., $f_4$. **(b)** Frequency-dependent stitching weights, obtained from the conditions $W_\mu(f) \propto |\tilde{\underline{a}}_{G,\mu}(f)|^{-2}$ and $\sum_\mu W_\mu(f) = 1$. This choice of weights maximizes the signal-to-noise ratio (SNR) of the stitched signal. **(c)** Stitched signal $\tilde{\underline{a}}_S^{(\text{est})}(f)$ and stitched receiver noise floor $\tilde{a}_G(f)$.

As the bandwidth $B$ of the IQ receivers is larger than half the FSR of the LO comb, $B > f_{\text{FSR}}/2$, the back-shifted slices $\tilde{\underline{a}}_{S,1}(f), ..., \tilde{\underline{a}}_{S,M}(f)$ are spectrally overlapping, Fig. S2. The redundant signal information in these overlap regions (OR) can be exploited to estimate the exact FSR of the LO comb and to digitally compensate for phase drifts of the optical system and for phase- and amplitude fluctuations of the LO comb, see Sections 1.3 and 1.4 below. Once these fluctuations are compensated, the slices are stitched by calculating a weighted average using a frequency-dependent weight function $W_\mu(f)$,

$$\tilde{\underline{a}}_S^{(est)}(f) = \sum_{\mu=1}^{M} W_\mu(f) \tilde{\underline{a}}_{S,\mu}^{(est)}(f) = \tilde{\underline{a}}_S(f) + \tilde{\underline{a}}_G(f) + \tilde{\underline{a}}_X(f), \quad \text{with} \sum_{\mu=1}^{M} W_\mu(f) = 1 \ \forall f \ . \quad (S21)$$

The stitching weights $W_\mu(f)$ are chosen in proportion to the inverse of the reconstructed receiver noise $\tilde{\underline{a}}_{G,\mu}(f)$ of the respective spectral slice, $W_\mu(f) \propto |\tilde{\underline{a}}_{G,\mu}(f)|^{-2}$, while keeping the sum $\sum_\mu W_\mu(f) = 1$ for all frequencies $f$. The power spectral densities of the various receiver noise contributions $\tilde{\underline{a}}_{G,\mu}(f)$ are known from noise measurements, see Sect. 8 for details. Note that calculating the weighted average in Eq. (S21) reduces the resulting receiver noise $\tilde{\underline{a}}_G(f) = \sum_\mu W_\mu(f) \tilde{\underline{a}}_{G,\mu}(f)$ in the stitched signal spectrum $\tilde{\underline{a}}_S^{(est)}(f)$ by up to 3 dB. As can be shown, the choice $W_\mu(f) \propto |\tilde{\underline{a}}_{G,\mu}(f)|^{-2}$ leads to the most pronounced noise reduction and maximizes the overall signal-to-noise ratio (SNR) of the stitched signal. The time-domain waveform is finally obtained from $\tilde{\underline{a}}_S^{(est)}(f)$ by an inverse Fourier transformation. Figure S2 (a) shows an example for the reconstructed and frequency-shifted signal slices $\tilde{\underline{a}}_{S,\mu}^{(est)}(f)$ (red) along with the separately recorded and processed receiver noise contributions $\tilde{\underline{a}}_{G,\mu}(f)$ (gray). Because the photodetector responses are equalized, the receiver noise is increased for high frequencies. Figure S2 (b) shows the frequency-dependent stitching weights $W_\mu(f)$, and Fig. S2(c) displays the resulting signal $\tilde{\underline{a}}_S^{(est)}(f)$ and noise $\tilde{\underline{a}}_G(f)$ after stitching.

### 1.3. Estimation of the free spectral range (FSR) of the LO comb

The reconstruction of the signal $\tilde{\underline{a}}_S^{(est)}(f)$ from its spectral slices $\tilde{\underline{a}}_{S,1}^{(est)}(f+f_1)$, ..., $\tilde{\underline{a}}_{S,M}^{(est)}(f+f_M)$ according to the previous paragraph requires precise knowledge of the free-spectral rage (FSR) $f_{FSR}$ of the LO comb, because the signal slices need to be frequency-shifted back relative to their original position before being stitched. Since our experiment currently relies on a free-running dissipative Kerr soliton comb as an LO, the FSR is not locked to the ADC clock and must thus be estimated for each recording. To this end, we exploit the fact that the bandwidth of the IQ receivers exceeds half the FSR of the comb, $B > f_{FSR}/2$, such that spectral components inside the overlap regions, see "OR" in Fig. S2, $f \in \left[ f_{\mu+1} - B, f_\mu + B \right]$ are down-converted by both adjacent comb lines at $f_\mu$ and $f_{\mu+1}$. These down-converted spectral components appear at both edges of the measured baseband spectra $\tilde{I}_\nu(f)$ and $\tilde{Q}_\nu(f)$, close to $\pm f_{FSR}/2$ with a frequency shift of $f_{FSR}$. This exact frequency shift can be extracted by calculating the autocorrelation $\tilde{I}_\nu(f) \otimes \tilde{I}_\nu(f) = \int_{-\infty}^{\infty} \tilde{I}_\nu(f_0) \tilde{I}_\nu^*(f_0 - f) \, df_0$ and by detecting a peak close to the expected FSR. Figure S3 illustrates this concept for data recorded with an LO comb FSR of $f_{FSR} = 150.8$ GHz. In this example, the optical input signal $\tilde{\underline{a}}_S(f)$ consisted of seven data signals that were generated from only two independent modulators, see Fig. 2 in the main manuscript, leading to additional peaks in the autocorrelation $\tilde{I}_\nu(f) \otimes \tilde{I}_\nu(f)$ of Fig. S3 (b). Note that the signal spectrum is obtained from a discrete Fourier transform (DFT) with discrete frequency points spaced by $\Delta f = 1/T_{obs}$, where $T_{obs}$ is the observation time. Consequently, the FSR $f_{FSR}$ of the free-running comb may fall between two such points. For a precise estimation of the exact FSR, we therefore use a sinc-type interpolation of the discrete measurement points in the spectrum and detect the peak of the interpolation, see Fig. S3 (c). The resulting frequency shifts that are applied to the various slices prior to stitching, see Fig. S2(a), do hence not correspond to an integer multiple of $\Delta f$ and cannot be straightforwardly implemented in the frequency domain. Instead, we switch to the time domain and multiply each slice $\mu$ with the complex exponential $\exp\{j2\pi(f_\mu - f_{ref})\}$. The reference frequency $f_{ref}$ corresponds to a global frequency shift of the entire signal spectrum and is chosen to keep the sampling rate in the DSP algorithms manageable.

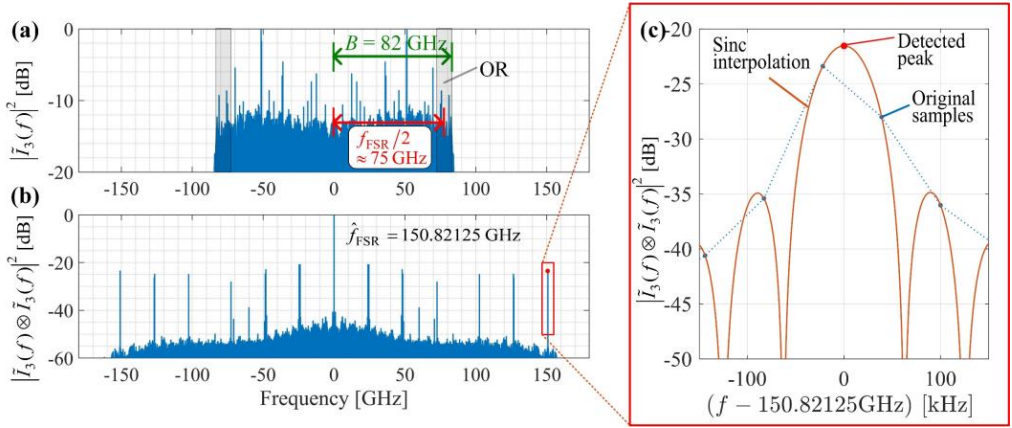

**Fig. S3.** FSR estimation of the free-running LO comb. **(a)** Received baseband spectrum $\tilde{I}_3(f)$ of the in-phase component of IQR 3. We indicate the receiver bandwidth $B$, half the free spectral range $f_{\mathrm{FSR}}/2$, and the down-converted overlap region (OR). **(b)** Squared modulus of the autocorrelation $\tilde{I}_3(f) \otimes \tilde{I}_3(f)$ of the spectrum $\tilde{I}_3(f)$ shown in (a), exhibiting a peak at $f_{\mathrm{FSR}}$. Additional peaks are caused by the fact that our test signal contains data channels that were generated by the same electro-optic modulator, see Fig. 2 in the main manuscript. **(b)** Zoom-in of the peak around $\hat{f}_{\mathrm{FSR}} = 150.82125\,\mathrm{GHz}$. We interpolate the data to estimate the $f_{\mathrm{FSR}}$ with high precision.

## 1.4. Phase-drift compensation

For a practical implementation of the OAWM system, all transfer functions $\tilde{\underline{H}}_{\nu\mu}^{(\mathrm{I,t})}(f)$ and $\tilde{\underline{H}}_{\nu\mu}^{(\mathrm{Q,t})}(f)$ as used in Eqs. (S12) and (S13) must be determined, which requires a calibration of the system. Importantly, the calibrated transfer functions might be impaired by amplitude and phase drifts of the comb lines, see Eqs. (S10) and (S11), and by random phase drifts in the setup. The amplitude and phase changes occur on a time scale of hundreds of microseconds, and the transfer functions $\tilde{\underline{H}}_{\nu\mu}^{(\mathrm{I,t})}(f)$ and $\tilde{\underline{H}}_{\nu\mu}^{(\mathrm{Q,t})}(f)$ can therefore be considered constant during one recording with a typical length of a few microseconds but may still vary between different recordings. We thus can separate the transfer functions $\tilde{\underline{H}}_{\nu\mu}^{(\mathrm{I,t})}(f)$ and $\tilde{\underline{H}}_{\nu\mu}^{(\mathrm{Q,t})}(f)$ into a time-invariant, but frequency-dependent part $\tilde{\underline{H}}_{\nu\mu}^{(\mathrm{I})}(f)$ and $\tilde{\underline{H}}_{\nu\mu}^{(\mathrm{Q})}(f)$, and a time-variant but frequency-independent complex-valued factor $\underline{H}_{\mathrm{F},\nu}^{(\mathrm{t})} \times \underline{H}_{\mathrm{LO},\mu}^{(\mathrm{t})}$ that is constant during one recording, but varies between recordings,

$$\tilde{\underline{H}}_{\nu\mu}^{(\mathrm{I,t})}(f) = \underline{H}_{\mathrm{F},\nu}^{(\mathrm{t})} \times \underline{H}_{\mathrm{LO},\mu}^{(\mathrm{t})} \times \tilde{\underline{H}}_{\nu\mu}^{(\mathrm{I})}(f),$$
$$\tilde{\underline{H}}_{\nu\mu}^{(\mathrm{Q,t})}(f) = \underline{H}_{\mathrm{F},\nu}^{(\mathrm{t})} \times \underline{H}_{\mathrm{LO},\mu}^{(\mathrm{t})} \times \tilde{\underline{H}}_{\nu\mu}^{(\mathrm{Q})}(f).$$
(S22)

Note that the complex-valued factor $\underline{H}_{\mathrm{F},\nu}^{(\mathrm{t})} \times \underline{H}_{\mathrm{LO},\mu}^{(\mathrm{t})}$ is assumed to be the same for the in-phase and quadrature detection channel, both of which contain the same fiber path and are fed by the same LO signal. The frequency-dependent parts $\tilde{\underline{H}}_{\nu\mu}^{(\mathrm{I})}(f)$ and $\tilde{\underline{H}}_{\nu\mu}^{(\mathrm{Q})}(f)$ are determined during a one-time calibration procedure, see Sect. 4 in this document and Fig. 3 in the main paper, while the parameters $\underline{H}_{\mathrm{F},\nu}^{(\mathrm{t})}$ and $\underline{H}_{\mathrm{LO},\mu}^{(\mathrm{t})}$ are estimated for each recording from redundant spectral components. As the absolute optical phase of the signal is not of importance, we set $\underline{H}_{\mathrm{F},1}^{(\mathrm{t})} = \underline{H}_{\mathrm{LO},1}^{(\mathrm{t})} = 1$ and determine all remaining parameters relative to the first channel $\nu = 1$ and the first comb line $\mu = 1$, respectively. We merge the frequency-independent complex-valued factors $\underline{H}_{\mathrm{F},\nu}^{(\mathrm{t})}$ and $\underline{H}_{\mathrm{LO},\mu}^{(\mathrm{t})}$ into a parameter vector $\underline{\mathbf{p}}$, which can be written as $\underline{\mathbf{p}} = \left(\underline{H}_{\mathrm{F},2}^{(\mathrm{t})}, \ldots, \underline{H}_{\mathrm{F},N}^{(\mathrm{t})}, \underline{H}_{\mathrm{LO},2}^{(\mathrm{t})}, \ldots, \underline{H}_{\mathrm{LO},M}^{(\mathrm{t})}\right)$ in the most general case. This parameter vector is estimated for reach recording by minimizing a cost function $J(\underline{\mathbf{p}})$ that quantifies the deviation between redundantly reconstructed spectral signal portions in the overlap region (OR) of neighboring slices. In our implementation, we choose a cost function of the form

$$J(\underline{\mathbf{p}}) = \sum_{\mu=1}^{M-1} \frac{\int_{\text{OR}} \left| \tilde{\underline{a}}_{S,\mu}^{(\text{est})}(f,\underline{\mathbf{p}}) - \tilde{\underline{a}}_{S,\mu+1}^{(\text{est})}(f,\underline{\mathbf{p}}) \right|^2 df}{\int_{\text{OR}} \left| \tilde{\underline{a}}_{S,\mu}^{(\text{est})}(f,\underline{\mathbf{p}}) + \tilde{\underline{a}}_{S,\mu+1}^{(\text{est})}(f,\underline{\mathbf{p}}) \right|^2 df} , \qquad (S23)$$

which was found to lead to stable convergence. This cost function is minimized using the quasi-Newton method. For efficient implementation of the parameter estimation, it is imperative to reduce the number of free elements in the parameter vector $\underline{\mathbf{p}}$ as much as possible and to minimize the computational effort of calculating the cost according to Eq. (S23) in each iteration. This can be accomplished by using additional assumptions, which allow for certain mathematical simplifications as explained in the following sections.

To reduce the number of free parameters, we first make use of the fact that the optical fibers contribute phase drifts, but do not cause any amplitude fluctuations. We may thus reduce the number of free parameters by assuming $\underline{H}_{F,\nu}^{(t)} = \exp(j\varphi_{F,\nu}^{(t)})$, $\left| \underline{H}_{F,\nu}^{(t)} \right| = 1$. As an additional simplification, we may use the fact that the comb tones are strictly phase-locked and that the pulse shape of the LO comb is thus stable, rendering the delay $\tau$ of the pulse with respect to the trigger point $t = 0$ of the oscilloscope the only free parameter of the LO comb, where $\underline{H}_{\text{LO},\mu+1}^{(t)} / \underline{H}_{\text{LO},\mu}^{(t)} = \exp(j 2\pi f_{\text{FSR}} \tau)$, i.e., $\underline{\mathbf{p}} = \left( \varphi_{F,2}^{(t)}, ..., \varphi_{F,N}^{(t)}, \tau \right)$.

We further want to reduce the computational effort for evaluating the cost function $J(\underline{\mathbf{p}})$ according to Eq. (S23) in each iteration. The evaluation of $J(\underline{\mathbf{p}})$ implies the reconstruction of the redundant signal portions $\tilde{\underline{a}}_{S,\mu}^{(\text{est})}(f,\underline{\mathbf{p}})$ using a reconstruction matrix $\tilde{\underline{\mathbf{H}}}_{\text{rec}}^{(t)}(f)$ based on the respective parameter vector $\underline{\mathbf{p}}$. To minimize the computational effort, we reformulate the associated reconstruction relation according to Eq. (S19) such that all processing steps that are independent of the parameter vector $\underline{\mathbf{p}}$ can be moved outside the iteration loop and hence only need to be performed once. The various steps of this reformulation are described in the following sections.

In a first step, we need to overcome the problem that the general system model according to Eq. (S14) does not allow to separate the time-variant complex factors $\underline{H}_{F,\nu}^{(t)}$ and $\underline{H}_{\text{LO},\mu}^{(t)}$ from the time-invariant transfer functions $\tilde{\underline{H}}_{\nu\mu}^{(I)}(f)$ and $\tilde{\underline{H}}_{\nu\mu}^{(Q)}(f)$ when calculating the pseudo-inverse according to Eq. (S18). Consequently, for each evaluation of the cost function, we would need to compute an updated reconstruction matrix $\tilde{\underline{\mathbf{H}}}_{\text{rec}}^{(t)}(f)$ by first constructing the transfer matrix $\tilde{\underline{\mathbf{H}}}_{\text{IQ}}^{(t)}(f)$ using the parameters in $\underline{\mathbf{p}}$, see Eqs. (S14),(S15), and (S22), and by then computing the pseudo inverse, Eq. (S18), which is finally used for the reconstruction of the redundant components $\tilde{\underline{a}}_{S,\mu}^{(\text{est})}(f,\underline{\mathbf{p}})$ in the overlap regions. This would render the parameter optimization computationally expensive and slow. In contrast to the general model according Eq. (S14), the simplified model according to Eq. (S16) would allow to overcome this problem. In this notation, the time-variant parameters $\underline{H}_{F,\nu}^{(t)}$ and $\underline{H}_{\text{LO},\mu}^{(t)}$, see Eq. (S22), can be written in form of diagonal matrices $\underline{\mathbf{H}}_F^{(t)}$ and $\underline{\mathbf{H}}_{\text{LO}}^{(t)}$ that are multiplied from the left and the right to the frequency-dependent but time-invariant transfer matrix $\tilde{\underline{\mathbf{H}}}^{(f)}(f)$ – this will be detailed in the context of Eq. (S29) below. However, Eq. (S16) relied on the assumption that the in-phase and quadrature transfer functions $\tilde{\underline{H}}_{\nu\mu}^{(I,t)}(f)$ and $\tilde{\underline{H}}_{\nu\mu}^{(Q,t)}(f)$ are identical and have an ideal 90° phase relationship, $\tilde{\underline{\mathbf{H}}}^{(Q,t)}(f) = -j \tilde{\underline{\mathbf{H}}}^{(I,t)}(f)$, see Eq. (S16). This is not fulfilled for our experimental setup, which can, e.g., be inferred from Fig. S13 and Fig. S14 in Sect. 4.3, which show distinct differences in the amplitudes and phases of the measured frequency-dependent transfer functions for in-phase and quadrature phase detection channels of the same IQR.

To overcome this problem, we re-formulate the system model to arrive at a representation that requires weaker assumptions than the simplified model in Eq. (S16), but still leads to a relation of the form of Eq. (S16) between the signal vector $\tilde{\underline{\mathbf{A}}}_S(f)$ and a composite measurement vector $\tilde{\underline{\mathbf{U}}}(f)$, $\tilde{\underline{\mathbf{U}}}(f) = \tilde{\underline{\mathbf{H}}}(f)\tilde{\underline{\mathbf{A}}}_S(f)$. This is accomplished by defining the components $\tilde{\underline{U}}_\nu(f)$ of the measurement vector $\tilde{\underline{\mathbf{U}}}(f)$ as pre-corrected composite baseband spectra that are constructed from the in-phase $\tilde{I}_\nu(f)$ and quadrature $\tilde{Q}_\nu(f)$ components in several steps: In a first step, we divide the measured baseband spectra $\tilde{I}_\nu(f)$ and $\tilde{Q}_\nu(f)$ obtained from balanced detector BPD-I$_\nu$ and BPD-Q$_\nu$, Fig. S1, by the complex-conjugate of the mirrored frequency-dependent time-invariant transfer functions $\tilde{\underline{H}}_{\nu m}^{(I)*}(-f)$ and $-\mathrm{j}\tilde{\underline{H}}_{\nu m}^{(Q)*}(-f)$ of the respective detectors. These transfer functions are obtained for a certain LO line $\mu = m$, which can be chosen arbitrarily in the range $1 \leq m \leq M$ and which is assumed to represent the transfer function of the detector for all comb lines. We then add the results using the factor j to obtain the corresponding pre-corrected composite baseband spectrum $\tilde{\underline{U}}_\nu(f)$,

$$\tilde{\underline{U}}_\nu(f) = \left(\frac{1}{\tilde{\underline{H}}_{\nu m}^{(I)*}(-f)}\right)\tilde{I}_\nu(f) + \mathrm{j}\left(\frac{\mathrm{j}}{\tilde{\underline{H}}_{\nu m}^{(Q)*}(-f)}\right)Q_\nu(f). \tag{S24}$$

Introducing Eqs. (S12) and (S13) into Eq. (S24) leads to a condition for which $\tilde{\underline{U}}_\nu(f)$ becomes independent of the complex-conjugate frequency-inverted signal components $\tilde{\underline{a}}_S^*(-f + f_\mu)$, as in the representation of Eq. (S16),

$$\tilde{\underline{U}}_\nu(f) = \sum_{\mu=1}^{M}\left[\underbrace{\left(\frac{\tilde{\underline{H}}_{\nu\mu}^{(I,t)}(f)}{\tilde{\underline{H}}_{\nu m}^{(I)*}(-f)} - \frac{\tilde{\underline{H}}_{\nu\mu}^{(Q,t)}(f)}{\tilde{\underline{H}}_{\nu m}^{(Q)*}(-f)}\right)}_{\tilde{\underline{H}}_{\nu\mu}(f)}\tilde{\underline{a}}_S(f+f_\mu) + \underbrace{\left(\frac{\tilde{\underline{H}}_{\nu\mu}^{(I,t)*}(-f)}{\tilde{\underline{H}}_{\nu m}^{(I)*}(-f)} - \frac{\tilde{\underline{H}}_{\nu\mu}^{(Q,t)*}(-f)}{\tilde{\underline{H}}_{\nu m}^{(Q)*}(-f)}\right)}_{\approx 0}\tilde{\underline{a}}_S^*(-f+f_\mu)\right]$$

$$+ \underbrace{\left(\frac{\tilde{G}_\nu^{(I)}(f)}{\tilde{\underline{H}}_{\nu m}^{(I)*}(-f)} - \frac{\tilde{G}_\nu^{(Q)}(f)}{\tilde{\underline{H}}_{\nu m}^{(Q)*}(-f)}\right)}_{\tilde{\underline{G}}_\nu(f)} \tag{S25}$$

$$\approx \sum_{\mu=1}^{M}\left[\tilde{\underline{H}}_{\nu\mu}(f)\tilde{\underline{a}}_S(f+f_\mu)\right] + \tilde{\underline{G}}_\nu(f),$$

for

$$\frac{\tilde{\underline{H}}_{\nu\mu}^{(I,t)*}(-f)}{\tilde{\underline{H}}_{\nu m}^{(I)*}(-f)} - \frac{\tilde{\underline{H}}_{\nu\mu}^{(Q,t)*}(-f)}{\tilde{\underline{H}}_{\nu m}^{(Q)*}(-f)} = 0. \tag{S26}$$

It turns out that the condition (S26) is fulfilled if we assume that the wavelength-dependent optical characteristics $\tilde{\underline{h}}_{\nu\mu}^{(t)}(f)$ of the in-phase and quadrature channels are identical, while still allowing for individual electrical responses $\tilde{\underline{H}}_\nu^{(I)}(f)$ and $\tilde{\underline{H}}_\nu^{(Q)}(f)$ for the in-phase and quadrature components, respectively,

$$\tilde{\underline{H}}_{\nu\mu}^{(I,t)}(f) \approx \tilde{\underline{H}}_\nu^{(I)}(f)\tilde{\underline{h}}_{\nu\mu}^{(t)}(f),$$
$$\tilde{\underline{H}}_{\nu\mu}^{(Q,t)}(f) \approx \tilde{\underline{H}}_\nu^{(Q)}(f)\tilde{\underline{h}}_{\nu\mu}^{(t)}(f). \tag{S27}$$

For the devices used in our experiment, this assumption is fulfilled to a very good approximation. Reformulating the last line of Eq. (S25) in matrix-vector format leads to a relationship between the vector $\tilde{\underline{\mathbf{U}}}(f) = (\tilde{\underline{U}}_1(f)\ldots\tilde{\underline{U}}_N(f))^\mathrm{T}$ of measured composite baseband spectra, pre-corrected according to Eq. (S24), and the accordingly modified transfer matrix $\tilde{\underline{\mathbf{H}}}(f)$ with elements $\tilde{\underline{H}}_{\nu\mu}(f)$ according to Eq. (S25),

$$\tilde{\underline{\mathbf{U}}}(f) = \tilde{\underline{\mathbf{H}}}(f)\tilde{\underline{\mathbf{A}}}_S(f) + \tilde{\underline{\mathbf{G}}}(f) \tag{S28}$$

In a next step, we can reformulate this relation such that processing steps that are independent of the parameter vector $\underline{\mathbf{p}}$ can be moved outside the optimization loop and hence only need to be performed once. Specifically, this applies to the computation of the frequency-dependent pseudo inverse $\underline{\tilde{\mathbf{H}}}^{(f)+}(f)$ and the multiplication with the measurement vector $\underline{\tilde{\mathbf{U}}}(f)$, which do not depend on the parameter vector $\underline{\mathbf{p}}$. To this end, we decompose the modified transfer matrix $\underline{\tilde{\mathbf{H}}}(f)$ in Eq. (S28) into a frequency-dependent but time-invariant transfer matrix $\underline{\tilde{\mathbf{H}}}^{(f)}(f)$, that is multiplied from both sides with diagonal matrices $\underline{\mathbf{H}}_F^{(t)}$ and $\underline{\mathbf{H}}_{LO}^{(t)}$ that contain the time-variant factors $\underline{H}_{F,\nu}^{(t)}$ and $\underline{H}_{LO,\mu}^{(t)}$ according to Eq. (S22),

$$\underbrace{\begin{bmatrix} \tilde{U}_1(f) \\ \vdots \\ \tilde{U}_N(f) \end{bmatrix}}_{\underline{\tilde{\mathbf{U}}}(f)} = \underbrace{\begin{bmatrix} \underline{H}_{F,1}^{(t)} & \cdots & 0 \\ \vdots & \ddots & \vdots \\ 0 & \cdots & \underline{H}_{F,N}^{(t)} \end{bmatrix}}_{\underline{\mathbf{H}}_F^{(t)}} \underbrace{\begin{bmatrix} \underline{\tilde{H}}_{11}^{(f)}(f) & \cdots & \underline{\tilde{H}}_{1M}^{(f)}(f) \\ \vdots & \ddots & \vdots \\ \underline{\tilde{H}}_{N1}^{(f)}(f) & \cdots & \underline{\tilde{H}}_{NM}^{(f)}(f) \end{bmatrix}}_{\underline{\tilde{\mathbf{H}}}^{(f)}(f)} \underbrace{\begin{bmatrix} \underline{H}_{LO,1}^{(t)} & \cdots & 0 \\ \vdots & \ddots & \vdots \\ 0 & \cdots & \underline{H}_{LO,M}^{(t)} \end{bmatrix}}_{\underline{\mathbf{H}}_{LO}^{(t)}} \underbrace{\begin{bmatrix} \tilde{\underline{a}}_S(f+f_1) \\ \vdots \\ \tilde{\underline{a}}_S(f+f_M) \end{bmatrix}}_{\underline{\tilde{\mathbf{A}}}_S(f)} + \underbrace{\begin{bmatrix} \tilde{\underline{G}}_1(f) \\ \vdots \\ \tilde{\underline{G}}_N(f) \end{bmatrix}}_{\underline{\tilde{\mathbf{G}}}(f)}.$$

$$\underbrace{\phantom{}}_{\underline{\tilde{\mathbf{H}}}(f)}$$

(S29)

Note that this separation is possible because the matrix elements of the modified transfer matrix $\underline{\tilde{\mathbf{H}}}(f)$ in Eq. (S28) can be written as a product $\underline{\tilde{H}}_{\nu\mu}(f) = \underline{H}_{F,\nu}^{(t)} \times \underline{H}_{LO,\mu}^{(t)} \times \underline{\tilde{H}}_{\nu\mu}^{(f)}(f)$ of the corresponding time-invariant element $\underline{\tilde{H}}_{\nu\mu}^{(f)}(f)$ and the time-variant factors $\underline{H}_{F,\nu}^{(t)}$ and $\underline{H}_{LO,\mu}^{(t)}$, which are identical for the in-phase and quadrature detection path according to Eq. (S22).

The formulation according to Eq. (S29) allows to separate the computation of the pseudo inverse $\underline{\tilde{\mathbf{H}}}^{(f)+}(f)$ of the time-invariant transfer matrix $\underline{\tilde{\mathbf{H}}}^{(f)}(f)$, that is obtained from a one-time calibration measurement, from the multiplication with the inverse $\left(\underline{\mathbf{H}}_F^{(t)}\right)^{-1}$ and $\left(\underline{\mathbf{H}}_{LO}^{(t)}\right)^{-1}$ of the time-variant diagonal matrices $\underline{\mathbf{H}}_F^{(t)}$ and $\underline{\mathbf{H}}_{LO}^{(t)}$,

$$\begin{aligned}\underline{\tilde{\mathbf{H}}}^+ = \left(\underline{\tilde{\mathbf{H}}}^\dagger \underline{\tilde{\mathbf{H}}}\right)^{-1} \underline{\tilde{\mathbf{H}}}^\dagger &= \left(\mathbf{H}_{LO}^{(t)\dagger}\ \underline{\tilde{\mathbf{H}}}^{(f)\dagger}\ \underbrace{\mathbf{H}_F^{(t)\dagger} \mathbf{H}_F^{(t)}}_{\text{identity, as } \left|\underline{H}_{F,\nu}^{(t)}\right|=1}\ \underline{\tilde{\mathbf{H}}}^{(f)}\ \mathbf{H}_{LO}^{(t)}\right)^{-1} \mathbf{H}_{LO}^{(t)\dagger}\ \underline{\tilde{\mathbf{H}}}^{(f)\dagger}\ \mathbf{H}_F^{(t)\dagger} \\ &= \left(\mathbf{H}_{LO}^{(t)}\right)^{-1} \left(\underline{\tilde{\mathbf{H}}}^{(f)\dagger} \underline{\tilde{\mathbf{H}}}^{(f)}\right)^{-1} \underbrace{\left(\mathbf{H}_{LO}^{(t)\dagger}\right)^{-1} \mathbf{H}_{LO}^{(t)\dagger}}_{\text{identity matrix}}\ \underline{\tilde{\mathbf{H}}}^{(f)\dagger}\ \mathbf{H}_F^{(t)\dagger} \\ &= \left(\mathbf{H}_{LO}^{(t)}\right)^{-1} \underbrace{\left(\underline{\tilde{\mathbf{H}}}^{(f)\dagger} \underline{\tilde{\mathbf{H}}}^{(f)}\right)^{-1} \underline{\tilde{\mathbf{H}}}^{(f)\dagger}}_{\text{pseudo inverse } \underline{\tilde{\mathbf{H}}}^{(f)+}(f)} \left(\mathbf{H}_F^{(t)}\right)^{-1} \\ &= \left(\mathbf{H}_{LO}^{(t)}\right)^{-1} \underline{\tilde{\mathbf{H}}}^{(f)+} \left(\mathbf{H}_F^{(t)}\right)^{-1}.\end{aligned}$$

(S30)

In this relation, we dropped the frequency argument of the transfer matrix $\underline{\tilde{\mathbf{H}}}(f)$ and $\underline{\tilde{\mathbf{H}}}^{(f)}(f)$ for better readability and use the fact that $\left|\underline{H}_{F,\nu}^{(t)}\right|=1$ and that consequently the diagonal matrix $\underline{\mathbf{H}}_F^{(t)}$ is unitary, $\underline{\mathbf{H}}_F^{(t)\dagger} = \left(\underline{\mathbf{H}}_F^{(t)}\right)^{-1}$. Based on Eq. (S30), the reconstruction relation according to Eq. (4) of the main manuscript can be written as

$$\underline{\tilde{\mathbf{A}}}_S^{(\mathrm{est})}(f,\underline{\mathbf{p}}) = \left(\underline{\mathbf{H}}_{LO}^{(t)}(\underline{\mathbf{p}})\right)^{-1} \underline{\tilde{\mathbf{H}}}^{(f)+}(f) \left(\underline{\mathbf{H}}_F^{(t)}(\underline{\mathbf{p}})\right)^{-1} \underline{\tilde{\mathbf{U}}}(f) \tag{S31}$$

In this relation, we explicitly indicate the dependence of the diagonal matrices $\underline{\mathbf{H}}_F^{(t)}(\underline{\mathbf{p}})$ and $\underline{\mathbf{H}}_{LO}^{(t)}(\underline{\mathbf{p}})$, and of the resulting signal vector $\underline{\tilde{\mathbf{A}}}_S^{(\mathrm{est})}(f,\underline{\mathbf{p}})$ on the parameters $\underline{\mathbf{p}}$.

We now use Eq. (S31) for estimating the redundant signal portions $\tilde{\underline{a}}_{S,\mu}^{(\mathrm{est})}(f,\underline{\mathbf{p}})$, $\mu=1,...,M$, in the overlap regions, $f \in \left[f_{\mu+1}-B, f_\mu+B\right]$, $\mu=1,...,M-1$. The estimated redundant signal portions $\tilde{\underline{a}}_{S,\mu}^{(\mathrm{est})}(f,\underline{\mathbf{p}})$ then serve as an input for evaluating the cost function $J(\underline{\mathbf{p}})$ according to Eq. (S23). In this step, the computational effort can be further reduced by merging the measured baseband spectra $\tilde{\underline{\mathbf{U}}}(f)$, which do not depend on the parameters $\mathbf{p}$, with the $\mathbf{p}$-independent pseudo-inverse $\tilde{\underline{\mathbf{H}}}^{(\mathrm{f})+}(f)$. To this end, we start from a component-wise notation of Eq. (S31) and swap the on-diagonal elements of $\left(\underline{\mathbf{H}}_{\mathrm{F}}^{(\mathrm{t})}(\underline{\mathbf{p}})\right)^{-1}$ with elements of the vector $\tilde{\underline{\mathbf{U}}}(f) = \left(\tilde{\underline{U}}_1(f)\ldots\tilde{\underline{U}}_N(f)\right)^{\mathrm{T}}$ containing the measured pre-corrected baseband spectra $\tilde{\underline{U}}_\nu(f)$,

$$
\underbrace{\begin{bmatrix} \tilde{\underline{a}}_S^{(\mathrm{est})}(f+f_1,\underline{\mathbf{p}}) \\ \vdots \\ \tilde{\underline{a}}_S^{(\mathrm{est})}(f+f_M,\underline{\mathbf{p}}) \end{bmatrix}}_{\tilde{\underline{\mathbf{A}}}_S^{(\mathrm{est})}(f,\underline{\mathbf{p}})} = \underbrace{\begin{bmatrix} H_{\mathrm{LO},1}^{(\mathrm{t})} & \cdots & 0 \\ \vdots & \ddots & \vdots \\ 0 & \cdots & H_{\mathrm{LO},M}^{(\mathrm{t})} \end{bmatrix}}_{\underline{\mathbf{H}}_{\mathrm{LO}}^{(\mathrm{t})}(\underline{\mathbf{p}})} \underbrace{\begin{bmatrix} \tilde{\underline{H}}_{11}^{(\mathrm{f})}(f) & \cdots & \tilde{\underline{H}}_{1M}^{(\mathrm{f})}(f) \\ \vdots & \ddots & \vdots \\ \tilde{\underline{H}}_{N1}^{(\mathrm{f})}(f) & \cdots & \tilde{\underline{H}}_{NM}^{(\mathrm{f})}(f) \end{bmatrix}}_{\tilde{\underline{\mathbf{H}}}^{(\mathrm{f})}(f)} \underbrace{\begin{bmatrix} H_{\mathrm{F},1}^{(\mathrm{t})} & \cdots & 0 \\ \vdots & \ddots & \vdots \\ 0 & \cdots & H_{\mathrm{F},N}^{(\mathrm{t})} \end{bmatrix}}_{\underline{\mathbf{H}}_{\mathrm{F}}^{(\mathrm{t})}(\underline{\mathbf{p}})} \underbrace{\begin{bmatrix} \tilde{\underline{U}}_1(f) \\ \vdots \\ \tilde{\underline{U}}_N(f) \end{bmatrix}}_{\tilde{\underline{\mathbf{U}}}(f)}
$$
$$
= \underbrace{\begin{bmatrix} H_{\mathrm{LO},1}^{(\mathrm{t})} & \cdots & 0 \\ \vdots & \ddots & \vdots \\ 0 & \cdots & H_{\mathrm{LO},M}^{(\mathrm{t})} \end{bmatrix}}_{\underline{\mathbf{H}}_{\mathrm{LO}}^{(\mathrm{t})}(\underline{\mathbf{p}})} \underbrace{\begin{bmatrix} \tilde{\underline{H}}_{11}^{(\mathrm{f})}(f) & \cdots & \tilde{\underline{H}}_{1M}^{(\mathrm{f})}(f) \\ \vdots & \ddots & \vdots \\ \tilde{\underline{H}}_{N1}^{(\mathrm{f})}(f) & \cdots & \tilde{\underline{H}}_{NM}^{(\mathrm{f})}(f) \end{bmatrix} \begin{bmatrix} \tilde{\underline{U}}_1(f) & \cdots & 0 \\ \vdots & \ddots & \vdots \\ 0 & \cdots & \tilde{\underline{U}}_N(f) \end{bmatrix}}_{\tilde{\underline{\mathbf{B}}}(f)} \begin{bmatrix} H_{\mathrm{F},1}^{(\mathrm{t})} \\ \vdots \\ H_{\mathrm{F},N}^{(\mathrm{t})} \end{bmatrix}.
$$
(S32)

The auxiliary matrix $\tilde{\underline{\mathbf{B}}}(f)$ is independent of the parameter vector $\underline{\mathbf{p}}$ and can thus be computed outside the optimization loop. For each frequency in the overlap regions, the redundant signal portions $\tilde{\underline{a}}_{S,\mu}^{(\mathrm{est})}(f,\underline{\mathbf{p}})$ are then obtained with relatively little effort by multiplying the auxiliary matrix $\tilde{\underline{\mathbf{B}}}(f)$ with the time-variant factors $H_{\mathrm{LO},\mu}^{(\mathrm{t})}$ and $H_{\mathrm{F},\nu}^{(\mathrm{t})}$, organized in a diagonal matrix and in a column vector, respectively. These signal portions are then used to evaluate and iteratively minimize the cost function in Eq. (S23).

### 1.5. Reconstruction accuracy

The reconstruction of the signal vector $\tilde{\underline{\mathbf{A}}}_S(f)$ from Eq. (S28) requires the pseudo-inverse of the transfer matrix $\tilde{\underline{\mathbf{H}}}(f)$, and the reconstruction accuracy thus depends on how well $\tilde{\underline{\mathbf{H}}}(f)$ is conditioned. In the following, we investigate the impact of the time delays $\tau_\nu$, see Fig. 1 of the main manuscript, on the condition of the transfer matrix $\tilde{\underline{\mathbf{H}}}(f)$. Focusing on the practically relevant case presented in the main manuscript, we assume that the number $M$ of LO comb lines is equal to the number $N$ of IQ receivers, $M = N$, leading to a square $N \times N$ matrix. For such matrices, the condition number $\mathrm{cond}_2(\tilde{\underline{\mathbf{H}}}(f))$ with respect to the $L^2$-norm, $1 \leq \mathrm{cond}_2(\tilde{\underline{\mathbf{H}}}(f)) < \infty$, is a widely used metric for the matrix condition. To illustrate the idea, we reformulate Eq. (S28),

$$\tilde{\underline{\mathbf{H}}}(f)\ \tilde{\underline{\mathbf{A}}}_S(f) = \left[\tilde{\underline{\mathbf{U}}}(f) - \tilde{\underline{\mathbf{G}}}(f)\right]. \tag{S33}$$

In this relation, the unknown noise $\tilde{\underline{\mathbf{G}}}(f)$ represents a perturbation, rendering the measured baseband signals $\tilde{\underline{\mathbf{U}}}(f)$ different from their ideal, but unknown counterparts $\tilde{\underline{\mathbf{U}}}(f) - \tilde{\underline{\mathbf{G}}}(f)$. This will lead to a corresponding deviation of the reconstructed signal vector $\tilde{\underline{\mathbf{A}}}_S^{(\mathrm{est})}(f) = \tilde{\underline{\mathbf{H}}}^{-1}(f)\tilde{\underline{\mathbf{U}}}(f)$ and its ideal counterpart $\tilde{\underline{\mathbf{A}}}_S(f)$. The condition number $\mathrm{cond}_2(\tilde{\underline{\mathbf{H}}}(f))$ now corresponds to the maximum ratio of the relative error $\|\tilde{\underline{\mathbf{A}}}_S^{(\mathrm{est})}(f) - \tilde{\underline{\mathbf{A}}}_S(f)\| / \|\tilde{\underline{\mathbf{A}}}_S(f)\|$ of the reconstructed signal vector and the relative error $\|\tilde{\underline{\mathbf{G}}}(f)\| / \|\tilde{\underline{\mathbf{U}}}(f) - \tilde{\underline{\mathbf{G}}}(f)\|$ of the measured baseband signals [2],

$$\frac{\|\tilde{\underline{\mathbf{A}}}_S^{(\mathrm{est})}(f) - \tilde{\underline{\mathbf{A}}}_S(f)\|}{\|\tilde{\underline{\mathbf{A}}}_S(f)\|} \leq \mathrm{cond}_2(\tilde{\underline{\mathbf{H}}}(f)) \times \frac{\|\tilde{\underline{\mathbf{G}}}(f)\|}{\|\tilde{\underline{\mathbf{U}}}(f) - \tilde{\underline{\mathbf{G}}}(f)\|}. \tag{S34}$$

In these relations $\|\cdot\|$ indicates the $L^2$-norm of the corresponding column vector. The square of the expression on the right hand-side of Eq. (S34), $\|\tilde{\underline{\mathbf{G}}}(f)\|^2 / \|\tilde{\underline{\mathbf{U}}}(f) - \tilde{\underline{\mathbf{G}}}(f)\|^2$, can be interpreted as the ratio of the sum of the noise powers $\|\tilde{\underline{\mathbf{G}}}(f)\|^2 = \sum_1^N |\tilde{\underline{G}}_\nu(f)|^2$ in the various channels and the sum of the corresponding noise-free signal powers $\|\tilde{\underline{\mathbf{U}}}(f) - \tilde{\underline{\mathbf{G}}}(f)\|^2 = \sum_1^N |\tilde{\underline{U}}_\nu(f) - \tilde{\underline{G}}_\nu(f)|^2$ and may be understood as a quality metric of the measured signals. Note, however, that this ratio cannot be simply related to the frequency-dependent SNR of the various receiver channels, which is defined by the ratio of the individual signal and noise powers, $\mathrm{SNR}_\nu(f) = \overline{|\tilde{\underline{U}}_\nu(f)|^2} / \overline{|\tilde{\underline{G}}_\nu(f)|^2}$, where the overbar indicates an ensemble average over all noise realizations. Still, we may infer from Eq. (S34) the heuristic notion that a smaller condition number $\mathrm{cond}_2(\tilde{\underline{\mathbf{H}}}(f))$ leads to a more accurate estimation $\tilde{\underline{\mathbf{A}}}_\mathrm{S}^{(\mathrm{est})}(f)$. The condition number cannot be smaller than one, corresponding to the somewhat intuitive fact that the reconstructed signal cannot be more accurate than its measured counterpart in case of a square transfer matrix $\tilde{\underline{\mathbf{H}}}(f)$.

For estimating how the condition number $\mathrm{cond}_2(\tilde{\underline{\mathbf{H}}}(f))$ of the transfer matrix $\tilde{\underline{\mathbf{H}}}(f)$ depends on the LO time delays $\tau_\nu$, we first assume that the time-invariant frequency responses $\tilde{\underline{H}}_{\nu\mu}^{(\mathrm{f})}(f)$ as defined in Eq. (S29) are identical for all receivers, except for the phase factors associated with the various LO tones,

$$\tilde{\underline{H}}_{\nu\mu}^{(\mathrm{f})}(f) = \tilde{\underline{H}}^{(\mathrm{f})}(f) \exp(\mathrm{j}\varphi_{\nu\mu}) \quad \forall \nu, \mu = 1,\ldots,N, \quad \varphi_{\nu\mu} = 2\pi(\mu-1)\frac{\tau_\nu}{T_\mathrm{LO}}, \tag{S35}$$

where $T_\mathrm{LO} = 1/f_\mathrm{FSR}$ is the repetition period of the LO comb. This assumption renders the associated matrix condition number frequency independent. Inserting Eq. (S35) in Eq. (6) of the main manuscript, we obtain the matrix elements $\tilde{\underline{H}}_{\nu\mu}(f)$ of the overall transfer matrix $\tilde{\underline{\mathbf{H}}}(f)$,

$$\tilde{\underline{H}}_{\nu\mu}(f) = \underline{H}_{\mathrm{F},\nu}^{(\mathrm{t})} \times \tilde{\underline{H}}^{(\mathrm{f})}(f) \exp(\mathrm{j}\varphi_{\nu\mu}) \times \underline{H}_{\mathrm{LO},\mu}^{(\mathrm{t})}. \tag{S36}$$

The time-variant complex-valued factors $\underline{H}_{\mathrm{F},\nu}^{(\mathrm{t})}$, $\underline{H}_{\mathrm{LO},\mu}^{(\mathrm{t})}$ in Eq. (S36) do not significantly affect the matrix condition of $\tilde{\underline{\mathbf{H}}}(f)$, because their absolute value is close to one, $|\underline{H}_{\mathrm{F},\nu}^{(\mathrm{t})}| \approx 1$ and $|\underline{H}_{\mathrm{LO},\mu}^{(\mathrm{t})}| \approx 1$, and because they may be written in form of diagonal matrices, see $\underline{\mathbf{H}}_\mathrm{F}^{(\mathrm{t})}$ and $\underline{\mathbf{H}}_\mathrm{LO}^{(\mathrm{t})}$ in Eq. (S29). Consequently, the condition number of $\tilde{\underline{\mathbf{H}}}(f)$ is approximately independent of the fluctuations of the LO tones and the various receiver-path phase fluctuations, and according to Eq. (S35), predominantly depends on the relative delays $\tau_\nu / T_\mathrm{LO}$. For equidistant delays,

$$\tau_\nu = \frac{T_\mathrm{LO}(\nu-1)}{N}, \tag{S37}$$

the phase $\varphi_{\nu\mu}$ according to Eq. (S35) is given by $\varphi_{\nu\mu} = 2\pi(\mu-1)(\nu-1)/N$. In that case the condition number assumes its minimum possible value of one. Specifically, for $N = 4$ the condition number of the matrix $\tilde{\underline{\mathbf{H}}}^{(\mathrm{f})}(f)$ then reads,

$$\mathrm{cond}_2\{\tilde{\underline{\mathbf{H}}}(f)\} \approx \mathrm{cond}_2\{\exp(\mathrm{j}\varphi_{\nu\mu})\} = \mathrm{cond}_2 \begin{pmatrix} 1 & 1 & 1 & 1 \\ 1 & \mathrm{j} & -1 & -\mathrm{j} \\ 1 & -1 & 1 & -1 \\ 1 & -\mathrm{j} & -1 & \mathrm{j} \end{pmatrix} = 1. \tag{S38}$$

Conversely, if the relative time delays $\tau_\nu / T_\mathrm{LO}$ deviate from their ideal counterparts, the matrix condition number is increased. This can be seen from Fig. S4 (a), where the matrix condition number $\mathrm{cond}_2(\tilde{\underline{\mathbf{H}}}(f))$ of the transfer matrix $\tilde{\underline{\mathbf{H}}}(f)$ according to Eq. (S36) for $N = 3$ is shown as a function of $\tau_2$ and $\tau_3$, while $\tau_1 = 0$ without loss of generality. Close to the ideal equidistant delays, $\tau_2 = \frac{1}{3}T_\mathrm{LO}$ and $\tau_3 = \frac{2}{3}T_\mathrm{LO}$ or $\tau_2 = \frac{2}{3}T_\mathrm{LO}$ and $\tau_3 = \frac{1}{3}T_\mathrm{LO}$, the system is well conditioned. If any two time delays come close to each other, $\tau_\nu \approx \tau_n$, for $\nu = 1,\ldots,N$, $n = 1,\ldots,N$, and $\nu \neq n$, the condition number approaches infinity, and $\tilde{\underline{\mathbf{H}}}(f)$ becomes a singular matrix.

If we relax the condition formulated in Eq. (S35) allowing for transfer functions $\underline{\tilde{H}}_{\nu\mu}^{(f)}(f)$ with non-identical frequency dependencies, then the condition number will become frequency-dependent. We illustrate this effect in Fig. S4 (b) by plotting the condition number of the actually measured transfer matrix $\underline{\tilde{\mathbf{H}}}_{IQ}(f)$ according to Eq. (S14) as a function of frequency $f$. Since the photodetectors used in our experiment have quite different amplitude responses, see Fig. S13 in Sect. 4.3, the relative magnitude of the elements within the same column of the transfer matrix $\underline{\tilde{\mathbf{H}}}_{IQ}(f)$ varies with frequency, see Eqs. (S14) and (S15). This leads to a frequency-dependent condition number, $\mathrm{cond}_2\left(\underline{\tilde{\mathbf{H}}}_{IQ}(f)\right)$, blue trace in Fig. S4 (b), despite the time delays $\tau_\nu$ being close to their optimum values. If we eliminate the frequency-dependent magnitudes of the measured transfer-matrix elements by normalizing them, $\tilde{H}_{IQ,\nu\mu}^{(\mathrm{norm})}(f) = \tilde{H}_{IQ,\nu\mu}(f)/|\tilde{H}_{IQ,\nu\mu}(f)|$, we can again observe an almost frequency-independent condition number $\mathrm{cond}_2\left(\underline{\tilde{\mathbf{H}}}_{IQ}^{(\mathrm{norm})}(f)\right)$, orange trace in Fig. S4 (b). This confirms that the delays $\tau_\nu$ in our experiments have been chosen close to their optimum values according to Eq. (S37).

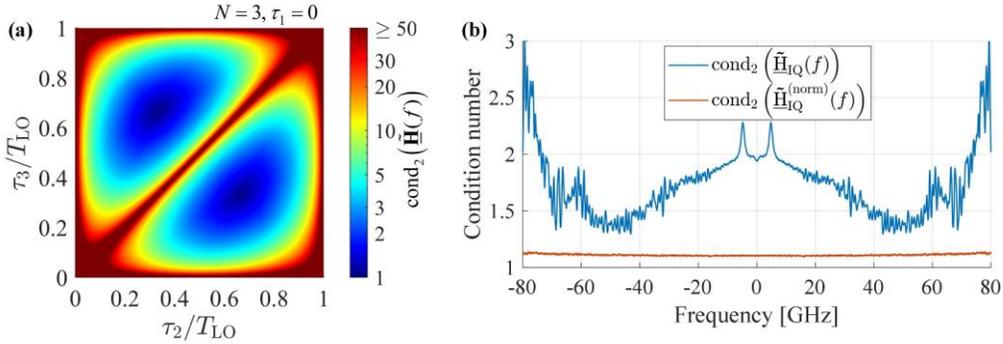

**Fig. S4.** Dependence of the matrix condition number on the time delays $\tau_\nu$ and the frequency $f$. **(a)** Condition number $\mathrm{cond}_2\left(\underline{\tilde{\mathbf{H}}}(f)\right)$ of the square matrix $\underline{\tilde{\mathbf{H}}}(f)$ as a function of the delay $\tau_2$ and $\tau_3$ according to Eq. (S36) for $N = 3$ IQ receivers and $\tau_1 = 0$. Close to the ideal equidistant delays, $\tau_2 = \tfrac{1}{3}T_{LO}$ and $\tau_3 = \tfrac{2}{3}T_{LO}$ or $\tau_2 = \tfrac{2}{3}T_{LO}$ and $\tau_3 = \tfrac{1}{3}T_{LO}$, the system is well conditioned. Note that the logarithmic color scale is clipped at $\mathrm{cond}_2\left(\underline{\tilde{\mathbf{H}}}(f)\right) = 50$ for better readability. **(b)** Condition number of the measured transfer matrix $\underline{\tilde{\mathbf{H}}}_{IQ}(f)$ according to Eq. (S14) for $N = 4$. Since the photodetectors used in the experiment have different amplitude responses, see Fig. S13 in Sect. 4.3, we observe frequency-dependent magnitudes of matrix elements within a column of the transfer matrix $\underline{\tilde{\mathbf{H}}}_{IQ}(f)$. Therefore, the condition number, $\mathrm{cond}_2\left(\underline{\tilde{\mathbf{H}}}_{IQ}(f)\right)$, varies with frequency, blue trace. If we eliminate the frequency-dependent magnitudes of the measured transfer-matrix elements by normalizing them, $\tilde{H}_{IQ,\nu\mu}^{(\mathrm{norm})}(f) = \tilde{H}_{IQ,\nu\mu}(f)/|\tilde{H}_{IQ,\nu\mu}(f)|$, we can observe an almost frequency-independent condition number close to one, orange trace. This confirms that the delays in our experiment have been chosen close to their optimum values according to Eq. (S37).

## 2. Comparison of our slice-less OAWM and time-interleaved optical sampling

From a hardware perspective, the concept of time-interleaved parallel optical sampling [3-6] is similar to our slice-less OAWM technique. In the following, we briefly summarize the concept of time-interleaved optical sampling and discuss the differences to slice-less OAWM. In time-interleaved parallel optical sampling, a pulse train emitted by a mode-locked laser with repetition period $T_{\mathrm{rep}}$ is split equally into $N$ paths and delayed by $\tau_\nu = (\nu-1)T_{\mathrm{rep}}/N$ in path $\nu$, leading to equidistantly interleaved sampling pulses, see Fig. S5 for an illustration of an associated setup using $N = 2$ parallel IQ receivers [6]. Assuming that the optical sampling pulses are sufficiently narrow, the complex amplitude of the beat between the signal and the sampling pulses is proportional to the signal field at the center of the sampling pulse [4]. Consequently, the signal is optically sampled in each of the $N$ parallel channels. After digitizing all signals, an estimate of the original signal is reconstructed by interleaving the complex-valued samples that have been acquired by the $N$ parallel IQ receivers beforehand. However, as the optical phase in the different receiver paths is not stable, a digital phase-drift compensation is required before interleaving the acquired samples. In [6], this problem is addressed by running a multidimensional optimization procedure [7] during signal reconstruction, exploiting a-priori information about the measured QPSK signals. Time-interleaved optical sampling of arbitrary signals with unknown time dependence has not been demonstrated so far.

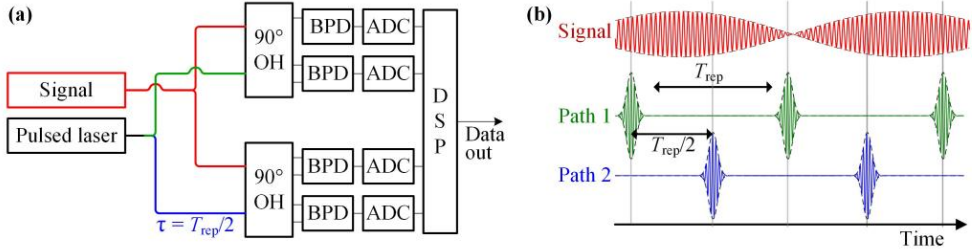

**Fig. S5**. Concept of time-interleaved optical sampling for $N = 2$ IQ receivers. **(a)** Hardware setup, comprising a pulsed laser source and $N = 2$ IQ receivers. **(b)** Sketch of interleaved optical sampling pulses. (Adapted from [6]).

This problem can be overcome by slice-less OAWM, which does not require a-priori knowledge about the signal. Instead, we exploit spectral overlap regions that allow for estimating all relevant phase parameters for arbitrary waveforms see Sect. 1.4. We further rely on a calibrated frequency-domain system model to compensate not only for imperfect delays $\tau_\nu$ but also for the frequency dependent characteristics of the various receiver channels. Consequently, we can significantly suppress the crosstalk among different spectral slices that would otherwise impair the reconstructed waveform according to Eq. (S20).

## 3. Comparison to spectrally sliced OAWM

The description of the slice-less OAWM system can also be related to that of spectrally sliced OAWM systems [8-10]. By including slicing filters for the signal and LO, the frequency-dependent time-invariant transfer matrix $\underline{\tilde{\mathbf{H}}}^{(\mathrm{f})}(f)$ in Eq. (S29) essentially simplifies to a diagonal matrix, and the time-variant diagonal matrices $\underline{\mathbf{H}}_{\mathrm{F}}^{(\mathrm{t})}$ and $\underline{\mathbf{H}}_{\mathrm{LO}}^{(\mathrm{t})}$ can be combined into a single diagonal matrix $\underline{\mathbf{H}}^{(\mathrm{t})}$,

$$\underbrace{\begin{bmatrix} \tilde{U}_1(f) \\ \vdots \\ \tilde{U}_N(f) \end{bmatrix}}_{\underline{\tilde{\mathbf{U}}}(f)} = \underbrace{\begin{bmatrix} \tilde{\underline{H}}_1^{(\mathrm{f})}(f) & \cdots & 0 \\ \vdots & \ddots & \vdots \\ 0 & \cdots & \tilde{\underline{H}}_N^{(\mathrm{f})}(f) \end{bmatrix}}_{\underline{\tilde{\mathbf{H}}}^{(\mathrm{f})}(f)} \underbrace{\begin{bmatrix} H_1^{(\mathrm{t})} & \cdots & 0 \\ \vdots & \ddots & \vdots \\ 0 & \cdots & H_N^{(\mathrm{t})} \end{bmatrix}}_{\underline{\mathbf{H}}^{(\mathrm{t})}} \underbrace{\begin{bmatrix} \tilde{a}_\mathrm{S}(f+f_1) \\ \vdots \\ \tilde{a}_\mathrm{S}(f+f_N) \end{bmatrix}}_{\underline{\tilde{\mathbf{A}}}_\mathrm{S}(f)} + \underbrace{\begin{bmatrix} \tilde{G}_1(f) \\ \vdots \\ \tilde{G}_N(f) \end{bmatrix}}_{\underline{\tilde{\mathbf{G}}}(f)}. \quad \text{(S39)}$$

Note that in this case it is mandatory to match the number of receivers and LO comb tones, $M = N$, as each LO tone is strictly assigned to one receiver. Further note that non-zero off-diagonal elements of $\underline{\tilde{\mathbf{H}}}^{(\mathrm{f})}(f)$ occur if the filter suppression is not sufficient. In this case, the slice-less system model according to Eq. (S29) or (S14) can be used for reconstructing the signal from the imperfectly separated spectral slices, which would otherwise be subject to inter-slice crosstalk.

## 4. Calibration

The signal reconstruction in slice-less OAWM crucially relies on a precise calibration of the receiver system, which is obtained in several steps. In a first step, we measure and digitally compensate the IQ-skew of the IQ receivers, i.e., the time-delay differences of the in-phase and quadrature signals on their way from the optical hybrids to the respective BPD, see Sect. 4.1. In a second step, we tune the various delays in our system to obtain a well-conditioned transfer matrix according to Eq. (S37), see Sect. 4.2. Based on these pre-calibration steps, we finally measure the time-invariant frequency-dependent transfer functions $\underline{\tilde{H}}_{\nu\mu}^{(I)}(f)$, Eq. (S10), and $\underline{\tilde{H}}_{\nu\mu}^{(Q)}(f)$, Eq. (S11), by using a well-known optical reference waveform (ORW) generated by a highly stable femtosecond mode-locked laser (Menhir 1550), see Section 4.3 below.

### 4.1. IQ skew calibration and measurement of amplitude response

For the IQ skew calibration, we connect two external-cavity lasers (ECL) to the IQ receiver array and sweep their frequency difference $f$ in discrete steps – an exemplary sketch for IQR ν is shown in Fig. S6 (a). This leads to sinusoidal output signals with frequency $f$, $I_\nu(t)$ and $Q_\nu(t)$ at BPD-ν, Fig. S6 (b). From these output signals we extract the phase difference $\varphi_{IQ}$ of the in-phase and quadrature components as well as the respective amplitudes $A_{I\nu}$ and $A_{Q\nu}$ by fitting sinusoidal model functions to the measurement data. In Fig. S6 (c) we plot the extracted IQ phases as a function of frequency. The time delay differences between the I and Q signal, are obtained from the slope of the phase difference over frequency, accessible through a linear fit. The skews are then compensated by shifting the recorded time-domain waveforms accordingly. Note that the measurement technique used here only allows to extract the group delay differences between the I and the Q component of the same IQ receiver, but not the group delay differences between different IQ receivers or the associated phase transfer functions – these measurements require a broadband optical reference waveform (ORW) with a precisely known frequency-dependent phase, see Sect. 1.5. Still, the frequency-dependent amplitudes $A_{I,\nu}(f)$ and $A_{Q,\nu}(f)$ for IQR ν can be compared to the amplitude responses $\left|\underline{\tilde{H}}_{\nu\mu}^{(I)}(f)\right|$ and $\left|\underline{\tilde{H}}_{\nu\mu}^{(Q)}(f)\right|$ that we measured with the ORW based calibration procedure, see Fig. S13 below and in Fig. 3 in the main paper.

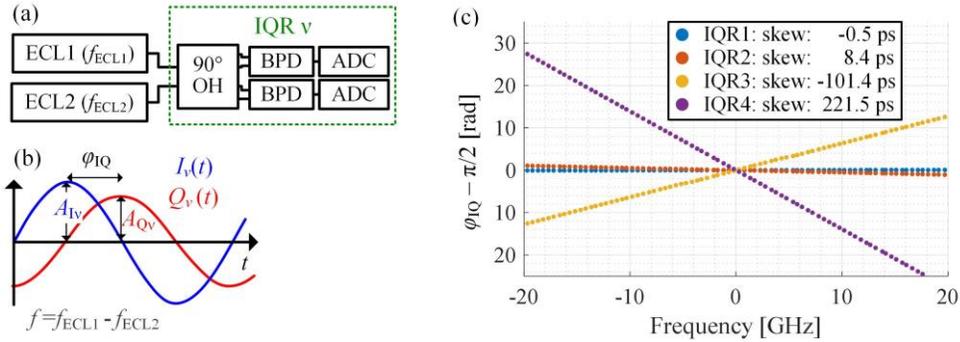

**Fig. S6.** Measurement of the time-delay differences (IQ skew) between in-phase and quadrature signals. **(a)** Measurement setup: Two external-cavity lasers (ECL) are connected to an IQ receiver (IQR), consisting of a 90° optical hybrid (OH) with balanced photodetectors (BPD) and analog-to-digital converters (ADC). The frequency difference between ECL 1 and ECL 2 is swept in discrete steps, leading to sinusoidal output signals of the respective difference frequency at the two photodetectors. **(b)** Sketch of the time-domain trace for the in-phase $I(t)$ and quadrature-phase $Q(t)$ components. The frequency $f$, the amplitudes $A_{I\nu}$ and $A_{Q\nu}$, and the IQ phases $\varphi_{IQ}$ are extracted by fitting a sinusoidal model function to the measurement data. **(c)** IQ phase error $\varphi_{IQ} - \pi/2$ as a function of frequency $f$ for all four IQ receivers. The time delay differences between the I and Q signal, are obtained from the slope of the phase difference over frequency, accessible through a linear fit.

## 4.2. Optical delay tuning and compensation

The goal of the second step is to measure and numerically equalize the time-delay difference $\tau_{S,\nu} - \tau_{S,n}$, $\nu, n = 1,...,N$ of the various signals on their way from the power splitter to the optical hybrid and to tune the corresponding LO delays $\tau_{LO,\nu}$, thus obtaining approximately equidistant relative delays $\tau_\nu = \tau_{LO,\nu} - \tau_{S,\nu}$ according to Eq. (S37), as required for a well-conditioned transfer matrix. To this end, we first generate a random broadband data signal using an IQ modulator and send it through the signal paths, while an external-cavity laser (ECL2) is connected to the LO input, switch position A in Fig. S7 (a). We extract the relative signal delays $\tau_{S,\nu} - \tau_{S,1}$ by detecting the peak of the modulus of the temporal complex cross-correlation between the received baseband signals $\underline{U}_\nu(t) = I_\nu(t) + jQ_\nu(t)$ and $\underline{U}_1(t) = I_1(t) + jQ_1(t)$. For a sufficiently large unambiguity range, the signal must not repeat within twice the expected maximum path delay mismatch. We numerically compensate the time delay differences in the signal path so that $\tau_{S,\nu} - \tau_{S,1} = 0$, $\forall \nu = 1,...,N$. Next, we send the broadband signal through the LO paths and connect ECL2 to the signal input, switch position B in Fig. S7 (a). The goal now is to adjust the relative delays $\tau_\nu = \tau_{LO,\nu} - \tau_{S,\nu}$ associated with the various IQ receivers as required by Eq.(S37). To this end, we repeat the cross-correlation procedure for measuring $\tau_{LO,\nu} - \tau_{LO,1}$. Making use of the fact that the signal delays were already equalized, $\tau_{S,\nu} - \tau_{S,1} = 0$ $\forall \nu = 1,...,N$, and that thus $\tau_{LO,\nu} - \tau_{LO,1} = \tau_\nu - \tau_1$, we can now manually adjust the optical delay lines (DL) for the required time differences $\tau_{LO,\nu} - \tau_{LO,1} = (\nu - 1)T_{LO}/N$. Figure S7 (b) shows the measured delay differences $\tau_{LO,\nu} - \tau_{LO,1}$ as a function of the recording number having set the optimal equidistant delays for an LO comb with $f_{FSR} \approx 150\,\text{GHz}$. The delays associated with receiver three (yellow) and receiver four (purple) in Fig. S7 (b) show a common fluctuation relative to receiver one (blue) and two (orange). This fluctuation is presumably related to an imperfect synchronization of the two oscilloscopes (OSC 1 and OSC 2) and does not represent drifts of the optical delays. We do not expect these fluctuations to have any significant impact on the measurements presented in this work.

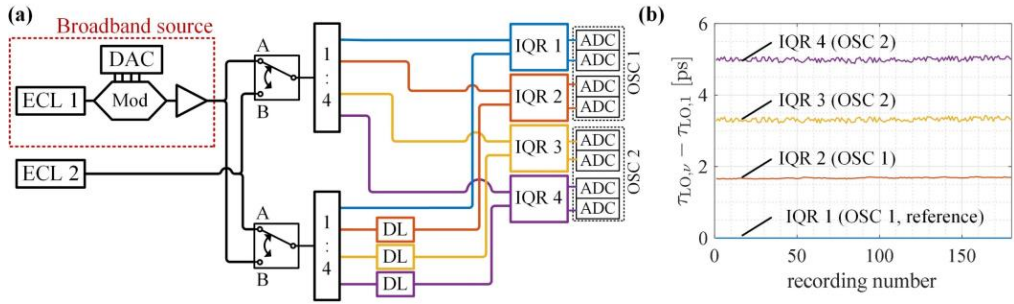

**Fig. S7**. Tuning and compensation of optical delays. **(a)** Setup for measuring path-delay differences. Switch position A is used for measuring differences of the signal-path delays, while switch position B allows to measure differences of the LO-path delays. **(b)** Consecutive measurements of the LO-path differences after equalizing the signal delays and after tuning the delay lines (DL) according to Eq. (S37) for an LO with an FSR of 150 GHz. The delays associated with receiver three (yellow) and receiver four (purple) show a common fluctuation relative to receiver one (blue) and two (orange), which we attribute to an imperfect synchronization of the two oscilloscopes (OSC 1 and OSC 2).

*4.3. Calibration with femtosecond laser as optical reference waveform (ORW) source*

Having tuned all delays in our setup, we can measure the time-invariant frequency-dependent transfer functions $\underline{\tilde{H}}_{\nu\mu}^{(I)}(f)$, Eq. (S10), and $\underline{\tilde{H}}_{\nu\mu}^{(Q)}(f)$, Eq. (S11), using a known optical reference waveform (ORW) generated by a highly stable femtosecond mode-locked laser (Menhir 1550, Menhir Photonics AG, Glattbrugg, Switzerland). The mode-locked laser features a repetition rate of 250 MHz and a highly stable pulse shape that was independently measured by the manufacturer using frequency-resolved optical gating (FROG), see Fig. S8. Notably, the spectral amplitude and the phase profile are smooth and approximately flat within the wavelength range (1555 mm to 1560 nm) used in our experiments, rendering the laser well suited for calibrating our OAWM system.

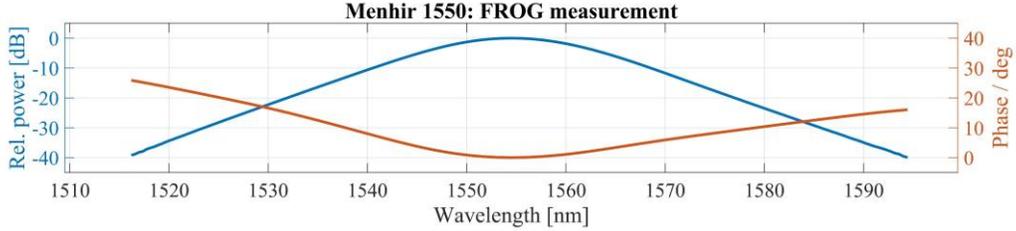

**Fig. S8**. Spectral amplitude and phase of the optical reference waveform (ORW) as obtained from a frequency-resolved optical gating (FROG) measurement. The ORW comb has an FSR of 250 MHz and was generated by a femtosecond mode-locked laser (Menhir 1550, Menhir Photonics AG).

Fig. S9 illustrates the calibration principle for a system comprising only $M = 2$ LO tones, spaced by the FSR $f_{\text{FSR}}$, using a multitude of ORW tones with FSR $f_{\text{ORW}}$. Figure S9 (a) shows the optical spectra of the ORW tones (red) and of the LO tones at frequencies $f_1$ (purple) and $f_2$ (green), while Fig. S9 (b) shows an exemplary baseband spectrum $\tilde{I}_1(f)$ obtained from the in-phase component of IQR 1, where the colors of the spectral lines are chosen according to the associated LO tone. Because $I_\nu(t)$ and $Q_\nu(t)$ are real-valued signals, their spectra are symmetric $\tilde{I}_\nu(f) = \tilde{I}_\nu^*(-f)$ and $\tilde{Q}_\nu(f) = \tilde{Q}_\nu^*(-f)$. Thus, the mixing of the ORW spectrum $\underline{\tilde{a}}_{\text{ORW}}(f)$ (red) with each LO tone at frequency $f_\mu$ generates two symmetric RF sub-combs in the baseband, $\underline{\tilde{H}}_{\nu\mu}^{(I)}(f)\underline{\tilde{a}}_{\text{ORW}}(f + f_\mu)$ (solid lines), and $\underline{\tilde{H}}_{\nu\mu}^{(I)*}(-f)\underline{\tilde{a}}_{\text{ORW}}^*(-f + f_\mu)$ (dotted lines), see Eqs. (S12) and (S13) for $\underline{\tilde{a}}_S(f) = \underline{\tilde{a}}_{\text{ORW}}(f)$. We extract all transfer functions by finding the amplitude and phase of the corresponding (solid) comb lines and by referring them to the known amplitude and phase of the ORW as obtained from the FROG measurement. The transfer function $\underline{\tilde{H}}_{11}^{(I)}(f)$ is extracted from the RF comb $\underline{\tilde{H}}_{11}^{(I)}(f)\underline{\tilde{a}}_{\text{ORW}}(f + f_1)$, solid purple lines in Fig. S9 (b), and the transfer function $\underline{\tilde{H}}_{12}^{(I)}(f)$ is obtained from the RF comb $\underline{\tilde{H}}_{12}^{(I)}(f)\underline{\tilde{a}}_{\text{ORW}}(f + f_2)$, solid green lines. Note that this is only possible if the tones of the four RF combs do not fall on top of one another, which is ensured by proper choice of the ORW's spectral position and its FSR.

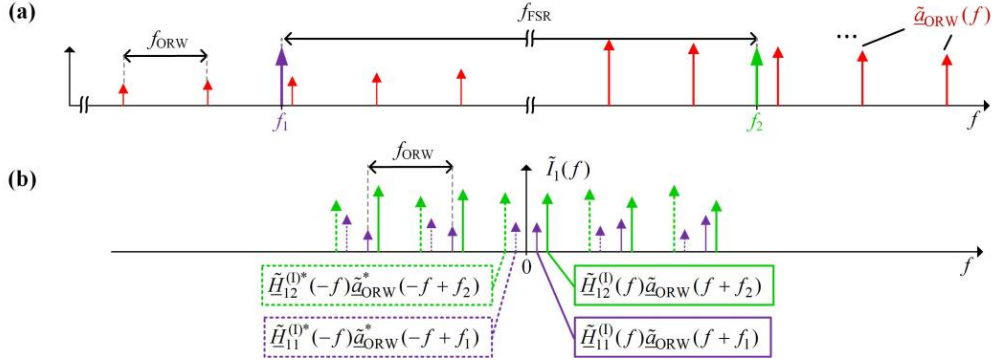

**Fig. S9.** Calibration concept, illustrated for a simple system using only two LO tones at frequencies $f_1$ (purple) and $f_2$ (green), separated by $f_{\mathrm{FSR}} = f_2 - f_1$ along with a known optical reference waveform (ORW) that is fed to the OAWM system. **(a)** The spectrum $\tilde{a}_{\mathrm{ORW}}(f)$ (red) of the ORW corresponds to a comb with FSR $f_{\mathrm{ORW}}$. Note the abscissa is interrupted – the LO-comb FSR typically amounts to tens of GHz (110 GHz and 150 GHz for the experiments shown in the main manuscript), whereas the ORW FSR is chosen much smaller (250 MHz in the experiments shown in the main manuscript) to obtain densely spaced sampling points for the measured transfer functions. **(b)** Exemplary baseband spectrum $\tilde{I}_1(f)$ obtained from the in-phase component of IQR 1. The spectrum consists of the superimposed mixing products of each LO tone with the ORW, depicted as purple and green in Subfigure (a). Each LO tone leads to two symmetric RF sub-combs in the baseband (solid and dotted lines), see Eq. (S13) for $\tilde{a}_S(f) = \tilde{\underline{a}}_{\mathrm{ORW}}(f)$. The transfer functions $\tilde{\underline{H}}_{1,1}^{(I)}(f)$ and $\tilde{\underline{H}}_{1,2}^{(I)}(f)$ are measured by extracting amplitudes and phases of the corresponding (solid) RF-comb lines, and by referring the results to the known amplitudes and phases of the ORW as obtained from the FROG measurement. Finally, we interpolate the transfer functions between the frequency-domain sampling points obtained from the various ORW comb lines.

The software implementation of this calibration procedure requires several processing steps that are illustrated in Fig. S10. In the following, we give a more detailed explanation of the signal processing techniques that are used in each of these steps.

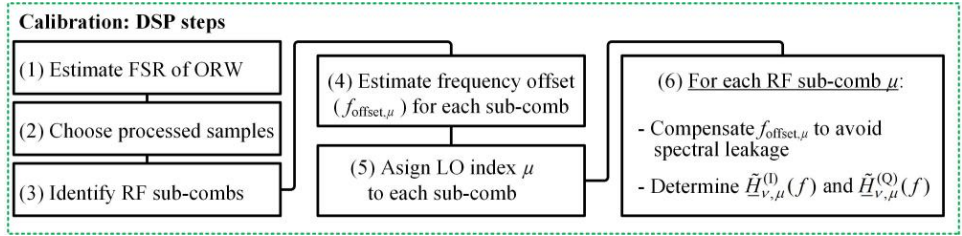

**Fig. S10.** Block diagram of processing steps performed to extract the frequency-dependent transfer functions $\tilde{\underline{H}}_{\nu,\mu}^{(I)}(f)$ and $\tilde{\underline{H}}_{\nu,\mu}^{(Q)}(f)$ from the associated calibration measurements with a known optical reference waveform (ORW).

### Step (1): Estimation of FSR of ORW

First, we estimate the FSR $f_{\mathrm{ORW}}$ of the ORW using the clock of the recording oscilloscopes (sampling interval $t_s = 1/f_s$). To this end, we calculate the autocorrelation function of one of the baseband spectra measured at the in-phase or quadrature output of any the IQ receivers, see Fig. S11 for the autocorrelation e.g., $\tilde{I}_1(f) \otimes \tilde{I}_1(f)$. The autocorrelation function shows peaks at integer multiples of $f_{\mathrm{ORW}}$. By extracting the frequency of several peaks for higher orders, e.g., at $300 \times f_{\mathrm{ORW}}$, $315 \times f_{\mathrm{ORW}}$, …, an accurate FSR estimate is found.

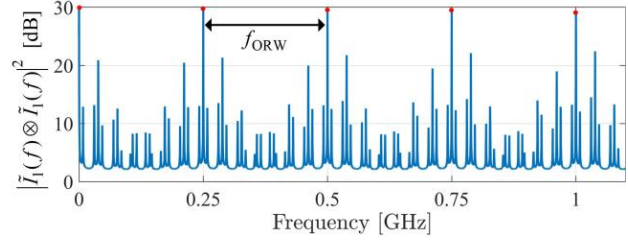

**Fig. S11.** Estimation of the FSR $f_{\text{ORW}} = 1/T_{\text{ORW}}$ from the periodically repeated peaks of the autocorrelation function $\tilde{I}_1(f) \otimes \tilde{I}_1(f)$ of the baseband spectrum $\tilde{I}_1(f)$.

### Step (2): Choice of processed samples

In Step (2), we use the estimated FSR of the ORW to select a suitable number of samples $N_{\text{samples}}$ for further processing such that the corresponding observation time interval $T_{\text{obs}} = N_{\text{samples}} t_s$ is as close as possible to an integer multiple $N_{\text{periods}}$ of the ORW comb repetition period $T_{\text{ORW}} = 1/f_{\text{ORW}}$. This is important to avoid spectral leakage, that occurs in case RF sub-comb tones fall between the frequency points that are associated with the discrete Fourier transform (DFT), see Step (6) for further details. In the experiment described in the main manuscript, the FSR of the ORW, $f_{\text{ORW}} = 250\,\text{MHz}$, and the sampling frequency of the ADC, $f_s = 1/t_s = 256\,\text{GHz}$, are defined with very high precision and form an integer multiple of one another, $T_{\text{ORW}}/t_s = f_s/f_{\text{ORW}} = 1024$. In this case, we may simply choose an integer number $N_{\text{periods}}$ of ORW periods as the observation time, $T_{\text{obs}} = N_{\text{periods}} T_{\text{ORW}}$, corresponding to $N_{\text{samples}} = N_{\text{periods}} \times f_s/f_{\text{ORW}} = 1024 \times N_{\text{periods}}$ samples. In case the sampling frequency $f_s$ of the ADC does not naturally form an integer multiple of the FSR $f_{\text{ORW}}$ of the ORW, the number of observed periods $N_{\text{periods}}$ needs to be chosen to minimize the residual temporal offset $\Delta T = \left| N_{\text{samples}} t_s - N_{\text{periods}} T_{\text{ORW}} \right|$.

### Step (3): Identification of RF sub-combs associated with different LO tones

In Step (3), we identify the $M$ sets of RF sub-combs generated by the LO tones at frequencies $f_\mu$, $\mu = 1,...,M$, see solid and dotted comb lines in Fig. S9 (b) for an example with $M = 2$ LO tones. To this end, we construct for IQR ν the composite baseband spectra $\tilde{U}_\nu(f) = \tilde{I}_\nu(f) + j\tilde{Q}_\nu(f)$ and assume $\tilde{\underline{H}}_{\nu\mu}^{(Q)}(f) = -j\tilde{\underline{H}}_{\nu\mu}^{(I)}(f)$, see Eq. (S16), such that the mirrored RF sub-combs $\tilde{\underline{H}}_{\nu\mu}^{(I)*}(-f)\tilde{\underline{a}}_{\text{ORW}}^*(-f + f_\mu)$, corresponding to the dotted comb lines in Fig. S9 (b), do not appear in $\tilde{U}_\nu(f)$. Using Eq. (S16) and disregarding noise, we obtain

$$\tilde{I}_\nu(f) + j\tilde{Q}_\nu(f) = \sum_{\mu=1}^{M}\left[\tilde{\underline{H}}_{\nu\mu}^{(I)}(f)\tilde{\underline{a}}_{\text{ORW}}(f + f_\mu) + \tilde{\underline{H}}_{\nu\mu}^{(I)*}(-f)\tilde{\underline{a}}_{\text{ORW}}^*(-f + f_\mu)\right]$$
$$+ j\sum_{\mu=1}^{M}\left[-j\tilde{\underline{H}}_{\nu\mu}^{(I)}(f)\tilde{\underline{a}}_{\text{ORW}}(f + f_\mu) + j\tilde{\underline{H}}_{\nu\mu}^{(I)*}(-f)\tilde{\underline{a}}_{\text{ORW}}^*(-f + f_\mu)\right] \quad (S40)$$
$$= \sum_{\mu=1}^{M} 2\tilde{\underline{H}}_{\nu\mu}^{(I)}(f)\tilde{\underline{a}}_{\text{ORW}}(f + f_\mu).$$

If the assumptions made for Eq. (S16) do not hold exactly, the construction of the composite spectrum $\tilde{U}_\nu(f) = \tilde{I}_\nu(f) + j\tilde{Q}_\nu(f)$ will nevertheless notably suppress one set of RF sub-combs, dashed combs in Fig. S9 (b), relative to the respective other set, solid combs in Fig. S9 (b). This is sufficient for the identification of the respective set of lines, leading to one identified RF sub-comb per LO tone, i.e., $M$ identified RF sub-combs $\sum_{\mu=1}^{M} \tilde{\underline{H}}_{\nu\mu}^{(I)}(f)\tilde{\underline{a}}_{\text{ORW}}(f + f_\mu)$ in total. Only these identified sub-combs are considered further.

## Steps (4) and (5): Estimation of frequency offsets and assignment of LO index $\mu$ for each sub-comb

Each of the $M$ superimposed RF combs identified in Step (3) now needs to be assigned to the corresponding LO comb line $f_\mu$, $\mu = 1,...,M$. The tones of each of the $M$ RF combs appear at equidistant frequencies $f_{\text{RF},\mu,\zeta}$ and are subject to a frequency offset $f_{\text{offset},\mu}$,

$$f_{\text{RF},\mu,\zeta} = f_{\text{offset},\mu} + \zeta \times f_{\text{ORW}}, \qquad \zeta \in \mathbb{Z}, \mu = 1,..M. \tag{S41}$$

Since the ORW comb (FSR $f_{\text{ORW}}$) spans the entire measurement range covered by the LO comb (FSR $f_{\text{FSR}}$) with a uniform FSR, the various offsets $f_{\text{offset},\mu}$ differ by the same increment $\Delta f_{\text{offset}} = f_{\text{offset},\mu} - f_{\text{offset},\mu-1}$,

$$f_{\text{offset},\mu} = f_{\text{offset},1} + (\mu - 1)\Delta f_{\text{offset}}, \qquad \forall \mu = 2,...,M, \tag{S42}$$

where

$$\Delta f_{\text{offset}} = \left(\left\lceil \frac{f_{\text{FSR}}}{f_{\text{ORW}}} \right\rceil f_{\text{ORW}} - f_{\text{FSR}}\right), \qquad \Delta f_{\text{offset}} < f_{\text{ORW}}. \tag{S43}$$

These criteria allow to unambiguously assign each of the $M$ RF combs to the respective LO tone $f_\mu$, assuming $f_{\text{FSR}}$ and $f_{\text{ORW}}$ are known. Note that, without loss of generality, we may define $f_{\text{offset},1}$ as the frequency difference between the first LO comb line $f_1$ and the closest ORW comb line appearing at a frequency larger than $f_1$, such that $f_{\text{offset},1} \in [0, f_{\text{ORW}}[$, leading to the offset frequencies $f_{\text{offset},\mu}$ indicated in Fig. S12 below.

## Step (6): Compensation of frequency offset and determination of transfer functions

Having identified the subscripts of the $M$ RF sub-combs, the complex-valued amplitudes of the spectral lines divided by their respective counterparts $\tilde{\underline{a}}_{\text{ORW}}(f + f_\mu)$ obtained from the FROG measurement of the ORW would represent the transfer functions $\tilde{\underline{H}}_{\nu,\mu}^{(\text{I})}(f)$ and $\tilde{\underline{H}}_{\nu,\mu}^{(\text{Q})}(f)$, see Eq. (S12) and (S13). However, the frequency offsets $f_{\text{offset},\mu}$ of the various sub-combs are not integer multiples of the frequency bin width $1/(N_{\text{samples}}^{(\text{opt})} t_s)$ of the underlying DFT, which is dictated by the choice of $N_{\text{samples}}^{(\text{opt})}$ in Step 2. Therefore, the lines of the associated RF sub-comb suffer from spectral leakage, i.e., they are spread out over several frequency bins such that it is not directly possible to extract a single complex-valued envelope. For each of the $M$ RF sub-combs, this problem can be overcome by multiplying the time-domain baseband signals $I_\nu(t)$ and $Q_\nu(t)$ by $\exp(-j2\pi f_{\text{offset},\mu} t)$ before calculating the DFT, corresponding to a spectral down-shift of the RF sub-comb of interest by its specific offset frequency $f_{\text{offset},\mu}$. After the DFT, we then undo this frequency shift by relabeling the frequency axis accordingly. Note that the frequency shift resolves the leakage problem for all lines of the respective RF sub-comb $\mu$ since the observation time $T_{\text{obs}}$ has already been adjusted to an integer multiple of the ORW period $T_{\text{ORW}}$ in Step 2, rendering the frequency bin width of the DFT an integer fraction of the sub-comb FSR $f_{\text{ORW}}$. For each IQR $\nu$, we thus obtain $M$ specific spectra $\tilde{I}_{\nu,\mu}(f)$ and $\tilde{Q}_{\nu,\mu}(f)$, $\mu = 1,...,M$, for which leakage is minimized for the RF sub-comb associated with the LO comb line $f_\mu$. In Fig. S12, we show as an example the $M = 4$ power spectra $\left|\tilde{I}_{\nu,\mu}(f)\right|^2$, $\mu = 1,...,4$, for IQR 1 ($\nu = 1$), where the leakage is minimized for the $\mu$-th RF sub-comb, respectively. We also indicate the frequency $f_{\text{offset},\mu}$ that has been taken care of by the frequency shift $\exp(-j2\pi f_{\text{offset},\mu} t)$ before calculating the DFT. The dashed red lines represent an interpolation between the corresponding RF sub-comb lines. Because only a frequency range up to $0.6\,\text{GHz}$ is shown, the interpolated curve looks like a straight line. If a closer frequency spacing than $f_{\text{ORW}}$ is required for capturing finer features of the transfer functions, multiple calibrations can be combined in a post-processing step, see Fig. 3 (b) in the main paper. In this case, we tune the optical frequency offset of the ORW from the LO comb between several calibration recordings.

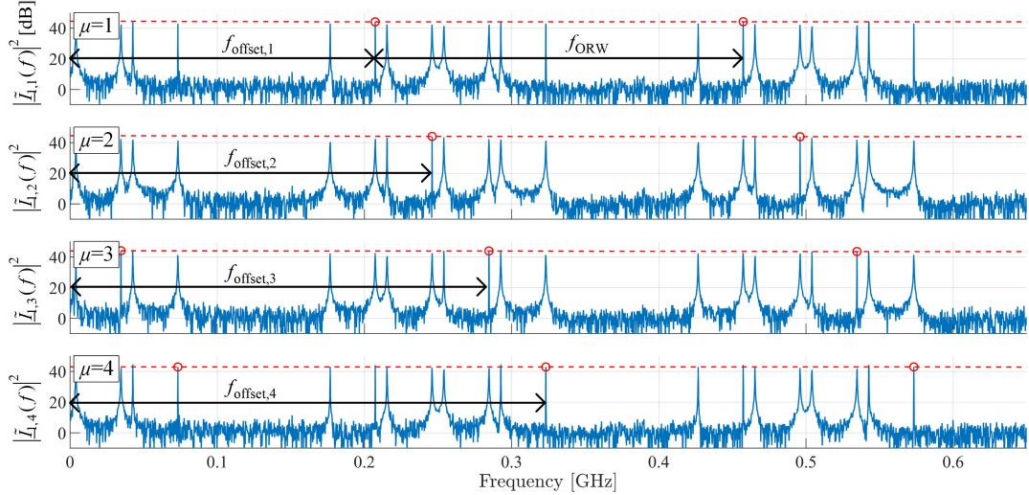

**Fig. S12.** Example power spectra $\left|\tilde{I}_{1,\mu}(f)\right|^2$ for IQ receiver $\nu = 1$. Before calculating the DFT, we multiply $I_1(t)$ by a phase factor $\exp\left(-\mathrm{j}2\pi f_{\mathrm{offset},\mu} t\right)$ in the time domain, thereby introducing a frequency down-shift by $f_{\mathrm{offset},\mu}$ and mitigating spectral leakage for the $\mu$-th RF sub-comb. This frequency shift is undone by relabeling of the frequency axis after calculating the DFT. The transfer functions $\underline{\tilde{H}}_{1,\mu}^{(\mathrm{I})}(f)$ are obtained by dividing the complex-valued amplitudes of the leakage-free RF sub-comb by their respective counterparts $\tilde{a}_{\mathrm{ORW}}(f + f_\mu)$ obtained from the FROG measurement of the ORW. For frequencies between the ORW comb tones, the transfer functions $\underline{\tilde{H}}_{1,\mu}^{(\mathrm{I})}(f)$ are obtained by either an interpolation of the complex-valued amplitudes extracted from the marked comb lines (dashed red lines), or by repeating the measurement with different optical frequency offsets of the ORW comb from the LO comb, see Fig. 3 (b) of the main manuscript.

To validate the calibration results obtained with the ORW, we compare the amplitude transfer functions $\left|\underline{H}_{\nu,\mu}^{(\mathrm{I})}(f)\right|$ and $\left|\underline{H}_{\nu,\mu}^{(\mathrm{Q})}(f)\right|$ with the frequency-dependent amplitudes $A_{\mathrm{I},\nu}(f)$ and $A_{\mathrm{Q},\nu}(f)$ extracted from the beat note of two ECLs according to Fig. S6 above. As shown in Fig. 3 in the main paper, we observe almost identical results for both measurement procedures for the first IQ receiver (IQR 1). However, we record a slight discrepancy between the two measurements for some of the 100 GHz BPDs, see, e.g., $\underline{\tilde{H}}_{22}^{(\mathrm{Q})}(f)$ in Fig. S13. This discrepancy is caused by the fact that the photodetectors saturate under the relatively high optical peak power produced by the mode-locked laser used to generate the ORW, see Sect. 4.4, which leads to nonlinear saturation behavior and thus a reduction of the bandwidth [11], see also Section 4.4 below. The problem can be avoided in future experiments by dispersing the ORW, i.e., by widening its pulses prior to feeding them to OAWM system. The use of a dispersed ORW is preferable as compared to a simple power attenuation as the dispersion does not sacrifice optical power but simply reduces the peak-to-average power ratio (PAPR) of the pulse train. Note that for a more compact display we choose to show in Fig. S13 positive frequencies only, even though the transfer functions $\underline{\tilde{H}}_{\nu,\mu}^{(\mathrm{I})}(f)$ and $\underline{\tilde{H}}_{\nu,\mu}^{(\mathrm{Q})}(f)$ comprise optical and electrical frequency-dependent transfer characteristics and thus are not Hermitian symmetric, $\underline{\tilde{H}}_{\nu,\mu}^{(\mathrm{I})}(f) \neq \underline{\tilde{H}}_{\nu,\mu}^{(\mathrm{I})*}(-f)$, $\underline{\tilde{H}}_{\nu,\mu}^{(\mathrm{Q})}(f) \neq \underline{\tilde{H}}_{\nu,\mu}^{(\mathrm{Q})*}(-f)$. Still, the amplitude responses are dominated by the photodetectors such that $\left|\underline{\tilde{H}}_{\nu,\mu}^{(\mathrm{I})}(f)\right| \approx \left|\underline{\tilde{H}}_{\nu,\mu}^{(\mathrm{I})}(-f)\right|$ and $\left|\underline{\tilde{H}}_{\nu,\mu}^{(\mathrm{Q})}(f)\right| \approx \left|\underline{\tilde{H}}_{\nu,\mu}^{(\mathrm{Q})}(-f)\right|$ hold approximately and the characteristics at negative frequencies can be inferred from their positive-frequency counterparts. The phase transfer functions for both I and Q components are shown in Fig. S14.

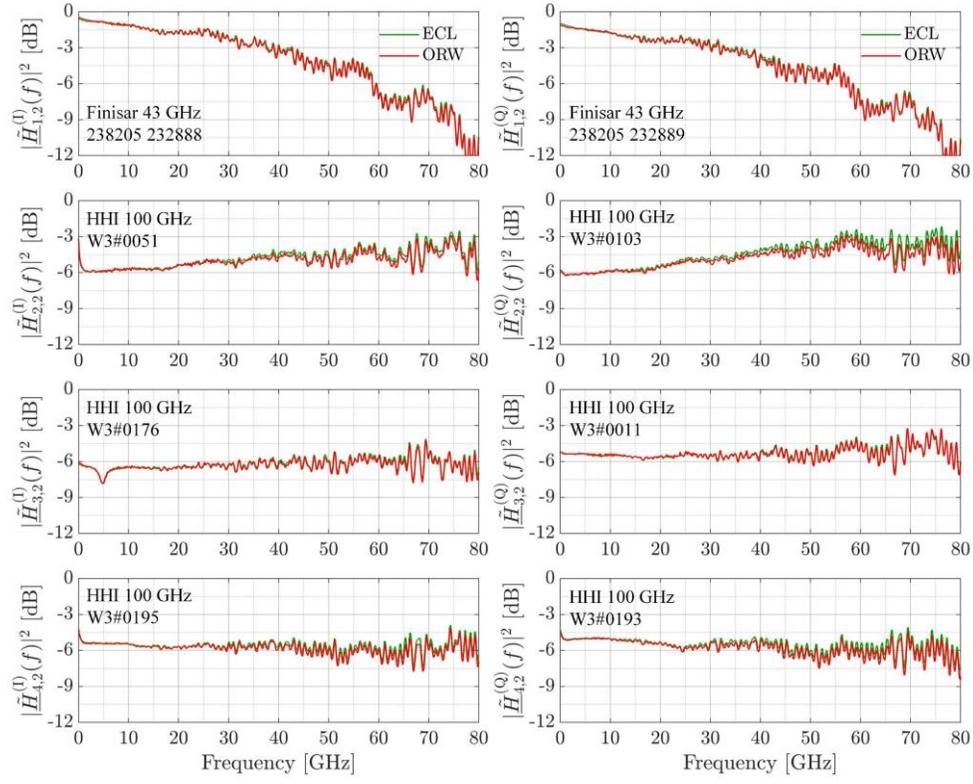

**Fig. S13**. Comparison of the amplitude transfer function for $f > 0$ measured with a pair of external-cavity lasers (ECL), see Sect. 4.1, and with the known optical reference waveform (ORW). The bandwidth measured with the pulsed ORW is reduced for some BPDs, because the photodetectors are saturated by the high peak power of the short ORW pulses. This can be avoided by dispersing the pulse trains prior to feeding them to the OAWM system.

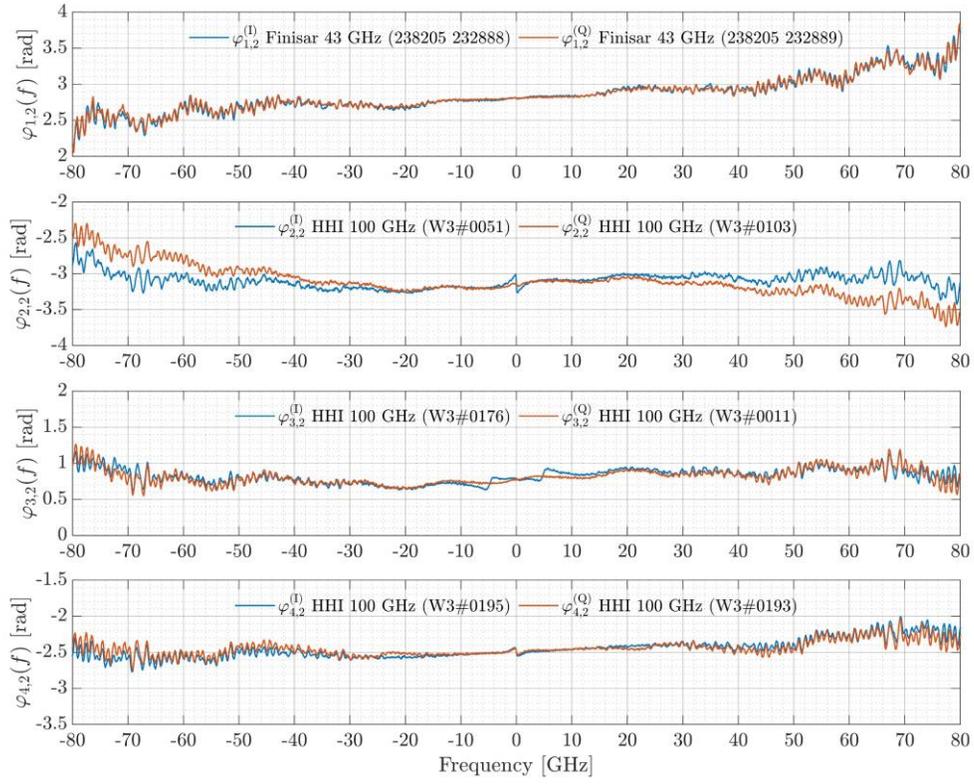

**Fig. S14.** Phase transfer functions for IQR 1, IQR 2 IQR 3, and IQR 4 and LO tone, measured using the optical reference waveform (ORW). For a better comparison, we removed the π/2 phase difference between the in-phase and quadrature transfer functions.

*4.4. Saturation of photodetectors under illumination with a fs-laser*

We investigate the saturation behavior of different photodetectors by sending the ORW to a single detector of the BPD, and by recording the pulse amplitudes $A$ of the generated pulse train, Fig. S15 (a), (b). We observe a significantly lower saturation input power for the 100 GHz BPDs (Fraunhofer Heinrich-Hertz Institute, HHI; Berlin, Germany) as compared to their 43 GHz counterparts (Finisar 43 GHz Balanced Photodetector BPDV21x0R), see Fig. S15 (c). The BPD with the lowest saturation power (#W0103) also shows the highest discrepancy of the amplitude transfer function measured with a pair of external-cavity lasers (ECL) and with the known optical reference waveform (ORW), Fig. S13.

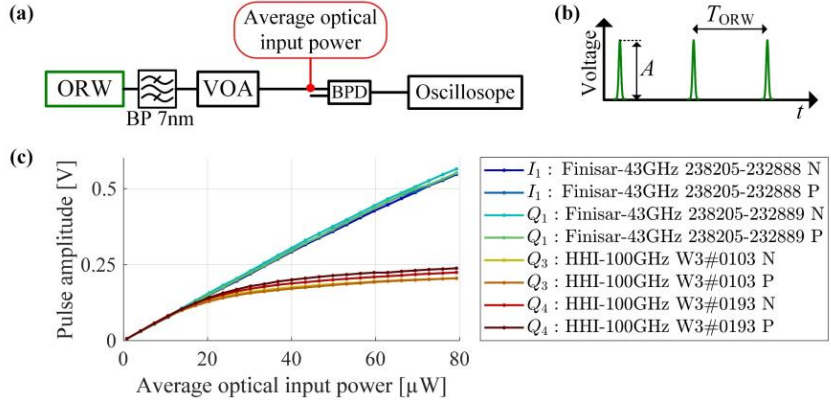

**Fig. S15.** Characterization of the saturation behavior of the photodetectors used in our experiments under pulsed illumination  **(a)** Setup comprising a variable optical attenuator (VAO), an optical bandpass (BP) filter, and a balanced photodetector (BPD), of which only a single photodiode named N or P is illuminated. The ORW is a femtosecond laser with repetition rate 250 MHz ($T_{ORW} = 4$ ns).  **(b)** Sketch of the measured pulse train. **(c)** Measurement result showing the average pulse amplitude as a function of the average optical input power. We observe a significantly lower saturation input power for the 100 GHz BPDs (Fraunhofer Heinrich-Hertz Institute, HHI; Berlin, Germany) as compared to their 43 GHz counterparts (Finisar 43 GHz Balanced Photodetector BPDV21x0R).

## 5. Local-oscillator comb

The four-tone local oscillator (LO) used for the experiment discussed in the main manuscript is derived from a dissipative Kerr soliton (DKS) comb that is generated using the setup depicted in Fig. S16 (a), see [12] for a more detailed description on how to tune into a low-phase-noise soliton state. A single-tone laser is amplified and injected into a high Q micro-resonator to generate the DKS. At the output, the remaining pump line is suppressed by a notch filter. The optical spectrum for the 110 GHz comb before (Ⓐ) and after (Ⓑ) a 5 nm optical bandpass (BP) is shown in Fig. S16 (b). The limited output power per line at point Ⓑ directly translates into an ASE-noise-limited optical carrier-to-noise ratio (OCNR) after amplification Ⓒ. For the comb with a 110 GHz FSR shown in Fig. S16 (b), the OCNR in a reference bandwidth of 12.5 GHz (0.1 nm at a center wavelength of 1550 nm) was measured to be 23.3 dB, 24.3 dB, 24.2 dB, 23.5 dB for the four LO tones at frequencies $f_1 = 192.269$ THz ($\lambda_1 = 1559.236$ nm), $f_2 = 192.379$ THz ($\lambda_2 = 1558.344$ nm), $f_3 = 192.489$ THz ($\lambda_3 = 1557.452$ nm), and $f_4 = 192.599$ THz ($\lambda_4 = 1556.564$ nm), respectively. The four tones ($f_1,...,f_4$) are isolated using a wavelength-selective switch (WSS). Since the WSS features a fixed grid of 12.5 GHz-wide switchable frequency "pixels" that does not perfectly coincide with the 110 GHz FSR of the comb, it was necessary to "open" two pixels for the outer comb lines at frequency $f_1$ and $f_4$, leading to wider transfer functions for the underlying spectrally white noise. The rightmost plot in Fig. S16 (b) shows the four comb tones of interest along with the measured filter function of the WSS (green). Due to the low OCNR of the LO comb, the width of the individual filter functions for each of the four lines is important for the overall system performance. The transmitted LO noise imposes an upper limit on the SNR that can be achieved after coherent detection, see discussion of signal distortions in Sect. 7. When measuring data signals, the multiplicative noise originating from the LO increases the noise for the outer constellation points more than for the inner ones and is most noticeable for the part of the signal that was down-converted with lines at frequencies $f_1$ or $f_4$, for which two pixels had to be "opened" in the WSS, see constellation diagrams recorded from channels L1 and L4 in Fig. 4(a) in the main manuscript. In Fig. S16 (c), we show a photograph of the micro-resonator and the associate optical package that was used for generation of the 110 GHz Kerr comb in our experiment. The LO comb with a 150 GHz FSR was derived from a native comb with a 50 GHz line spacing. In this case, the WSS was configured to select only every third tone. Since the 150 GHz FSR is well aligned with the 12.5 GHz grid of the WSS, it is now sufficient to "open" only a single WSS pixel per line. As a consequence, the outermost WDM channels received with the 150 GHz LO comb are less impacted by multiplicative noise than the ones received with the 110 GHz comb, see Fig. 4 (b) of the main manuscript. For the 150 GHz comb, the OCNR at point Ⓒ, i.e., at the output of EDFA 2, amounts to 25.6 dB, 26.8 dB, 26.9 dB, and 24.8 dB for the comb-tone frequencies $f_1 = 192.340$ THz ($\lambda_1 = 1558.660$ nm), $f_2 = 192.491$ THz ($\lambda_2 = 1557.436$ nm), $f_3 = 192.641$ THz ($\lambda_3 = 1556.220$ nm), and $f_4 = 192.792$ THz ($\lambda_4 = 1555.006$ nm), respectively.

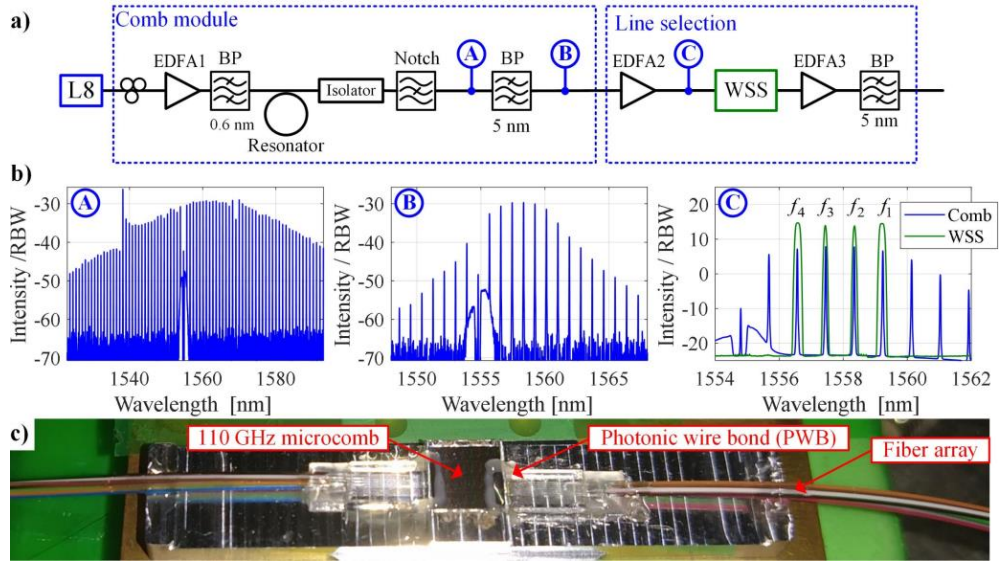

**Fig. S16**. Generation of Kerr soliton (DKS) combs. **(a)** Experimental setup used to generate the local-oscillator (LO) combs. EDFA = erbium-doped fiber amplifier; BP = bandpass filter; WSS = wavelength selective switch **(b)** Optical spectrum of the 110 GHz LO comb at different positions Ⓐ, Ⓑ, Ⓒ in the comb-generation setup. All spectra are recorded with a resolution bandwidth of RBW = 0.02 nm. **(c)** Photograph of a 110 GHz microcomb chip which is connected to a fiber array via photonic wire bonds (PWB) [13,14].

## 6. Benchmarking of the OAWM scheme

We evaluate the performance of the OAWM system by measuring broadband test signals with bandwidths of approximately 490 GHz and 610 GHz, both consisting of several WDM channels with different modulation formats and symbol rates, which are simultaneously received. While this offers the possibility to receive signals with ultra-high symbol rates along with utmost flexibility and agility in terms of software-defined wavelength assignment, it is also important to understand which impairments are introduced by the OAWM scheme and how OAWM-based detection compares to channel-by-channel reception of WDM signals using a series of independent IQ receivers with dedicated LO laser tones. To investigate these effects, we use the 490 GHz-wide test signal, which comprises seven wavelength-division multiplexing (WDM) data channels, generated by modulating four optical carriers (L1 to L4) using a first IQ modulator and three additional optical carriers (L5 to L7) using a second IQ modulator, see Fig. S17 (a) below as well as Sect. 3 in the main manuscript for more details on the underlying experimental setup. The optical signal-to-noise power ratio (OSNR) of these signals can be artificially reduced by adding spectrally white noise, generated by an amplified spontaneous emission (ASE) noise source. The power level of the added ASE noise is controlled using a variable optical attenuator (VOA). The optical spectrum of the test signal at point Ⓐ is shown in Fig. S17 (c).

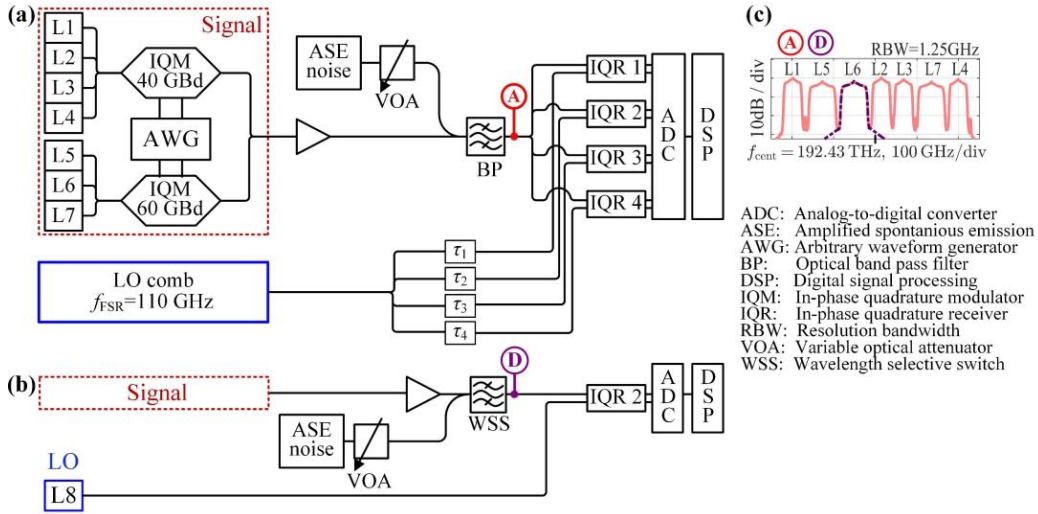

**Fig. S17.** Benchmarking the performance of the broadband multi-channel OAWM receiver system against a channel-by-channel reception using a conventional single-channel IQ receiver (IQR). **(a)** OAWM setup: A broadband multi-channel optical WDM signal is generated by modulating seven spectrally interleaved lasers (L1 to L7) with two different IQ modulators (IQM) that are driven with different signals, derived from a common arbitrary-waveform generator (AWG, Keysight M8194A). The output signals of the IQM are combined, and the resulting broadband optical test signal is then amplified. Spectrally white noise is added, and the signal is routed through a band-pass filter (BP) to the IQR array of the OAWM system. The LO comb with $f_{FSR} = 110\,GHz$ is split into four paths, delayed by different time intervals $\tau_1, \ldots \tau_4$, and routed to the IQR array. **(b)** Reference measurement with a conventional single-channel IQR: The various WDM channels contained in the broadband optical test signal are individually received with IQR 2, using the external-cavity laser L8, that previously served as a pump for the Kerr comb. The WDM channel of interest is isolated by a WSS, see spectrum Ⓓ in Subfigure (c). **(c)** Exemplary optical spectra of the broadband multi-channel WDM test signal at point Ⓐ and of an exemplary isolated WDM channel modulated onto the optical carrier generated by Laser L6 at point Ⓓ.

In our experiments, we benchmark our OAWM system against a single-channel IQ receiver. As the single IQ receiver is not sufficiently broadband to capture all seven data signals at once, we measure them sequentially by using a programmable WSS to isolate a single WDM channel along with a tunable external-cavity laser (L8) as LO for coherent detection, Fig. S17 (b). As an example, we additionally show in Fig. S17 (c) the optical spectrum recorded at point Ⓓ after isolating WDM channel L6, containing 60 GBd 16 QAM data modulated on a carrier generated by laser L6. After receiving the signal using the single-channel IQ receiver IQR 2, we obtain the power spectrum $\left|\tilde{U}_2^{(SCh)}(f)\right|^2$ of the composite baseband signal $\underline{U}_2^{(SCh)}(t) = I_2^{(SCh)}(t) + jQ_2^{(SCh)}(t)$, Fig. S18 (a). The lines apparent in the spectrum originate from ADC clock tones or higher harmonics at integer multiples of 16 GHz. Additional clock tones from the arbitrary waveform generator (AWG) at the transmitter appear at around 30 GHz and are hidden within the signal in Fig. S18 (a). In addition, we plot the corresponding acquisition noise $\tilde{G}_2^{(SCh)}(f)$, gray curve in Fig. S18 (a), as measured by disconnecting all optical receiver inputs, see Sect. 8 below for details. In Fig. S18 (b), we show the histogram of the digitized time-domain waveform $I_2^{(SCh)}(t) = \Re\{\underline{U}_2^{(SCh)}(t)\}$. The full-scale voltage $U_{FS}$ of the oscilloscope is set such that the signal $I_2^{(SCh)}(t)$ fills the full range without significant clipping. For the signal $I_2^{(SCh)}(t)$, we measure a peak-to-average power ratio (PAPR) of 10.3 dB. Figure S18 (c) shows the constellation diagram of the 60 GBd 16 QAM data modulated on the laser tone of L6, corresponding to the spectrum $\left|\underline{U}_2^{(SCh)}(f)\right|^2$ in Fig. S18 (a).

In a next step, we receive the full broadband signal consisting of all seven data channels using our OAWM system. In this case, each baseband signal $\underline{U}_\nu(t) = I_\nu(t) + jQ_\nu(t)$, $\nu = 1,...,4$, comprises superimposed contributions generated by mixing of signal components with the nearest of the four LO tones at frequencies $f_\mu$, $\mu = 1,...,4$, see Fig 1 (b) in the main manuscript. As an example, we again show the spectrum $\tilde{U}_2(f)$ obtained from IQR 2, Fig. S18 (d). We also show the corresponding histogram of the time-domain signal $I_2(t) = \Re\{\underline{U}_2(t)\}$ in Fig. S18 (e). For the best results, we set the full-scale voltage $U_{FS}$ of the respective ADC channel to $U_{FS} \approx 2 \times 4\sigma_S$, where $\sigma_S$ is the standard deviation of the associated signal, e.g. $\sigma_S^{(I2)} = \sqrt{\overline{(I_2 - \overline{I_2})^2}}$ for signal $I_2(t)$, where the overbar denotes a temporal average. This choice was found to lead to a good trade-off between signal clipping on the one hand and excessive impact of acquisition noise on the other hand, and limits the PAPR of signal $I_2(t)$ to 12.6 dB, see Fig. S26 and the related discussion in Sect. 9 below. In the following, we briefly discuss why the superposition of multiple spectral components, as it is the case for the slice-less OAWM system, can lead to an increased PAPR.

Assuming that the down-converted spectral portions originating from different WDM channels are statistically independent, we would expect the resulting superposition in Fig. S18 (e) to assume a more Gaussian-like histogram than an individual WDM channel, Fig. S18 (b), which is confirmed by our measurements. Disregarding noise and making use of the fact that the time-domain data signals in the various WDM channels have finite maximum amplitudes dictated by the respective transmitters, we would further expect that the PAPR of the superposition signal $I_2(t)$ is larger than that of the individual data signals observed in the output signal $I_2^{(SCh)}(t)$ of the single-channel IQ receiver: For $M$ statistically independent spectral slices having the same average power $P_{avg}$ and the same maximum amplitude $A_{max}$, the average power of the superposition signal is obtained as the sum $M \times P_{avg}$ of the average powers contributed by the various superimposed spectral slices, while the peak power corresponds to the square of the highest observed amplitude of the superposition signal, which may reach up to the sum $M \times A_{max}$ of the peak amplitude $A_{max}$ of the individual contributions. The upper bound for the PAPR of the superposition signal $I_2(t)$ is then given by the ratio $\left(M^2 A_{max}^2\right) / \left(M \times P_{avg}\right)$, i.e., the PAPR of the superposition signal is at most increased by a factor of up to $M = 4$ with respect to the PAPR of the individual contributions, see Sect. 9 for a more detailed mathematical formulation.

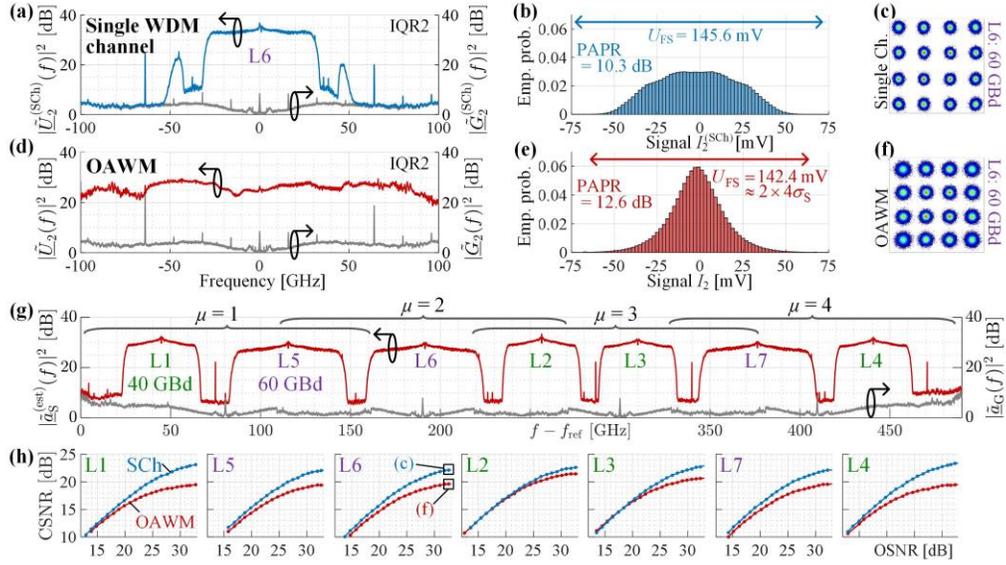

**Fig. S18.** Benchmarking the proposed OAWM system according to the setup in Fig. S17 (a) and channel-by-channel reception using a single IQ receiver (IQR 2) according to the setup in Fig. S17 (b). All spectra are plotted with a resolution bandwidth of 100 MHz. **(a)** Exemplary power spectrum $\left|\tilde{U}_2^{(SCh)}(f)\right|^2$ of the measured 60 GBd 16 QAM signal modulated onto laser L6, recorded with a single-channel IQ receiver (IQR 2, blue) along with the power spectrum of the acquisition noise $\left|\underline{G}_2^{(SCh)}(f)\right|^2$ (gray). **(b)** Histogram showing the empirical probability of the quantized voltages of the in-phase component $I_2^{(SCh)}(t) = \Re\{\underline{U}_2^{(SCh)}(t)\}$ recorded by IQR 2, corresponding to the spectrum in Subfigure (a). The bin width of the histogram amounts to 2 mV. **(c)** Constellation diagram of the 60 GBd signal modulated on L6 as measured with IQR 2. The constellation diagram was recorded at the maximum OSNR of 33 dB. **(d)** Power spectrum $\left|\tilde{U}_2(f)\right|^2$ obtained from IQR 2 when using the OAWM system to simultaneously receive all WDM channels L1, …, L7. **(e)** Histogram of quantized voltages of the in-phase component $I_2(t) = \Re\{\underline{U}_2(t)\}$, corresponding to the spectrum in Subfigure (c). Because of the addition of nearly independent signals, the histogram becomes narrower and more Gauss-like, and the peak-to-average power ratio (PAPR) increases, see Sect.9 below. **(f)** Constellation diagram of 60 GBd data L6 measured with the OAWM system at maximum OSNR of 33 dB. **(g)** Reconstructed signal spectrum $\underline{\tilde{a}}_S^{(est)}(f)$ (red) and stitched acquisition noise $\underline{\tilde{a}}_G^{(est)}(f)$ (gray) as obtained from the OAWM system. The horizonal axis indicates the offset of the optical frequency $f$ from a reference frequency $f_{ref}$, which, without loss of generality, was chosen to correspond to the lower-frequency edge of the first spectral slice, $f_{ref} = f_1 - B \approx 192.52$ THz. The overlapping braces $\mu = 1,...,4$ indicate the width of the reconstructed and frequency-shifted spectral slices $\underline{\tilde{a}}_{S,1}^{(est)}(f),...,\underline{\tilde{a}}_{S,M}^{(est)}(f)$. **(h)** CSNR of the various WDM channels L1, …, L7 as a function of the respective OSNR, measured with respect to a reference bandwidth of 12.5 GHz for the OAWM system (red) and for individual reception of each WDM channel L1, …, L7 (blue) using the same IQ receiver (IQR2) as in Subfigure (a). For high OSNR, we observe a slightly lower CSNR performance for the current implementation of our broadband OAWM system compared to the single-channel IQ receiver.

Fig. S18 (g) shows the power spectrum $\left|\underline{\tilde{a}}_S^{(est)}(f)\right|^2$ of the reconstructed signal. The overlapping spectral slices associated with the $\mu$-th LO tone are annotated with braces labeled $\mu = 1,…,4$. The data channels are labeled according to the optical carriers L1 to L4 (40 GBd 16 QAM data) and L5 to L7 (60 GBd 16 QAM data), see Fig. S17 (a). The acquisition noise $\tilde{\mathbf{G}}_{acq}^{(I)}(f)$ and $\tilde{\mathbf{G}}_{acq}^{(Q)}(f)$ contributed by the various ADC was separately recorded for disconnected optical receiver inputs and processed the same way as the data, leading to the gray curve in Fig. S18 (g). This processing introduces a pronounced frequency-dependence of the reconstructed noise floor, which is mainly caused by two effects: First, the photodetector responses are equalized by applying the reconstruction matrix $\tilde{\mathbf{H}}_{rec}^{(t)}(f)$, Eq (S19) in Sect. 1.2 above, which increases the noise for high frequencies, i.e., further away from the center frequency of each slice. Second, the uncorrelated complex-valued noise amplitudes of adjacent slices are weighted and averaged in the overlap regions of neighboring slices, see Eq. (S21) in Sect. 1.2, leading to a local reduction of the noise floor up to 3 dB, gray curve in Fig. S18 (g). Consequently, the resulting stitched receiver noise $|\tilde{\underline{a}}_G(f)|^2$ is highest at the lower and upper limit of the overall covered spectral range, where noise averaging from an adjacent slice does not occur. An exemplary constellation diagram for the 60 GBd signal L6 is given in Fig. S18 (f).

Finally, we compare the constellation SNR (CSNR) of the signals received with the OAWM system with those obtained from individual channel-by-channel IQ reception for various optical signal-to-noise ratios (OSNR), see red and blue curves in Fig. S18 (h). The CSNR corresponds to the square of the reciprocal of the error vector magnitude (EVM) normalized to the average signal power $\text{EVM}_m$ [15], $\text{CSNR}_{dB} = 10 \times \log_{10}(1/\text{EVM}_m^2)$. The OSNR is adjusted by artificially adding increasing levels of ASE noise, see Fig. S17 (a) and (b), which eventually dominates over the native noise of the receiver system. For low OSNR levels, the CSNR hence increases in proportion to the OSNR, while, for larger OSNR, the CSNR reaches a plateau because the combined effect of electrical transmitter noise (DAC and RF amplifier), of LO noise, and of electrical receiver noise (RF amplifier and ADC) is still present, even of no additional noise is introduced. For high OSNR, we observe a slightly lower CSNR performance for the broadband OAWM system compared to the single-channel IQ receiver, which can be explained by the following effects:

1. The 3dB bandwidth of 43 GHz of IQR 1 is insufficient for reception of a 55 GHz-wide spectral slice, requiring strong "digital" amplification of high-frequency components in the reconstruction process. This unavoidably leads to digital amplification of acquisition noise for high RF frequencies and thus reduces the resulting CSNR especially for the signals L1 and L4 as these are located in the roll-off region. This effect is not present for the single-channel IQ receiver (IQR 2), for which we used photodetectors with a bandwidth of 100 GHz in combination with an LO tone tuned to the center of the respective WDM channel such that the down-converted composite baseband spectrum appears approximately symmetrically around DC as shown in Fig. S18 (a).

2. Due to the larger PAPR, Fig. S18 (b) and (e), and the limited dynamic range of the ADC, the SNR of the digitized signals is reduced for the OAWM system as compared to the single-channel IQ receiver [16]. This effect is analyzed in more detail in Sect. 9 below.

3. The LO comb lines have a lower OCNR than the tone emitted by L8, which is used as an LO for channel-by-channel reception, see Sect. 5 above. When integrating the noise over the full filter bandwidth of the WSS, the LO comb lines have an carrier-to-noise ratio (CNR) of 25.5 dB for comb lines $\mu = 1$ and $\mu = 4$ and of 28 dB for $\mu = 2$ and $\mu = 3$, see Fig. S21 in Sec. 7 below. These relatively small CNR lead to higher multiplicative noise in the intradyne detection process.

4. Additional crosstalk $\tilde{\underline{a}}_X(f)$ results from calibration errors $\Delta\tilde{\mathbf{H}}_{\text{rec}}^{(t)}(f) \neq 0$ in Eq. (S20) These distortions are characterizes in more detail in the Sect. 7 below.

5. The OAWM system uses only $N = 4$ IQ receivers to measure the $N_{ch} = 7$ WDM data channels comprised in the broadband test signal, whereas the single-channel IQ receiver is dedicated to one specific WDM channel only. This leads to an additional penalty of approximately $10\log_{10}(N/N_{ch}) = 10\log_{10}(4/7) = -2.4 \text{dB}$ for the OAWM system, which can be best understood by considering the simplified scenario depicted in

Fig. S19. One receiver either detects a single WDM channel Fig. S19 (a), or simultaneously detects two WDM channels Fig. S19 (b). For simplicity, we assume that the overall waveforms acquired by the two receivers have the same PAPR – disregarding for the moment the fact that the superposition of two WDM signals usually leads to an increased PAPR in comparison to a single WDM channel, see Section 9 below. Consequently, the overall SNR of both waveforms is identical when integrating the signal and noise power spectral density over the full receiver bandwidth, see Eq. (S45) below. This result holds independent of the overall signal power $P$, because any increase of the overall signal power will lead to an increase of the acquisition noise by approximately the same amount, see Section 9. We thus may assume without loss of generality that that the two overall waveforms shown in Fig. S19 (a) and Fig. S19 (b) have also the same overall signal power $P$ and are impaired by acquisition noise with the same power spectral density. Consequently, we obtain a lower per-channel power, a lower per-channel SNR, and thus a lower CSNR for the two-channel reception, Fig. S19 (b), as compared to its single-channel counterpart, Fig. S19 (a). In general, the more WDM channels $N_{ch}$ are simultaneously received by a single IQ receiver, the lower the power associated with a single WDM channel will be, and the same applies to the resulting CSNR. Conversely, if more than one receiver is used to measure the same WDM channel, we can reduce the acquisition noise by averaging, and improve the resulting CSNR. In case of the OAWM system, both effects occur simultaneously as four receivers are used to measure seven WDM channels. Consequently, if the acquisition noise dominates over other noise sources, the CSNR is reduced by $10\log_{10}(4/7) = -2.4\,\text{dB}$ compared to a single-channel reception. Note that in case the number of received channels exceeds that of the receivers, $N_{ch} < N$, a performance improvement for the OAWM system compared a single-channel IQ receiver is expected. This was e.g., observed in [10,17], where OAWM systems receiving a single broadband data channel performed better than a single IQ receiver with the same total bandwidth.

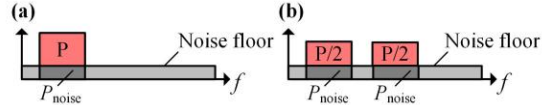

**Fig. S19**. Illustrative explanation of the performance penalty observed when detecting two WDM channels with a single IQ receiver. **(a)** Reception of single WDM channel with total power $P$. The SNR is given by the ratio $P/P_{noise}$, where $P_{noise}$ is obtained by integrating the noise power spectral density over the signal bandwidth. **(b)** Reception of two WDM channels with total power $P$ and individual power $P/2$. Assuming that the power spectral density of the noise is the same as in (a), the SNR of an individual WDM channel is 3 dB lower compared to the single-channel reception in (a).

## 7. Noise and distortion characterization

To further understand the impairments associated with the OAWM system, we perform measurements of well-defined single-frequency laser tones and identify the associated noise contributions and distortions. The setup for this experiment is depicted in Fig. S20(a). Note that a single-tone signal does not have spectral components in all overlap regions between neighboring spectral slices, see Fig. 1 (b) in the main manuscript for a visualization of the overlap regions, which is required to estimate the time dependent parameters $\underline{H}_{\text{LO},\mu}^{(t)}$ and $\underline{H}_{\text{F},\nu}^{(t)}$, see Sect. 1.4 above. To circumvent this issue, we add pilot tones, P1 to P4, to the signal prior to detection. An example for the optical spectrum after adding the pilot tones is shown in the inset of Fig. S20. The power of the pilot tones is low compared to the signal power.

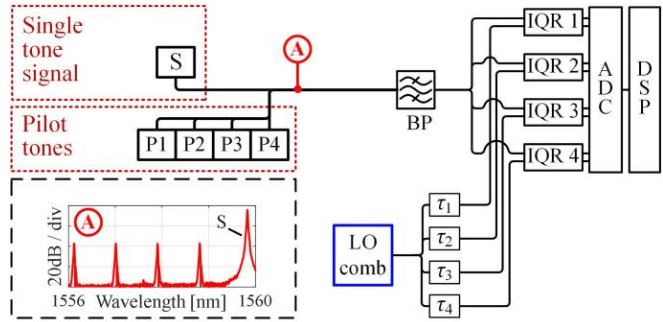

**Fig. S20**. Experimental setup for characterizing noise and distortions by recording a single-frequency laser tone with the OAWM system. Pilot tones are added to guarantee some signal in the overlap regions, which is necessary to determine the time variant parameters $\underline{H}_{\text{LO},\mu}^{(\text{t})}$ and $\underline{H}_{\text{F},\nu}^{(\text{t})}$ via the associated redundant information, see Sect. 1.4. The inset shows an exemplary optical signal spectrum comprising the strong signal tone (S) and four weaker pilot tones, P1, P2, P3, P4, three of which are located in the overlap regions between the four spectral slices.

Fig. S21 (a) and (b) show reconstructed power spectra for the single-tone signal (red) tuned to different frequencies. The horizontal axis indicates the offset of the optical frequency $f$ from a reference frequency $f_{\text{ref}}$, which was chosen to correspond to the lower-frequency edge of the first spectral slice, $f_{\text{ref}} = f_1 - B \approx 192.52$ THz. The displayed power spectra are normalized to the spectral peak of the signal and are smoothed with a 25 MHz-wide smoothing filter. The overlapping spectral slices $\underline{\tilde{a}}_{\text{S},\mu}^{(\text{est})}(f)$ for $\mu = 1,\ldots,4$ are indicated by braces. We identify different noise contributions and distortions, marked with different colors in Fig. S21, and calculate their integral power relative to the power of the single-tone signal. The following noise contributions and distortions can be identified:

1. The pilot tones (green)

2. The reconstruction crosstalk (magenta, cyan) corresponds to $\underline{\tilde{a}}_{\text{X}}(f)$ in Eq. (S21) in Sec. 1.2 above and is caused by an inaccurate calibration and an inaccurate estimation of the parameters $\underline{H}_{\text{F},\nu}^{(\text{t})}$, Eq. (S22), so that $\Delta\underline{\tilde{\mathbf{H}}}_{\text{rec}}^{(\text{t})}(f) \neq 0$, Eq. (S20). In Fig. S21 we separately highlight the IQ-crosstalk (cyan), which is a consequence of an imperfect compensation of the IQ-imbalance of the receivers, and the crosstalk between slices (magenta), which is also caused by errors in the reconstruction matrix.

3. The LO noise (blue) acts multiplicatively on the signal and creates a noise bump around the signal tone in Fig. S21 (a) and (b). This LO noise originates from the EDFA that is used to amplify the soliton comb and is spectrally shaped by the optical filter (WSS) that is configured to suppress most of this ASE noise, see Fig. S16 in Sect. 5. Note that due to the grid alignment of the WSS, the filter bandwidth and thus the width of the noise bump is wider for $\mu = 1$ and $\mu = 4$ than for $\mu = 2$ and $\mu = 3$, see Section 5 above.

4. The remaining noise (gray) is dominated by the reconstructed acquisition noise of the oscilloscopes, corresponding to $\underline{\tilde{a}}_{\text{G}}(f)$ in Eq. (S21). The remaining spectral tones have low power and are not separately treated. They originate, e.g., from the ADC clock or can be identified as reconstruction crosstalk associated with the low power pilot tones.

We calculate the signal-to-noise and distortion ratio (SINAD) by dividing the signal power (red) by the total power of the noise and the distortions. We obtain a SINAD of up to 21 dB for an acquisition bandwidth of 490 GHz. This SINAD would correspond to an effective number of bits

(ENOB) of approximately 3.2. We believe that there remains room for improving the signal quality, e.g., by

1. Using a high-power LO comb that provides higher OCNR upon amplification, thereby reducing the associated multiplicative noise (blue). Recently, dark [18] and bright [19] Kerr soliton combs with high conversion efficiency and thus higher output power have been demonstrated.

2. Increasing the calibration accuracy to reduce the crosstalk (magenta, cyan) according to Eq. (S20). Note that the current calibration recordings are impaired by saturating photodetectors and noise, see Sect. 4.4, which should leave room for further improvement.

3. Using high-bandwidth photodetectors in all channels to avoid digital amplification of the acquisition noise in the high-frequency regions of IQR 1.

4. Reducing the pilot tone powers. In further experiments we have found that the pilot-tone power can be reduced by more than 10 dB as compared to the current levels.

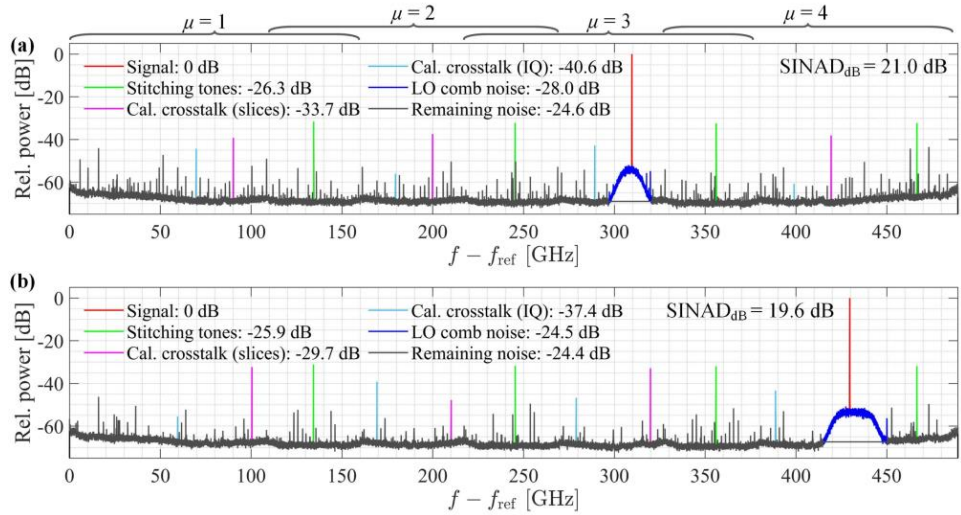

**Fig. S21**. Measurement of single-tone optical test signal generated by a narrowband external-cavity laser. (**a**) Power spectrum of a reconstructed single-tone signal located in spectral slice $\mu = 3$. The horizonal axis indicates the offset of the optical frequency $f$ from a reference frequency $f_{\text{ref}}$, which was chosen to correspond to the lower-frequency edge of the first spectral slice, $f_{\text{ref}} \approx 192.52$ THz. The spectrum is smoothed with a 25 MHz wide window. The signal (red) is surrounded by a noise bump (blue), which originates from the LO comb. An imperfect calibration and parameter estimation leads to crosstalk between the reconstructed slices (magenta) and to IQ crosstalk (cyan). The remaining noise (gray) is dominated by the acquisition noise of the receiver ADC. The measured signal-to-noise-and-distortion ratio (SINAD) of 21 dB would correspond to an effective number of bits (ENOB) of approximately 3.2, leaving room for further improvement. (**b**) Power spectrum of a reconstructed single-tone signal located in the spectral slice $\mu = 4$.

## 8. Oscilloscope noise

In this section, we quantify the noise contributions $G_{\text{acq},\nu}^{(I)}(t)$ and $G_{\text{acq},\nu}^{(Q)}(t)$ associated with the respective input channels of the Keysight UXR oscilloscopes used in our experiments [20]. To this end, we assume a sinusoidal input signal $S_{\text{in}}(t)$ and use a simplified noise model, Fig. S22, to describe the dependence of the noise level on the full-scale input voltage $U_{\text{FS}}$ of the oscilloscope. The full-scale voltage $U_{\text{FS}}^{(\text{ADC})}$ of the internal ADC is fixed. The variable power gain $\Gamma$ between the oscilloscope input and the internal ADC relates both quantities by $\sqrt{\Gamma} = U_{\text{FS}}^{(\text{ADC})}/U_{\text{FS}}$, so that the full-scale input voltage scales inversely with the amplitude gain $\sqrt{\Gamma}$. Note that the variable power gain $\Gamma = \Gamma_1 \times \Gamma_2$ is provided by an adjustable electro-mechanical attenuator of power gain $\Gamma_1 < 1$, followed by an electrical amplifier of power gain $\Gamma_2 > 1$. We assume a constant internally added noise power $\sigma_{\text{ADC}}^2$, which is effective at the input of the ADC. The variable input power gain is compensated digitally at the output for an overall gain of one. Therefore, the acquisition noise power $\sigma_{\text{acq}}^2$ after digital gain compensation scales inversely with the net gain $\Gamma$, and thus the acquisition noise amplitude $\sigma_{\text{acq}}$ scales linearly with the full-scale voltage $U_{\text{FS}}$,

$$\sigma_{\text{acq}}^2 = \frac{\sigma_{\text{ADC}}^2}{\Gamma}, \qquad \sigma_{\text{acq}} = \frac{\sigma_{\text{ADC}}}{\sqrt{\Gamma}} \propto U_{\text{FS}}. \tag{S44}$$

If the input voltage gain $\sqrt{\Gamma}$ is adjusted such that the peak amplitude $A$ of the input signal $S_{\text{in}}(t)$ corresponds to half of the full scale, $2A = U_{\text{FS}} = U_{\text{FS}}^{(\text{ADC})}/\sqrt{\Gamma}$, then the output noise power $\sigma_{\text{acq}}^2$ is proportionally to the peak power $A^2$, Eq. (S45). Consequently, the acquisition noise limited SNR of the output signal $S_{\text{out}}(t)$ is inversely proportional to the peak-to-average power ratio (PAPR),

$$\sigma_{\text{acq}} \propto U_{\text{FS}} \propto A \quad \Rightarrow \quad \sigma_{\text{acq}}^2 \propto A^2 \quad \Rightarrow \quad \text{SNR} = \frac{\overline{S_{\text{in}}^2}}{\sigma_{\text{noise}}^2} \propto \frac{\overline{S_{\text{in}}^2}}{A^2} = \frac{1}{\text{PAPR}}. \tag{S45}$$

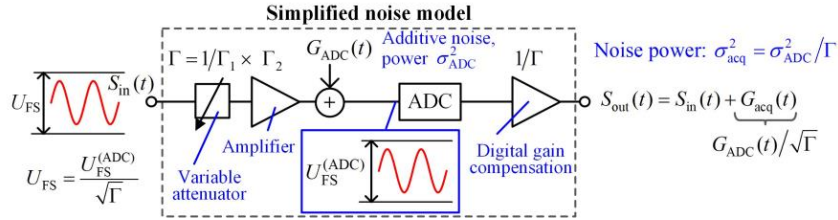

**Fig. S22.** Simplified noise model for an analog-to-digital conversion unit ("oscilloscope", dashed box). We assume that the added noise voltage $G_{\text{ADC}}(t)$ is independent of the input power gain $\Gamma = \Gamma_1 \times \Gamma_2$, which is provided by an adjustable electro-mechanical attenuator of power gain $\Gamma_1 < 1$, followed by an electrical amplifier of power gain $\Gamma_2 > 1$. The acquisition noise of the ADC is modeled by a constant internally added noise power $\sigma_{\text{ADC}}^2$, which is effective at the input of the ADC. As the net power gain $\Gamma$ is digitally compensated at the output for an overall equivalent gain of one, the noise power $\sigma_{\text{acq}}^2$ of the (digital) noise "voltage" $G_{\text{acq}}(t)$ in the output signal $S_{\text{out}}(t)$ scales inversely with the net gain $\Gamma$ of the input amplifier stage. $U_{\text{FS}}^{(\text{ADC})}$ is the internal full-scale voltage of the analog-to-digital converter, and $U_{\text{FS}}$ is the corresponding full-scale voltage at the oscilloscope's input.

We now compare this simplified model to actual measurements of the digitally represented acquisition noise $G_{\mathrm{acq}}(t)$ output of the oscilloscope. To this end, we disconnect all inputs from the oscilloscope such that $S_{\mathrm{in}}(t) = 0$, and we record $S_{\mathrm{out}}(t) = G_{\mathrm{acq}}(t)$ for different input full-scale voltages $U_{\mathrm{FS}}$. The noise amplitude $\sigma_{\mathrm{acq}}$ is then calculated from the discrete samples $S_{\mathrm{out}}[k]$ by

$$\sigma_{\mathrm{acq}} = \sqrt{\frac{1}{K-1}\sum_{k=1}^{K}\left|S_{\mathrm{out}}[k]-\overline{S}\right|^2}, \qquad \overline{S} = \frac{1}{K}\sum_{k=1}^{K}S_{\mathrm{out}}[k]. \tag{S46}$$

Fig. S23 (a) shows the power spectra $\overline{\left|\tilde{G}^{(\mathrm{I})}_{\mathrm{acq},1}(f)\right|^2}$ of the noise recorded with the first oscilloscope channel for full-scale voltages $U_{\mathrm{FS}} \propto 1/\sqrt{\Gamma}$ between 40 mV and 480 mV. The sharp decrease of the noise power spectral density above 100 GHz is caused by a digital filter that is built into the oscilloscope. Figure S23 (b) shows the relationship between $\sigma_{\mathrm{acq}}$ and $U_{\mathrm{FS}}$ for all eight oscilloscope channels, the first four channels having a nominal bandwidth of 100 GHz and the last four a nominal bandwidth of 80 GHz. The discrete steps result from discrete settings of the variable attenuator and lead to a stepwise approximation of $\sigma_{\mathrm{acq}} \propto U_{\mathrm{FS}}$ in accordance with Eq. (S44). In our experiments the oscilloscope channels 1, 3, 5, and 7 are used to measure the in-phase components $I_1, I_2, I_3$, and $I_4$, whereas the channels 2, 4, 6, and 8 are used for the respective quadrature components $Q_1, Q_2, Q_3$, and $Q_4$. Note that in our experiments, the acquisition noise dominates over the shot-noise and can therefore be used to approximate the overall receiver noise $G^{(\mathrm{I})}_\nu(t)$, $G^{(\mathrm{Q})}_\nu(t)$ as introduced in Eq. (S6) in Section 1.1 above, $G^{(\mathrm{I})}_\nu(t) \approx G^{(\mathrm{I})}_{\mathrm{acq},\nu}(t)$ and $G^{(\mathrm{Q})}_\nu(t) \approx G^{(\mathrm{Q})}_{\mathrm{acq},\nu}(t)$.

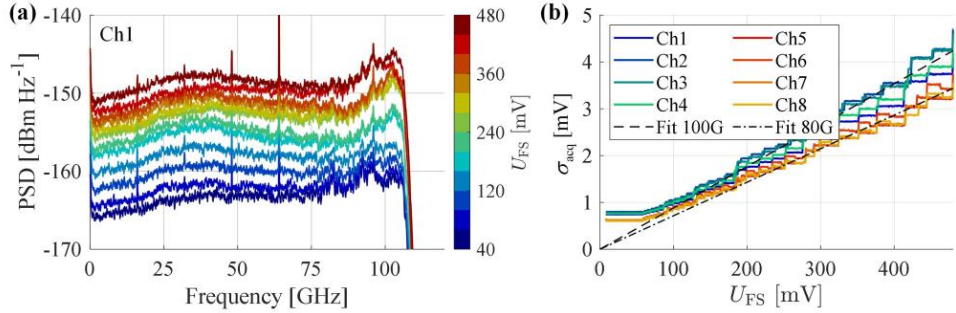

**Fig. S23**. Oscilloscope noise measurements. **(a)** Power spectrum of the acquisition noise of Channel 1 (Ch1) for different full-scale voltages $U_{\mathrm{FS}}$. The sharp decrease of the noise power spectral density above 100 GHz is caused by a digital filter that is built into the oscilloscope. **(b)** Noise amplitude $\sigma_{\mathrm{acq}}$ as a function of the full-scale voltage $U_{\mathrm{FS}}$. Note that Ch1 to Ch4 have a digitally limited nominal bandwidth of 100 GHz, whereas Ch5 to Ch8 have a digitally limited nominal bandwidth of 80 GHz leading to a smaller average slope. For $U_{\mathrm{FS}} < 58\,\mathrm{mV}$ the oscilloscope only applies digital magnification [20] and the acquisition range of the internal ADC is not fully used. Thus, $\sigma_{\mathrm{acq}}$ does not decrease further.

## 9. Considerations regarding the peak-to-average power ratio (PAPR)

The SNR of full-scale signals recorded with ADC sensitively depend on the PAPR of the captured signals, i.e., signals with lower PAPR can be measured with better SNR, see Eq. (S45). It is hence interesting to understand how the PAPR changes when superimposing $M$ spectral slices, as in the slice-less OAWM receiver, see Fig. 1 (b) in the main manuscript, compared to receiving a single slice or a single WDM channel with a single IQ receiver. In the following, we compare the resulting PAPR of the superposition signals $I_\nu(t)$ and $Q_\nu(t)$, $\nu=1,...,N$ with the PAPR of the individual time-domain waveforms associated with the individual spectral slices in a first step. In a second step, we then investigate signal-clipping as a method to reduce the PAPR and we explain how to choose an optimum acquisition range that provides an optimum tradeoff between clipping-related distortions and acquisition noise.

The spectra $\tilde{I}_\nu(f)$ and $\tilde{Q}_\nu(f)$ of the superposition signals $I_\nu(t)$ and $Q_\nu(t)$, $\nu=1,...,N$ can be calculated using Eqs. (S12) and (S13) in Sect. 1.1. As the calculations for $I_\nu(t)$ and $Q_\nu(t)$ are identical, we will limit the subsequent discussion to the in-phase components $I_\nu(t)$ – the relations for the quadrature components are obtained by simply replacing the symbol *I* by *Q*. If we neglect the noise spectra $\tilde{G}_\nu^{(I)}(f)$ in Eq (S12) and define the single-slice spectra $\underline{\tilde{L}}_{\nu\mu}(f) = \underline{\tilde{H}}_{\nu\mu}^{(I,t)}(f)\tilde{a}_S(f+f_\mu)$, we can simplify Eq (S12),

$$\tilde{I}_\nu(f) = \sum_{\mu=1}^{M}\left[\underline{\tilde{L}}_{\nu\mu}(f) + \underline{\tilde{L}}_{\nu\mu}^*(-f)\right], \qquad \underline{\tilde{L}}_{\nu\mu}(f) = \underline{\tilde{H}}_{\nu\mu}^{(I,t)}(f)\tilde{a}_S(f+f_\mu). \tag{S47}$$

By taking the inverse Fourier transform of Eq. (S47), we obtain the time-domain superposition signals $I_\nu(t)$ as the sum of the time-domain waveforms $I_{\nu\mu}(t) = \Re\{\underline{L}_{\nu\mu}(t)\}$ associated with the single-slice spectra $\underline{\tilde{L}}_{\nu\mu}(f)$,

$$I_\nu(t) = \sum_{\mu=1}^{M} I_{\nu\mu}(t), \qquad I_{\nu\mu}(t) = \Re\{\underline{L}_{\nu\mu}(t)\} = \Re\{\mathcal{F}^{-1}\left[\underline{\tilde{L}}_{\nu\mu}(f)\right]\}. \tag{S48}$$

To simplify the subsequent argument, we further assume that all time-domain waveforms $I_{\nu\mu}(t)$ related to the individual spectral slices have the same average power $P_{\text{avg}}$, that they are uncorrelated and noise-free, and that they have the same deterministic maximum amplitude $A_{\max}$. Under these assumptions, the average power $P_{\text{avg}}^{(\text{OAWM})}$ of the superposition signals $I_\nu(t)$ increases by a factor of $M$ in comparison to the average power of the signals $I_{\nu\mu}(t)$ associated with a single slice, $P_{\text{avg}}^{(\text{OAWM})} = M \times P_{\text{avg}}$. At the same time, if all peak amplitudes of the superimposed contributions $I_{\nu\mu}(t)$ occur simultaneously, then the peak amplitude $A_{\max}^{(\text{OAWM})}$ of the associated superposition signals $I_\nu(t)$ also increase by a factor of $M$. Consequently, the PAPR of the superposition signals $I_\nu(t)$ is at most $M$ times larger than the PAPR of the individual contributions $I_{\nu\mu}(t)$,

$$\text{PAPR}\{I_\nu(t)\} \leq M \times \text{PAPR}\{I_{\nu\mu}(t)\}. \tag{S49}$$

The increase of the PAPR is visualized in Fig. S24 (a) for a simple model signal comprising $M=4$ four phase-locked spectral tones with equidistantly increasing frequency offsets $\Delta f_\mu = \Delta f_0 + \mu \times \delta f$, $\mu=1,...,M$, from the respective LO tone $f_\mu$ - the respective cos-shaped signals are illustrated in Fig. S24 (b). In this case, the superposition signal, Fig. S24 (c), has an $M=4$ times larger peak amplitude, provided that the four tones have an appropriate phase relation. At the same time, the average power increases by a factor of $M=4$, and the same applies to the PAPR. Note that for other types of signals, a significantly lower PAPR increase can be observed. If, for example, the frequency offsets from the respective LO tone are identical, $\Delta f_1 = \Delta f_2 = \Delta f_3 = \Delta f_4$, then all sinusoidal tones are down-converted to the same baseband frequency. In this case, the superposition signals are also sinusoidal signals having the same PAPR = 2 as the individual contributions. In this special case $\text{PAPR}\{I_\nu(t)\} = \text{PAPR}\{I_{\nu\mu}(t)\}$ holds.

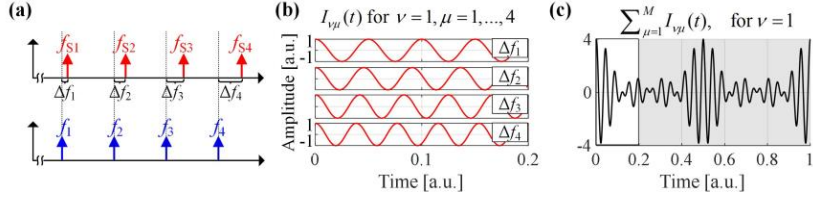

**Fig. S24.** Illustration of PAPR increase when superimposing phase-locked sinusoidal tones. **(a)** Signal spectrum (red) comprising four tones with equidistantly increasing frequency offsets $\Delta f_\mu = \Delta f_0 + \mu \times \delta f$, $\mu = 1,...,M$ from the respective LO tone $f_\mu$ (blue). **(b)** Time-domain signals associated with the individual contributions, $I_{\nu\mu}(t)$. The phase of the individual tones in (a) is chosen such that all cos-shaped contributions, $I_{\nu\mu}(t)$, have zero phase at $t = 0$, leading to a four-fold increased peak amplitude of the superposition signal. **(c)** Superposition of all cos-shaped contributions, leading to an $M = 4$ times larger peak amplitude. At the same time, the average power increases by a factor of $M = 4$, and the same applies to the PAPR; $\mathrm{PAPR}\{I_\nu(t)\} = M \times \mathrm{PAPR}\{I_{\nu\mu}(t)\}$.

So far, we have considered noise-free signals with deterministic maximum amplitude. However, practical systems are always subject to additive noise, which leads to theoretically infinitely large peak amplitudes and an undefined PAPR. For random variables it is therefore more instructive to consider the probability density functions (PDF) of the associated time-domain amplitudes and to quantify the tail of the PDF, e.g., by evaluating the fourth central moment (kurtosis) [22]. We can qualitatively associate a PDF with a strong tail and weak shoulders with a "higher PAPR" and a PDF with strong shoulders and weaker tail with a "lower PAPR". The central-limit theorem and generalizations thereof suggest that a properly normalized sum of statistically independent random variables tends to assume a Gaussian distribution even if the original random variables are not normally distributed [23]. Consequently, by superimposing several independent variables, the "PAPR" can either increase or decrease, depending on the shape of the original distributions. In our experiments, we measure noisy WDM data signals with root-raised cosine pulses (roll-off $\rho = 0.1$). In Fig. S25 (a), we show the empirical probability density functions (PDF), i.e., the properly centered and normalized histograms of the received waveforms. The histograms indicated in blue correspond to the seven recorded WDM data channels that were individually recorded with a single-channel IQ receiver using the setup shown in Fig. S17 (b) in Sect. 6 above, leading to a total of seven in-phase and seven quadrature signals and histograms. We then consider the four-slice OAWM reception of the broadband WDM test signal according to the setup Fig. S17 (a) in Sect. 6 above. In this experiment, we first analyze the histograms associated with the in-phase and the quadrature component of the 32 time-domain waveforms $I_{\nu\mu}^{(\mathrm{est})}(t)$, $Q_{\nu\mu}^{(\mathrm{est})}(t)$, $\nu = 1,...,4$, $\mu = 1,..,4$, that are obtained from the individual spectral slices, green traces in Fig. S25 (a). These waveforms are indicated by the superscript "est", since they were not directly measured but are obtained mathematically from the reconstructed signal spectrum $\tilde{\underline{a}}_S^{(\mathrm{est})}(f)$, $I_{\nu\mu}^{(\mathrm{est})}(t) = \Re\{\mathcal{F}^{-1}[\tilde{\underline{H}}_{\nu\mu}^{(\mathrm{I,t})}(f)\tilde{\underline{a}}_S^{(\mathrm{est})}(f+f_\mu)]\}$, $Q_{\nu\mu}^{(\mathrm{est})}(t) = \Re\{\mathcal{F}^{-1}[\tilde{\underline{H}}_{\nu\mu}^{(\mathrm{Q,t})}(f)\tilde{\underline{a}}_S^{(\mathrm{est})}(f+f_\mu)]\}$. The red traces in Fig. S25 (a) finally correspond to the in-phase and the quadrature components of the superposition signals $I_\nu(t)$ and $Q_\nu(t)$, $\nu = 1,...,4$ obtained from the four IQ receivers, IQR 1,…, IQR 4, see setup in Fig. S17(a) in Sect. 6 above. We can confirm that the superposition leads to a more Gaussian-like histogram (red), having longer tails and tentatively higher PAPR than the histogram associated with a single WDM data channel (blue). Each of the individual spectral slices comprises approximately two WDM channels or spectral parts thereof, which can also be interpreted as a superposition of independent signals. This also leads to more Gaussian-like histograms in comparison to the single WDM channel.

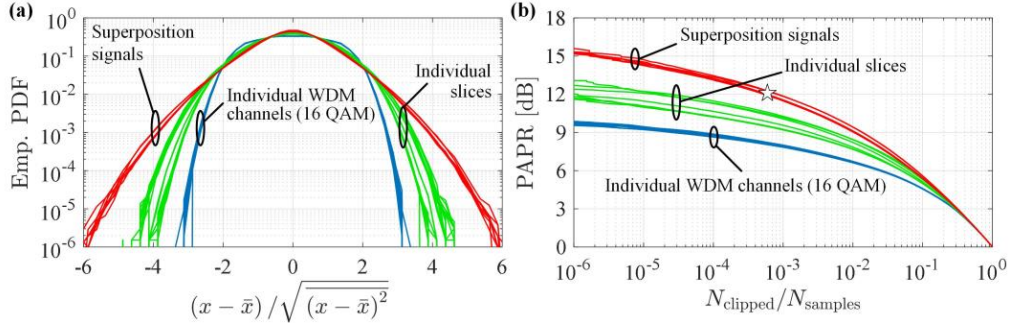

**Fig. S25.** Histogram and PAPR for the superposition signals $I_\nu(t)$, $Q_\nu(t)$, $\nu = 1,...,N$ measured with the OAWM receiver (red), the time domain waveforms $I_{\nu\mu}^{(\text{est})}(t)$, $Q_{\nu\mu}^{(\text{est})}(t)$, $\nu = 1,...,N$, $\mu = 1,...,M$, associated with the individual spectral slices (green), and the signals obtained when individually detecting 40 GBd and 60 GBd 16 QAM WDM channels using a single-channel IQ receiver (blue), see Fig. S17 in in Sect. 6 above for the measurement setups. **(a)** Histograms of the time-domain samples $x$ obtained from the various signals. The histograms are centered with respect to the average level $\bar{x}$ and normalized to the standard deviation $\sigma = \sqrt{\overline{(x-\bar{x})^2}}$, where the overbar indicates an expectation value. Comparing the red trace to the green or the green trace to the blue, we find that the superposition of several data signals or of parts thereof leads to more Gaussian-like histograms with longer tails and tentatively higher PAPR. **(b)** Peak-to-average power ratio (PAPR) of clipped signals as a function of the empirical clipping probability. The star indicates the optimum empirical clipping probability of 0.06% for the superposition signals, providing an ideal trade-off between acquisition noise on the one hand and clipping-induced distortions on the other hand, see Fig. S26.

We further measure the PAPR of the various signals, which, in presence of noise and other distortions in the experiment, is difficult, as the extracted PAPR strongly depends on rarely occurring outliers. To overcome this problem, we consider clipped signals, where the highest occurring amplitudes are simply cut off. Specifically, we assume a certain full-scale voltage $U_{\text{FS}}$ of the ADC and we clip the signal to the associated acquisition range $[-U_{\text{FS}}/2 \ \ U_{\text{FS}}/2]$. We then measure the PAPR of the clipped signals as a function of the empirical clipping probability, which is the ratio $N_{\text{clipped}}/N_{\text{sampels}}$ of the number of clipped samples $N_{\text{clipped}}$ and total number of samples $N_{\text{samples}}$. Figure S25 (b) shows the corresponding traces for the seven WDM data channels that were individually recorded with a single-channel IQ receiver (blue), the time-domain waveforms $I_{\nu\mu}^{(\text{est})}(t)$, $Q_{\nu\mu}^{(\text{est})}(t)$, $\nu = 1,...,4$, $\mu = 1,..,4$, associated with the individual spectral slices of the OAWM experiment (green), as well as the eight associated superposition signals $I_\nu(t)$ and $Q_\nu(t)$, $\nu = 1,...,4$ (red). We can observe that for low clipping probability, the PAPR associated with the superposition signals (red) is 5.5 dB higher than the PAPR of a single WDM channel (blue), which is in accordance with the fact that the normalized histograms of the superposition signals in Fig. S25 (a) (red) appears broader the than the ones for the individual WDM channels (blue). Similarly, for a give clipping ratio, the PAPR of the superposition signal is bigger than that of the individual slices, in accordance with the shape of associated the histograms in Fig. S25 (a). Due to the longer tail of the red histogram, clipping is expected to be more effective for the superposition signals, as a significant PAPR reduction can be obtained when clipping a only small number of samples.

Note that clipping does not only lead to a reduced PAPR and an associated higher SNR, see Eq. (S45) [21], but also introduces signal distortions. One should hence expect an optimum clipping ratio, leading to an ideal trade-off between clipping-induced signal distortions, which dominate at high clipping ratios, and excessive acquisition noise at low clipping ratios. We investigate this aspect for the superposition signals obtained in our OAWM experiment. To this end, we sweep the full-scale voltage $U_{FS}$ in the range from $2 \times 2\sigma_S$ to $12 \times 2\sigma_S$, where $\sigma_S$ is the standard deviation of the respective superposition signals $I_\nu(t)$ and $Q_\nu(t)$, $\nu = 1,...,4$ and we evaluate the constellation signal-to-noise-ratio (CSNR) of the seven received WDM signals. Figure S26 shows the CSNR of the four 40 GBd 16QAM and the three 60 GBd 16QAM WDM channels as a function of the normalized full-scale voltage, $U_{FS}/(2\sigma_S)$. The horizontal axis on top gives the associated average empirical clipping probability $N_{clipped}/N_{sampels}$ obtained by averaging the individual clipping probabilities of all eight superposition signals $I_\nu(t)$ and $Q_\nu(t)$, $\nu = 1,...,4$. As expected, we find a maximum CSNR for $\tfrac{1}{2}U_{FS} \approx 4\sigma_S$. Using this full-scale voltage, less than 0.06% of all recorded samples are clipped and the PAPR is limited to ~12 dB, indicated by a star in Fig. S25 (b).

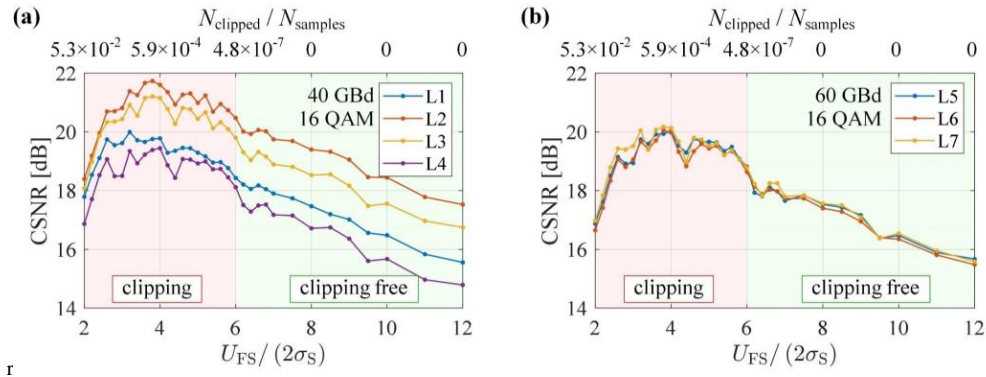

r

**Fig. S26.** Constellation signal-to-noise ratio (CSNR) for reconstructed data signals L1…L7 as a function of the normalized full-scale voltage $U_{FS}/(2\sigma_S)$, $\sigma_S$ being the standard deviation of the signal measured in the respective ADC channel. The optical spectrum comprising all seven data signals L1, …, L7 is shown in Fig. S18(g) in Sect. 6. Subfigure (a) refers to the four 40 GBd 16QAM channels (L1, …, L4), while Subfigure (b) displays the results for the remaining three 60 GBd 16QAM channels (L5, …, L7). The acquisition noise added by the ADC increases with $U_{FS}$, see Fig. S23(b) and Eq. (S45) in Sect. 8, which deteriorates the CSNR at high full-scale voltages $U_{FS}$. Inside the red-shaded region signal clipping occurs, and the associated distortions increase with decreasing $U_{FS}$, leading to a decrease of the CSNR for small full-scale voltages $U_{FS}$. An optimum trade-off leads to a maximum CSNR for $\tfrac{1}{2}U_{FS} \approx 4\sigma_S$. In this case less than 0.06% of all recorded samples of the superposition signals are clipped, and the PAPR amounts to ~12 dB, indicated by a star in Fig. S25 (b).

**Funding.** This work was supported by the ERC Consolidator Grant TeraSHAPE (# 773248), by the EU H2020 project TeraSlice (# 863322), by the DFG projects PACE (# 403188360) and GOSPEL (# 403187440) within the Priority Programme SPP 2111, by the joint DFG-ANR project HybridCombs (# 491234846), by the DFG Collaborative Research Center (CRC) WavePhenomena (SFB 1173, Project-ID 258734477), by the BMBF project Open6GHub (# 16KISK010), by the EU H2020 Marie Skłodowska-Curie Innovative Training Network MICROCOMB (# 812818), by the Alfried Krupp von Bohlen und Halbach Foundation, by the MaxPlanck School of Photonics (MPSP), and by the Karlsruhe School of Optics & Photonics (KSOP).